\documentclass[%
 reprint,
 amsmath,amssymb,
 aps,
prx,
]{revtex4-2}

\usepackage{graphicx}
\usepackage{dcolumn}
\usepackage{bm}
\usepackage{xcolor}
\usepackage{graphicx}
\usepackage[mathlines]{lineno}
\usepackage{amsthm} 
\usepackage{nameref}
\usepackage{hyperref}
\usepackage{algorithm}
\usepackage{algpseudocode}

\newcommand{\rev}[1]{\textcolor{black}{#1}}
\def\P{{\sf P}}
\def\E{{\mathbb{E}}}
\newtheorem{theorem}{Theorem}

\begin{document}
	\title{The Limits of Inference in Complex Systems: \\
    When Stochastic Models Become Indistinguishable}
    
    \author{Javier Aguilar}
	\affiliation{Dipartimento di Fisica e Astronomia Galileo Galilei, Universit{\`a} degli Studi di Padova, via Marzolo 8, 35131 Padova, Italy}
    \affiliation{Departamento de
		Electromagnetismo y F{\'\i}sica de la Materia and \\ Instituto Carlos I
		de F{\'\i}sica Te{\'o}rica y Computacional. Universidad de Granada.
		E-18071, Granada, Spain}

    \author{Miguel A. Mu{\~n}oz}
	\affiliation{Departamento de
		Electromagnetismo y F{\'\i}sica de la Materia and \\ Instituto Carlos I
		de F{\'\i}sica Te{\'o}rica y Computacional. Universidad de Granada.
		E-18071, Granada, Spain}
             \affiliation{Senior authors sharing the  leadership of the research}

              \author{Sandro Azaele}
        \affiliation{Dipartimento di Fisica e Astronomia Galileo Galilei, Universit{\`a} degli Studi di Padova, via Marzolo 8, 35131 Padova, Italy}
	\affiliation{Istituto Nazionale di Fisica Nucleare, 35131, Padova, Italy}
	\affiliation{National Biodiversity Future Center,
		Piazza Marina 61, 90133 Palermo, Italy}
    \affiliation{Senior authors sharing the  leadership of the research}
    
\begin{abstract}
Robust inference for stochastic dynamical systems is often hampered by sparse sampling and the absence of closed-form likelihoods. We introduce a Monte Carlo path-inference framework that leverages full-path statistics and bridge processes to deliver reliable parameter estimation and model selection from coarsely sampled time series, without requiring analytical solutions. Crucially, we couple mechanistic stochastic models with their inference procedures to quantify how experimental design —specifically, sampling frequency and dataset size— governs estimator precision and model distinguishability. This analysis reveals optimal sampling regimes and sharp, resolution-dependent limits beyond which competing models become empirically indistinguishable. We validate the approach across four disparate systems ---trajectories of optically trapped particles, human microbiome dynamics, social-media topic mentions, and forest population time series— recovering parameters and identifying when inference is fundamentally constrained by measurement resolution, thereby clarifying ongoing debates about dominant noise sources in these systems. Together, these results establish path-based Monte Carlo as a practical, general tool for inference and model discrimination in complex systems and provide principled guidelines for designing measurements that maximize information under real-world constraints.
\end{abstract}

\maketitle

\section{Introduction}

Minimal stochastic models have yielded fundamental insights into the dynamics of complex systems in ecology ~\cite{boyce1992,Lande2003,azaele2016statistical,Akjouj2024}, epidemiology~\cite{Newman2002,Ferguson2005,Pastor2015}, economics~\cite{chan_empirical_1992,rolski1999}, and climate science~\cite{Manabe1967,Hasselmann1976,Lucarini2019}. Their strength lies in interpretability: equations have terms with clear meanings, and parameters carry well-defined units within constrained ranges. Such models are particularly effective at capturing broad statistical patterns.
By contrast, data-driven methods—including machine learning and artificial intelligence—often excel at detecting dataset-specific features but typically sacrifice transparency. These two perspectives —model-based and data-driven— have thus emerged as complementary pillars of modern science~\cite{Ravishankara2022,Bianconi2023,McClelland2025}. Yet each comes with limitations: stochastic models may overlook idiosyncratic details, while data-driven approaches risk opacity. 

A promising way forward is to integrate the two~\cite{Cocco2009,Cocco2022,Lluis2024,Harris2024}. Inference techniques, grounded in a long tradition of stochastic modeling~\cite{sorensen2004,Iacus2008,Craigmile2023} and applied successfully across diverse domains~\cite{Guimera2020,Frishman2020,Picot2023}, provide a theoretical foundation for incorporating data into interpretable models. This synthesis strengthens robustness, enhances trustworthiness, and extends the applicability of models to real-world problems.

Still, important challenges remain. Different models can often fit the same data, making it hard to determine the best stochastic description. The difficulty of determining stochastic descriptions to data is twofold: first, parameter estimation, i.e., choosing parameter values that faithfully capture the statistical properties of the data; and second, model distinguishability, i.e, deciding which model, among several plausible candidates, best explains the observations. Theoretical ecology offers a paradigmatic example~\cite{Coyte2015,Hoshino2020,Pasqualini2025}: despite the ecological relevance of parameters, no standard estimation framework exists, and radically different models, based on demographic or environmental fluctuations, can all reproduce empirical patterns of species abundances~\cite{serGiacomi2018,grilli2020macroecological,deBuyl,Sireci,azaele2023growth,Nature1998}. Similar ambiguities arise in statistical physics. In optical tweezers experiments, the motion of a trapped particle is commonly described by an overdamped Langevin equation in a harmonic potential~\cite{florin_1998,berg-sorensen_2004}. Yet more refined models including  hydrodynamic memory~\cite{franosch_resonances_2011} and non-conservative forces~\cite{volpe_torque_2006,volpe_brownian_2007} can remain statistically consistent with the same observations. As in ecology, distinct stochastic descriptions may therefore appear equally plausible, and distinguishing among them requires sufficient resolution and careful inference. Such ambiguities are not confined to ecology and physics but arise as well in social dynamics, epidemics, and neuroscience, where distinct models may seem equally consistent with observed data~\cite{Morita2016,Odwyer2023universal,Aguilar2023a,didomenico2024}.

A common strategy for tackling parametric inference and model distinguishability is to compare readily accessible statistics ---such as selected moments or one- and two-point distributions--- that can be computed in both models and data~\cite{Fulcher2013,Fulcher2017}. By fitting these summary statistics, one can estimate parameters and identify the most plausible model. However, a more powerful approach is to combine full path statistics with information-theoretic and Bayesian inference methods~\cite{burnham_2010}. By leveraging the complete trajectory information ---rather than selected summary measures--- this framework enhances parameter estimation accuracy and improves model distinguishability. The drawback is that path statistics typically require either rich, high-frequency datasets (enabling evaluation of path-integral likelihoods) or models with known propagators in the case of sparse sampling~\cite{Craigmile2023}. Since such conditions are rare, full path-based inference has often remained impractical, leaving summary statistics as the standard tool.

The first contribution of this work is to expand the applicability of path-statistics–based inference. We achieve this by combining a known method for generating tractable, but approximate, estimates of parameters and likelihoods with a new numerical scheme, grounded in change-of-measure theory~\cite{Asmussen2007,Klebaner2012} and bridge processes~\cite{Majumdar2015,Aguilar2022,Aguilar2024}, that systematically reduces approximation errors. By applying these methods, we pursue the central aim of this work: to show that experimental protocols decisively shape both the accuracy of parameter estimates and the ability to distinguish between candidate stochastic models. Building on this insight, we formulate an optimal protocol for parameter estimation that minimizes errors and can be readily implemented in experimental setups. We further demonstrate that experimental design plays a critical role in determining whether competing models can be discriminated at all.

Finally, we apply our framework to four datasets: optical tweezers, human microbiome, topic mentions in social media, and forest datasets, quantifying the inference limits imposed by specific sampling protocols. These case studies reveal that even simple stochastic models, when paired with appropriate inference tools, can expose situations in which the very question of which model best describes the data is ill-posed because the data lack the richness needed for discrimination. Our results suggest that optimized experimental protocols ---designed both to minimize estimation errors and to account for fundamental limits of model identifiability--- are essential for ensuring that sampling is performed in regimes where reliable inference is possible.

The remainder of the paper is organized as follows. Section~\ref{sec:models} introduces the most widely used models of population dynamics. Section~\ref{sec:inference_in_stochastic_models} contrasts \rev{non-parametric} and parametric inference approaches, highlighting the limitations of the former. Sections~\ref{sec:inference_in_stochastic_models} and~\ref{sec:change of measure} then present our inference methodology, which combines a path integral approximation with a sampling scheme based on change-of-measure theory and bridge processes. Sections~\ref{sec:optimal_dt} and~\ref{sec:distinguishability} analyze how sampling resolution influences parameter estimation and model discriminability. Finally, Section~\ref{sec:data} applies the framework to real datasets using simple stochastic models equipped with inference tools.

\section{Modeling stochastic fluctuations around stable states}
\label{sec:models}

\begin{figure}
    \centering
\includegraphics[scale=1.0]{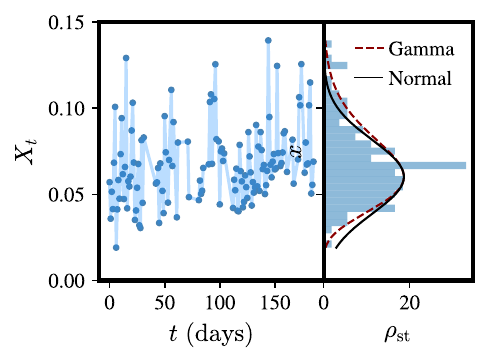}
    \caption{ \textbf{Population abundance fluctuations.} \emph{Left:} Example of a real trajectory showing the evolution of population density of a species belonging to the human gut microbiome~\cite{caporaso2011moving}. \emph{Right:} Empirical stationary distribution corresponding to the trajectory on the left. The dashed and solid lines represent Gaussian and Gamma fits, respectively.}
\label{fig:example_trajectory}
\end{figure}

Applications of inference to complex systems are diverse, ranging from estimating properties of the networks governing interactions—such as layers~\cite{lacasa_2018}, connections~\cite{martinez_survey_2017,scanagatta_survey_2019}, and communities~\cite{hastings_2006,peixoto_2019}—to learning previously unknown models that describe dynamics~\cite{Guimera2020,Frishman2020,gerardos2025principled,Harunari2022}. Here, we focus on the characterization of stationary fluctuations around stable states. The classic view of equilibrium fluctuations in physics describes a particle continuously attempting to return to a potential minimum after being perturbed by thermal noise~\cite{florin_1998}. However, many other instances of such behavior occur in nature. For example, similar dynamics are characteristic of ecological systems, where species abundances exhibit persistent and pronounced fluctuations around equilibrium~\cite{Lawton1999} (see Fig.~\ref{fig:example_trajectory}). Comparable patterns also appear in epidemiology and the social sciences, where stable states can be identified despite long-lasting variability~\cite{Aguilar2023a,Odwyer2023universal}. More broadly, this type of behavior characterizes much of population  stationary dynamics within complex systems, understood as the temporal evolution of the density of agents within a given class~\cite{Aguilar2024B} (e.g., individuals of the same species or sharing an epidemiological state).

Fluctuations around stable states are often modeled using Markovian stochastic differential equations (SDEs). Such models consist of deterministic drift terms, encoding predictable system features, together with stochastic noise terms that capture \rev{temporal} variability.  When fitted to experimental data, these models reveal key properties, including equilibrium values and characteristic fluctuation scales,  and enable predictions under perturbed conditions, providing a framework to explore the mechanisms underlying the emergence of robust macroscopic patterns across diverse environments~\cite{VanKampen1981,Gardiner2009,Risken1991,Frishman2020,grilli2020macroecological,Odwyer2023universal,Pasqualini2025}.

The minimal model for robust yet fluctuating states is the Ornstein–Uhlenbeck (OU) process, defined by its mean $\mu$, noise strength $D$, and autocorrelation time $k^{-1}$:
\begin{equation}
\label{eq:OU_process}
dX_t = k (\mu - X_t)\, dt + D\, dW_t \quad (\text{OU}).
\end{equation}
This is the canonical model of stochastic relaxation in thermal environments~\cite{florin_1998}. More broadly, it is widely used across disciplines because all of its statistical properties can be computed analytically~\cite{vankampen2007spp,Gardiner2009}. Even when the OU process does not provide a fully realistic description, it often serves as an effective approximation for a broad class of dynamics near an equilibrium point $x=\mu$, where linearization of the deterministic dynamics is valid. For example, in ecological modeling, although the OU process permits unphysical negative densities, it is nevertheless commonly employed to characterize fluctuations around ecological equilibria~\cite{Ross2006}. Owing to its simplicity and near-universal emergence as the linear approximation of stable fixed points, the OU process remains an invaluable framework for exploring fluctuations and temporal correlations without the additional complexity of fully nonlinear models.

Applications in which stationary fluctuations are strongly skewed and better described by Gamma rather than Gaussian distributions require alternatives to the OU process~\cite{azaele2016statistical,Odwyer2023universal} (see Fig.~\ref{fig:example_trajectory}). In such cases, the dynamics are more accurately captured by stochastic differential equations (SDEs) whose stationary distributions follow a Gamma form. Notably, the family of Itô SDEs
\footnote{We adopt Itô calculus to derive Eq.~\eqref{eq:general_model_theta}, but any interpretation of the noise could be used; the Stratonovich–Itô conversion~\cite{VanKampen1981} is absorbed into the definition of $\mu$.}
\begin{equation}\label{eq:general_model_theta}
dX_t = k X_t^{\theta} (\mu - X_t)\, dt + D X_t^{\frac{\theta+1}{2}}\, dW_t,
\end{equation}
where $W_t$ is a standard Wiener process~\cite{Wolfgang2014}, and with  $\theta < 2k \mu / D^2$, have gamma stationary distribution:
\begin{equation}\label{eq:st_distrib}
\rho(x) = \frac{\left(\frac{k}{2 D^2}\right)^{\frac{2k\mu}{D^2} - \theta}}{\Gamma!\left(\frac{2k\mu}{D^2} - \theta\right)} \, x^{\frac{2k\mu}{D^2} - \theta - 1} e^{-\frac{2kx}{D^2}},
\end{equation}
with $\rho(x)=\lim_{\substack{dx \to 0 \\ t \to \infty}}\frac{1}{dx}\P\left[X_t\in[x,x+dx)\right]$.

For specific values of $\theta$, Eq.~\eqref{eq:st_distrib} represents the two principal sources of multiplicative stochasticity: demographic and environmental fluctuations. Demographic noise captures random events at the individual level—birth, death, migration, predation—that introduce variability even in constant environments. Environmental noise reflects fluctuations in external factors that simultaneously affect all individuals, which can strongly shape system behavior~\cite{vankampen2007spp}, amplifying unpredictability across contexts ranging from ecology~\cite{Goldford2018} to epidemiology and social systems~\cite{CAI2018}.

In particular, for $\theta = 0$, Eq.~\eqref{eq:general_model_theta} becomes the Cox–Ingersoll–Ross model ~\cite{Feller1951}, 
\begin{equation}\label{eq:general_model_demographic}
    dX_t = k \, (\mu - X_t) \, dt + D \sqrt{X_t} \, dW_t, \quad(\text{DE})
\end{equation}
which includes a square-root noise term, typical of fluctuations in demographic processes such as a branching process \cite{diSanto}, a constant migration rate and spontaneous decay,
and serves as an approximate form of all one-step processes belonging to the directed percolation universality class~\cite{Henkel2008}.
Henceforth, we will refer to this model as the ``demographic-noise" model (DE). Since the drift in Eq.~\eqref{eq:general_model_demographic} is linear in  $X_t$, all its statistics can be derived analytically~\cite{Feller1951,Azaele2006,dornicPRL}. 

On the other hand, for $\theta = 1$, Eq.~\eqref{eq:general_model_theta} corresponds to the stochastic logistic model~\cite{grilli2020macroecological}:
\begin{equation}\label{eq:general_model_environmental}
    dX_t = k \, X_t (\mu - X_t) \, dt + D X_t \, dW_t.\quad(\text{EN})
\end{equation}
 The drift term in Eq.~\eqref{eq:general_model_environmental} represents logistic growth, which is commonly observed in isolated populations where small populations initially grow exponentially but eventually stabilize at the carrying capacity (here $\mu$)~\cite{Kiorbe_2009}, while fluctuations arise primarily from (environmental) variability affecting the linear growth rate \cite{VanKampen1981}. Henceforth, we will refer to this model as the ``environmental-noise'' model (EN).

Values of $\theta$ different from these limiting cases have been interpreted heuristically as effective forms of facilitation or cooperation in the context of population models~\cite{Hatton2024,aguade2025,Camacho2025}.
However, to the best of our knowledge, there is no microscopic model from which Eq.~\eqref{eq:general_model_theta} can be derived for general values of $\theta$. Accordingly, we do not assign a mechanistic interpretation to $\theta$. 
Rather, it should be viewed as a phenomenological parameter that interpolates between known limiting cases, allowing us to explore a family of statistical models with common stationary properties.

\section{Inference in stochastic processes}\label{sec:inference_in_stochastic_models}

Inference methods allow us to assess whether the models introduced above can reproduce the statistical patterns observed in real data and to identify which model performs best. In this study, we focus on time-series statistics, as temporal correlations often carry richer discriminatory power between competing models than static measures alone. More specifically, consider $M$ consecutive empirical measurements of an observable $x$  $x_{t_1}, x_{t_2}, \dots, x_{t_M}$, taken at times $t_1 < t_2 < \dots < t_M$. Through the text, the interpretation of the observable $x$ will vary depending on the application, being for example the position of an optically trapped particle or a population's density. For simplicity, assume these measurements are evenly spaced, so that $t_i = t_1 + (i-1)\Delta t$, where $\Delta t$ is the sampling time used in an experiment~\footnote{Throughout the text, $\Delta t$ denotes the sampling interval, i.e., the time between consecutive observations. This should not be confused with the temporal discretization used to generate sample paths from Eq.~\eqref{eq:general_model_theta}.}. These observations define a vector $\vec{x}$, with components $\vec{x}_i = x_{t_i}$. We aim to identify the model most likely to have generated the observed data \(\vec{x}\) and assess how the sampling time \(\Delta t\) and the number of measurements \(M\), which together define the experimental protocol and govern information content, jointly influence inference performance.

\subsection{Indistinguishability via moment estimators}\label{sec:quadratic_variation}

\begin{figure*}
    \centering
    \includegraphics[scale=0.85]{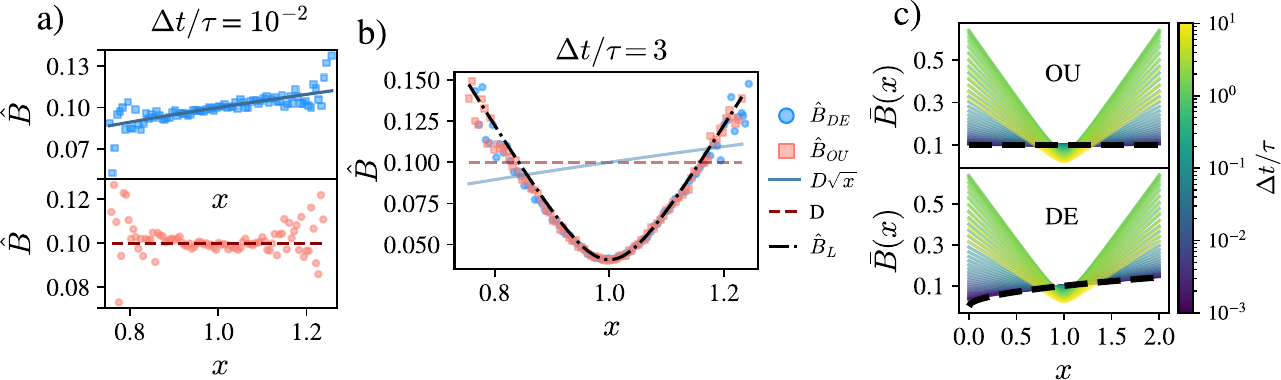}
    \caption{\textbf{Estimates of the diffusion functions signal limits of inference.} (a) Diffusion functions inferred from synthetic time series generated with the demographic noise (DE) model [Eq.\eqref{eq:general_model_demographic}] (blue, top) and the Ornstein–Uhlenbeck (OU) process [Eq.\eqref{eq:OU_process}] (red, bottom), for a small ratio of sampling interval to autocorrelation time ($\Delta t/\tau$, with $\tau = k^{-1}$). Although the OU and DE models have different stationary distributions, their parameters were tuned so that the stationary distributions share the same first and second centered moments. Dashed lines indicate the true diffusion function in the OU model, while solid lines indicate the true function in the DE model. (b) Same as (a), but for a larger $\Delta t/\tau$. The dash–dot line marks the limiting form of the diffusion estimator for large sampling intervals [Eq.~\eqref{eq:limiting_form_of_QV_estimator}]. (c) Mean of the diffusion estimator [Eq.\eqref{eq:mean_dif_estimator}] as a function of $x$ and $\Delta t$, for the OU model (top) and the DE model (bottom). Dashed lines represent the corresponding true diffusion functions. See Appendix\ref{AP_sec:B_err_df} for the analytical expression of Eq.\eqref{eq:mean_dif_estimator} and Appendix\ref{AP_sec:parameters} for parameter values.}
    \label{fig:Estimator_CV_B}
\end{figure*}

One of the preferred inference methodologies aims to identify the model that best describes the data, rather than estimating parameters within a constrained family of models (e.g., \cite{Frishman2020,Guimera2020}). This sort of \emph{non-parametric} inference applied to stochastic differential equations aims to estimate the drift and diffusion functions ($A(\cdot)$ and $B(\cdot)$ respectively) defining a generative process for the observed time series data,
\begin{equation}\label{eq:fully_general_1d_SDE}
    dX_t = A(X_t) \, dt+B(X_t)\,dW_t.
\end{equation}
Inferring the diffusion term is generally easier than inferring the drift, in the sense that estimators for the diffusion coefficient typically exhibit smaller errors than those for the drift.
 Moreover, accurately estimating the diffusion is key to identifying the correct model for population dynamics since the models in  
Eqs.~\eqref{eq:OU_process} and Eq.~\eqref{eq:general_model_theta}
 differ primarily in their diffusion functions. 
 
 Diffusion estimation relies on the relationship between the diffusion function and the second moments of the process increments. In particular, the diffusion function satisfies
 \cite{vankampen2007spp}:
\begin{equation}
    B^2(x) = \lim_{\Delta t\to0}  \frac{1}{\Delta t}\E\left[\left(X_{t+\Delta t}-x\right)^2\Big|X_t=x\right].
\end{equation}
Therefore, one can build a sample mean estimator for $B$ as follows. First,  discretize state space in bins of size $\Delta x$ such that $B_i = B(i \,\Delta x)$. Given a time series $\vec{x}$, data points will fall into the $i^\text{th}$ bin a total of $M_i\le M$ times. Defining the index set 
\begin{equation}
    J_i = \{\,j : x_{t_j}\in [\,x_i,\,x_{i}+\Delta x)\},
\end{equation}
the empirical estimator for the diffusion function  reads
\begin{equation}\label{eq:learned_diff_eq}
    \hat{B}_i= \sqrt{\frac{1}{\Delta t\, M_i}\sum_{j\in J_i}\left(x_{t_{j+1}}-x_{t_j}\right)^2},
\end{equation}
and fulfills
\begin{equation}\label{eq:errors_QV}
    B_i = \hat{B}_i \pm \sqrt{\frac{\sigma^2\left(\hat{B}_i\right)}{M_i}+\left[\epsilon_{\Delta t}\,\Delta t\right]^2+\left[\epsilon_{\Delta x}\,\Delta x\right]^2},
\end{equation}
where the three additive contributions to the error 
in Eq.~\eqref{eq:errors_QV} are independent and account for statistical uncertainty as described by the central limit theorem~\cite{Asmussen2007},
given by
$\sigma\left(\hat{B}_i\right)/\sqrt{M_i}$, along  with two sources of systematic error, $\epsilon_{\Delta t}$ and $\epsilon_{\Delta x}$: the first depending on the sampling time, $\Delta t$,
and the other arising from the state space discretization, $\Delta x$.

Importantly, we note that the error scaling shown in Eq.~~\eqref{eq:errors_QV} is valid only in the limit of small $\Delta t$. The behavior of biased errors in the estimator at any $\Delta t$ can be analyzed through its mean~\cite{Aguilar2024B},
\begin{equation}\label{eq:mean_dif_estimator}
    \bar{B}^2(x)=\E\left[\hat{B}^2(x)\right] = \frac{1}{\Delta t}\E\left[\left(X_{t+\Delta t}-x\right)^2\Big|X_t=x\right],
\end{equation}
which coincides with the empirical estimator in Eq.~\eqref{eq:learned_diff_eq} in the limit $M\to\infty$, i.e, when there are no statistical errors and all discrepancies with $B(x)$ are only due to the effect of discretizations.
As shown in Appendix~\ref{AP_sec:B_err_df}, as the sampling time $\Delta t$ increases, the mean of the estimator approaches a limiting (parabolic) form that depends only on the first two moments of the time series:
\begin{equation}
\label{eq:limiting_form_of_QV_estimator}
    \hat{B}_L =\lim_{\Delta t\to \infty} \hat{B}_i = \sqrt{\frac{\hat{\sigma}^2+\left(\hat{\mu}-x_i\right)^2}{\Delta t}},
\end{equation}
where $\hat{\mu}=\sum_i^Mx_{t_i}/M$, and $\hat{\sigma}^2=\sum_i^M( x^2_{t_i}-\hat{\mu}^2)/M$. 
Therefore, when the sampling interval $\Delta t$ is large compared to the process autocorrelation time $\tau$, the estimator in Eq.~\eqref{eq:learned_diff_eq} will poorly approximate the true diffusion function, and the error scaling described in Eq.~\eqref{eq:errors_QV} will no longer hold.

To illustrate the implications of 
Eq.\eqref{eq:limiting_form_of_QV_estimator}, we generate synthetic time series from two different models. First, we consider the Ornstein-Uhlenbeck process in Eq.~\eqref{eq:OU_process}, which has a Gaussian stationary distribution. Second, we examine the demographic-noise process described by Eq.~\eqref{eq:general_model_demographic}, which has a Gamma stationary distribution. The parameters of both models are chosen such that the first two moments of their stationary distributions coincide.

In Fig.\ref{fig:Estimator_CV_B}-(a), we show that the diffusion estimator accurately infers the diffusion functions of the generative models ---additive in one case and multiplicative in the other--- when the sampling times are sufficiently small relative to the autocorrelation time, as predicted by Eq.~\eqref{eq:errors_QV}. However, Fig.\ref{fig:Estimator_CV_B}-(b) shows that when the sampling times are too large (as compared with the autocorrelation time), the inferred diffusion term collapses into the limiting parabolic form of Eq.~\eqref{eq:limiting_form_of_QV_estimator}. As a result, inference applied to different datasets can yield the same (incorrect) outcome if their stationary distributions have similar first and second moments and the sampling times are exceedingly large.

In the OU and DE models, the mean of the diffusion estimator, Eq.~\eqref{eq:mean_dif_estimator}, can be computed analytically, offering information about the effect of biases at any sampling time $\Delta t$ (See appendix~~\ref{AP_sec:B_err_df}). In Fig.\ref{fig:Estimator_CV_B}-(c), we show how the estimator coincides with the true form of $B$ for small $\Delta t$, while it converges to the limiting form as $\Delta t$ increases. 

Therefore, when the sampling interval —set by the resolution of the experimental setup— is too large, processes with very similar stationary distributions can become indistinguishable using the estimator in Eq.\eqref{eq:learned_diff_eq}. This is noteworthy, given that this estimator is considered among the state-of-the-art methods in stochastic inference\cite{Iacus2008,Klebaner2012,Pavliotis2014,Frishman2020,Harris2024}. An open question remains: are the limits on distinguishability at large sampling intervals specific to this estimator, or do they reflect fundamental constraints, similar to other limits identified in the literature of statistical inference~\cite{Bernardi2020,Harunari2022,fajardo_2023} and signal processing~\cite{unser_2000,oppenheim_2014}? To address this, we next consider an alternative inference framework that leverages full path statistics, rather than relying solely on moment estimates as in Eq.~\eqref{eq:learned_diff_eq}.

\subsection{Maximum likelihood and Bayesian inference methods}

Hereafter, we focus on \emph{parametric inference}, which consists in using data to determine the parameters $\vec{\Theta}$ of a family of models defined by the stochastic differential equation
\begin{equation}\label{eq:general_SDE}
dX^{(\vec{\Theta})}_t = A(X^{(\vec{\Theta})}_t,\vec{\Theta})\,dt + B(X^{(\vec{\Theta})}_t,\vec{\Theta})\,dW_t,
\end{equation}
where both the drift $A$ and diffusion $B$ functions are known and depend on the state and the parameter vector $\vec{\Theta}$. 
For example, Eq.~\eqref{eq:general_model_theta} defines a family of models characterized by the parameter vector $\vec{\Theta} = (\mu, k, D, \theta)$. The aim of parametric inference is to determine the values of $\vec{\Theta}$ that make the general model most likely to reproduce the observed time series. 

Owing to the Markov property shared by the models under consideration, the probability density of observing a particular target time series, called the path likelihood, is computed as 
\begin{equation}\label{eq:prob_time_series}
L\left(\vec{x}\Big|\,\vec{\Theta}\,\right) = \rho_{t_1}\left(x_{t_1}\Big|\,\vec{\Theta}\,\right)\prod_{i=1}^{M-1}\rho_{_{t_i,t_{i+1}}} \left(x_{t_{i+1}}\Big|x_{t_{i}},\,\vec{\Theta}\,\right),
\end{equation}
where $\rho_{t_1}$ is the initial distribution of the data and $\rho_{_{s,t}} $ is the propagator of the process, namely, the probability density for the process to execute a specific increment from $y$ to $x$ between times $s$ and $t$:
\begin{equation}\label{eq:def_propagator}
    \rho_{_{s,t}} \left(x\,|\,y,\,\vec{\Theta}\,\right)\,dx = \P\left[X^{(\vec{\Theta})}_t\in[x,x+dx)\Bigg|X^{(\vec{\Theta})}_s =y\right],
\end{equation}
and is the solution of the Fokker-Planck equation associated with Eq.~\eqref{eq:general_model_theta} when using the It\^{o} prescription
\begin{align}\label{eq:FPK-equation}
    \partial_t & \, \rho_{s,t}\,\left(x|y,\vec{\Theta}\right) = \nonumber \\
     & -\partial_x  \left[A(x,\vec{\Theta}) \,\rho_{s,t}\left(x|y,\vec{\Theta}\right)\right] +\partial^2_x  \left[\frac{B^2(x,\vec{\Theta})}{2} \,\rho_{s,t}\left(x|y,\vec{\Theta}\right)\right],
\end{align}
with the initial condition $\rho_{s,s}\,(x|y)=\delta(x-y)$. 
If the likelihood $L\left(\vec{x}\Big|\,\vec{\Theta}\,\right)$ defined through Eq.~\eqref{eq:prob_time_series} can be evaluated, then the problem of estimating the most likely model parameters reduces to solving the equation~\cite{Iacus2008}
\begin{equation}\label{eq:gradient_of_likelihood}
    \nabla_{\vec{\Theta} }\,L\left(\vec{x}\Big|\,\vec{\Theta}\,\right) = \vec{0}.
\end{equation}

The solution to Eq.~\eqref{eq:gradient_of_likelihood}, here denoted as $\hat{\vec{\Theta}}$, is the maximum likelihood estimator (MLE) of the parameters.

It is well known that maximum likelihood estimation can be viewed as a special case of Bayesian inference under a uniform prior~\cite{gelman1995bayesian}. More generally, prior information about the parameters and penalties on model complexity can be incorporated through the specification of a prior distribution~\cite{burnham_2010}. Within this Bayesian framework, one may still work with single estimators that summarize the posterior distribution under suitable saddle-point assumptions~\cite{fajardo_2023}. In particular, the maximum likelihood estimator is replaced by the maximum a posteriori (MAP) estimator~\cite{bassett_2019}, defined as the solution to
\begin{equation}\label{eq:MAP} \nabla_{\vec{\Theta} }\left[\,L\left(\vec{x}\Big|\,\vec{\Theta}\,\right) \pi_o(\vec{\Theta})\right] = \vec{0}, \end{equation}
where $\pi_o$ denotes the prior distribution.

MLE estimators are typically used when comparing models of equal complexity, serving as a reference against which formulations with non-flat priors are evaluated~\cite{van_dongen_2006}.
On the other hand, when prior information is available or the models under comparison differ in parametric complexity, the MAP framework incorporates prior knowledge and regularizes parameter estimates to mitigate overfitting.
We focus on these estimators for simplicity of presentation, although more general posterior inference—where one works with probability distributions over parameters and models rather than single point estimates—is also possible, as all these formulations rely on the same central object of our analysis: the likelihood.

Parameter estimation through the MLE and MAP methods have been shown to converge to the true parameter values in the limit of infinitely many observations, provided that the time-series likelihood can be evaluated exactly~\cite{Craigmile2023,sorensen2004,Rossi2010}. However, analytical solutions to Eq.~\eqref{eq:FPK-equation} are known only for a few specific models. In general, the time-series likelihood is accessible only through approximations, which introduce biased errors in the inference framework. Generating estimates of the time-series likelihood in feasible computational times and assessing the errors introduced by these approximations are the main objectives of current research related to the inference in the context of SDEs~\cite{Craigmile2023}.

\subsection{Path integral likelihood}\label{sec:high_frec_inference}

In practice, empirical data are always sampled at discrete times, meaning that a complete continuous trajectory can never be observed. Nevertheless, it is useful to consider this hypothetical setting because, if continuous paths were available, evaluating the likelihood and inferring the model would become far more tractable. For instance, the diffusion function could be read off directly from the quadratic variation of the path~\cite{Klebaner2012}. Likewise, the drift parameters could be inferred via MLE without solving for the process propagator, since the likelihood of continuous paths is known. Importantly, this likelihood is not obtained as the naive $\Delta t \to 0$ limit of Eq.\eqref{eq:prob_time_series}, which is mathematically ill-defined (see Appendix~\ref{AP_sec:path_measures_Girsanov}). What is well-defined, instead, is the ratio of probabilities between two stochastic processes over the same path, i.e., the Radon–Nikodym derivative~\cite{Klebaner2012,Pavliotis2014,Wolfgang2014}. This derivative quantifies how much more likely a given trajectory is under one model relative to a reference process. A detailed discussion of why only probability ratios are meaningful for continuous paths is provided in Appendix~\ref{AP_sec:path_measures_Girsanov}, while here we introduce an illustrative example.

Consider an additive-noise process defined by
\begin{equation}\label{eq:general_lamperti_transformed_model}
dX_t = A(X_t,\vec{\Theta})\,dt + dW_t,
\end{equation}
where $W_t$ is a Wiener process. This process differs from pure Brownian motion only by the presence of a drift term $A(X_t,\vec{\Theta})$, which depends on both the state of the process $X_t$ and the parameters $\vec{\Theta}$, which we aim to estimate. Consider that data consists of a continuous path of duration $T$, noted $x_{[0,T]}$.  Girsanov’s theorem~\cite{Klebaner2012} allows one to compute the Radon-Nikodym derivative between the path measure $\mathbb{P}$ of the process in Eq.~\eqref{eq:general_lamperti_transformed_model} and the path measure $\mathbb{W}$ of the driftless Brownian motion~\cite{Klebaner2012,Pavliotis2014,Wolfgang2014}. The result --- as derived in Appendix~\ref{AP_sec:path_measures_Girsanov} and easily rationalized in terms of the Onsager-Machlup path integral representation ~\cite{Langouche1982,Wio2013}--- is:
\begin{equation}\label{eq:RN_der}
   L_T=\frac{d\mathbb{P}}{d\mathbb{W}} = e^{T\ell_T},
\end{equation}
with
\begin{equation}\label{eq:RN_der_2}
\ell_T\left(\vec{\Theta}\right) = \frac{1}{T} \int_0^T A\left[x_t,\vec{\Theta}\right]\,dx_t - \frac{1}{2T} \int_0^T A^2\left[x_t,\vec{\Theta}\right]\,dt,
\end{equation}
where $x_t$ is the path at time $t\in[0,T]$.  The quantity $\ell_T$ can be interpreted as a time-averaged log-likelihood ratio, quantifying how much more likely the observed path is under the drifted process than under its driftless counterpart~\cite{Frishman2020,Klebaner2012}. In this sense, it measures the statistical discrepancy between the two processes, although it does not define a distance. Because the reference measure (driftless Brownian motion) does not depend on model parameters, we can find the MLE by maximizing $\ell_T$ with respect to the model parameter vector $\vec{\Theta}$, i.e., by solving $\nabla_{\vec{\Theta}} \ell_T = 0$. Similarly, solving $\nabla_{\vec{\Theta}} \left[T\ell_T +\log(\pi_o)\right]= 0$ one obtains the MAP estimator when imposing prior knowledge. These optimization protocols can be done analytically in many relevant cases, finding explicit expressions for the parameters of the model as path functionals of trajectories (see Appendix~\ref{AP_sec:MLE_Girsanov}).

However, as noted earlier, continuous paths are never directly observed. Consequently, the integrals appearing in the expression for the likelihood (Eq.~\eqref{eq:RN_der_2}) and in subsequent formulas for parameter estimation (e.g., Eq.~\eqref{eq:MLE_estimator_continuous_paths}) must be approximated using discretizations. When the sampling interval is small (high-frequency data), the bias introduced by this approximation is negligible~\cite{Frishman2020}. In contrast, for coarsely sampled (low-frequency) data, the bias introduced by discretization can be significant and must be carefully addressed~\cite{Ferretti2020}.

The approximations entailed by using continuous frameworks to analyze discrete datasets highlight a fundamental trade-off: discrete-time data are directly measurable, but their likelihood cannot usually be expressed in terms of model parameters because the propagator in Eq.~\eqref{eq:prob_time_series} is unknown. Continuous-time path likelihoods, by contrast, admit closed-form expressions [e.g. Eq.~\eqref{eq:RN_der_2}]—but such paths are inherently unobservable. This tension gives rise to the inference conundrum: one must choose between tractable but unmeasurable path formulations and measurable but analytically intractable discrete observations.

Approximating likelihoods and parameters within the path-integral framework provides an inexpensive way to obtain an ``educated” first guess, since it relies on evaluating analytical formulas. This initial estimate becomes exact as the discretization step decreases and can still yield accurate parameter values even for finite sampling intervals, as we will see in the next section. However, these approximations prove insufficient in applications involving model differentiability, where precise likelihood estimates under coarse sampling are required. To address this limitation, our proposed strategy (developed in Sec.~\ref{sec:change of measure}) is to construct a Monte Carlo scheme that bridges the discrete and continuous inference perspectives, using continuous-path likelihoods to approximate the likelihood of discrete-time observations with arbitrary precision.

\section{Parametric inference through path integral approximation: optimal sampling times}\label{sec:optimal_dt}

Here, we show the applicability of the path-integral approximation outlined in Sec.~\ref{sec:high_frec_inference} to estimate parameters of stochastic differential equations. Specifically, we aim to compute the parameters $\mu$, $k$, and $D$ that appear in Eqs.~\eqref{eq:OU_process} and~\eqref{eq:general_model_theta} using time series data. For all models considered, the parameters are obtained as analytical path functionals: expressions for $k$ and $\mu$ are derived as MLEs in Appendix~\ref{AP_sec:MLE_Girsanov}, while $D$ is derived through the quadratic variation in Appendix~\ref{AP_sec:QV_calculations}.

The discrete nature of time series data, along with its finite length, introduces errors into any parametric estimation procedure. Such errors depend on the inference method, but also on the experimental protocol, which defines time series properties such as the sampling time and number of data points. In order to assess the magnitude of such errors, we study our path-integral estimators in synthetic datasets generated with simulations of the different considered models. In particular, we generate time series with varying sampling intervals $\Delta t$ and different numbers of measurements $M$ for the population dynamics models presented in Sec.\ref{sec:models}. The true values of $\mu$, $K$, and $D$ are known for each time series, allowing us to compare with the path-integral estimators, here noted as $\hat{\mu}$, $\hat{K}$, and $\hat{D}$. With this information, we can evaluate the relative errors of the inference method as 
\begin{equation}\label{eq:relative_errors}
\varepsilon_{\hat{\mu}} = \frac{\mathbb{E}(\hat{\mu}) - \mu}{\mu}, \quad
\varepsilon_{\hat{k}} = \frac{\mathbb{E}(\hat{k}) - k}{k}, \quad
\varepsilon_{\hat{D}} = \frac{\mathbb{E}(\hat{D}) - D}{D}.
\end{equation}

\begin{figure}
    \centering
    \includegraphics[scale=0.8]{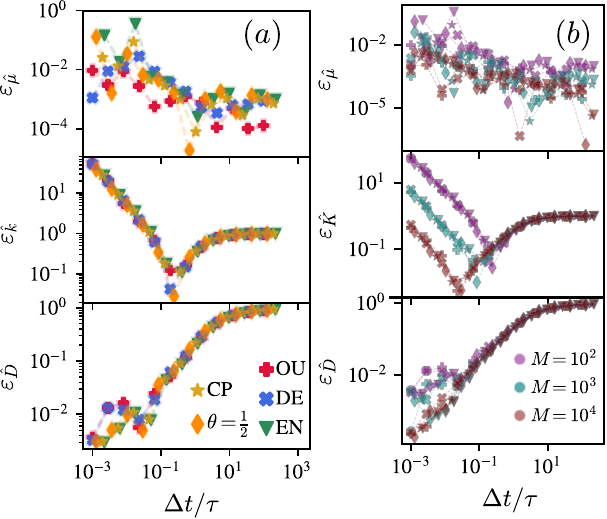}
    \caption{\textbf{Errors in parametric inference of $\mu$, $k$, and $D$ depend strongly on the experimental protocol.} Panel (a) reports the estimation errors for $\mu$, $k$, and $D$ across different models (see labels for symbols) and sampling times ($\Delta t$), using a fixed number of measurements ($M=100$). Panel (b) mirrors panel (a), but varies $M$ in addition to $\Delta t$. In (b), colors indicate different values of $M$, and symbols indicate the model.  The parameters $\mu$ and $k$ were estimated with the high-frequency method of Appendix~\ref{AP_sec:MLE_Girsanov}, while $D$ was obtained via the quadratic-variation approach of Appendix~\ref{AP_sec:QV_calculations}. Parameter values,values of $\tau$, and definition of CP model are given in Appendix~\ref{AP_sec:parameters}.}
    \label{fig:Estimation_mu_k_D}
\end{figure}

The expected value in Eq.~\eqref{eq:relative_errors} is computed by averaging over ensembles of time series generated with fixed values of $\mu$, $k$, $D$, $\Delta t$, and $M$. In Fig.~\ref{fig:Estimation_mu_k_D}, we generate time series with varying sampling intervals $\Delta t$ and different numbers of measurements $M$ for the models presented in Sec.\ref{sec:models}, and then evaluate the errors using Eqs.\eqref{eq:relative_errors}. Panel (a) of Fig.\ref{fig:Estimation_mu_k_D} shows the relative errors as a function of the  sampling interval $\Delta t$, while holding the total number of measurements constant at $M=100$, a value comparable to typical sample sizes encountered in the real datasets to be analyzed in Sec.~\ref{sec:data}. The trend of the relative errors with $\Delta t$ differs across parameters, but remains consistent across models when the sampling interval is normalized by the autocorrelation time of the process (see Appendix~\ref{AP_sec:correlation time}). Fig.~\ref{fig:Estimation_mu_k_D}-b) shows results for larger time series to shed light on the effect of increasing $M$. Once again, the relative errors vary across parameters but show consistent trends across models.

Hence, the experimental protocol, parameterized by $\Delta t$ and $M$, plays a fundamental role in tuning the errors of parametric inference (and thus the information that can be retrieved from time series analysis). Moreover, this effect is consistent across models: although the parameters $k$, $\mu$, and $D$ have different units and interpretations, they share common roles in shaping the temporal dynamics of the process. This explains the universal dependence of inference errors on $\Delta t$ and $M$. Specifically:

\begin{itemize}
    \item \textbf{$\mu$:} Relative errors in the estimation of $\mu$ decrease with increasing sampling interval $\Delta t$. This suggests that minimizing the error in $\mu$ requires sparsely sampled data with minimal temporal correlations. This aligns with the role of $\mu$ in tuning the mode of the stationary distribution, which is independent of temporal dynamics.

    \item \textbf{$D$:} In contrast to $\mu$, relative errors in estimating $D$ are minimized in the high-frequency regime and remain robust even when data points are highly correlated. This reflects the fact that $D$ governs the stochasticity of paths, which dominates the short-time-scale (highly correlated) behavior of the process.

    \item \textbf{$k$:} The relative errors in estimating $k$ are non-monotonic. Estimation is hindered both by temporal correlations at small $\Delta t$ and by systematic errors at large $\Delta t$. This is consistent with the dual role of $k$ in determining both the temporal dynamics (via the autocorrelation time) and the stationary distribution (via the balance between drift and diffusion). The optimal sampling time that minimizes the errors of this estimation greatly depends on the number of measures, as they tune the statistical errors that dominate in the high-frequency regime~\cite{Toral2014}.
\end{itemize}
 
From this information, we conclude that an effective experimental protocol for minimizing parameter estimation errors, within this approximation, should employ heterogeneous sampling strategies. High-frequency sampling over short time series improves the estimation of $D$, while low-frequency sampling enhances the accuracy of $\mu$ and $K$. If heterogeneous sampling is not feasible, a useful rule of thumb is to set $\Delta t$ close to the autocorrelation time, which offers a reasonable compromise for accurately estimating all parameters.

\section{Bridge Formalism for Accurate Propagator Estimation}
\label{sec:change of measure}

The path integral approximation of path likelihoods has been a valuable tool for parameter estimation. Yet, as we show in Sec.\ref{sec:distinguishability}, its inaccuracies limit its usefulness for model differentiation, a task highly sensitive to the discreteness of the data. Recall that evaluating discrete time-series likelihoods via Eq.\eqref{eq:prob_time_series} relies on the propagator. Beyond path integral approximations, one can obtain the propagator numerically by solving Eq.\eqref{eq:FPK-equation} or by constructing histograms from simulated trajectories \cite{Toral2014,Asmussen2007}. Both approaches, however, require state-space discretization, which introduces bias, inefficiency, and unnecessary computations across unobserved states. To overcome these issues, here we introduce an efficient Monte Carlo method that directly estimates the propagator at the observed increments—avoiding discretization altogether.

\begin{figure*}
    \centering
    \includegraphics[scale=0.9]{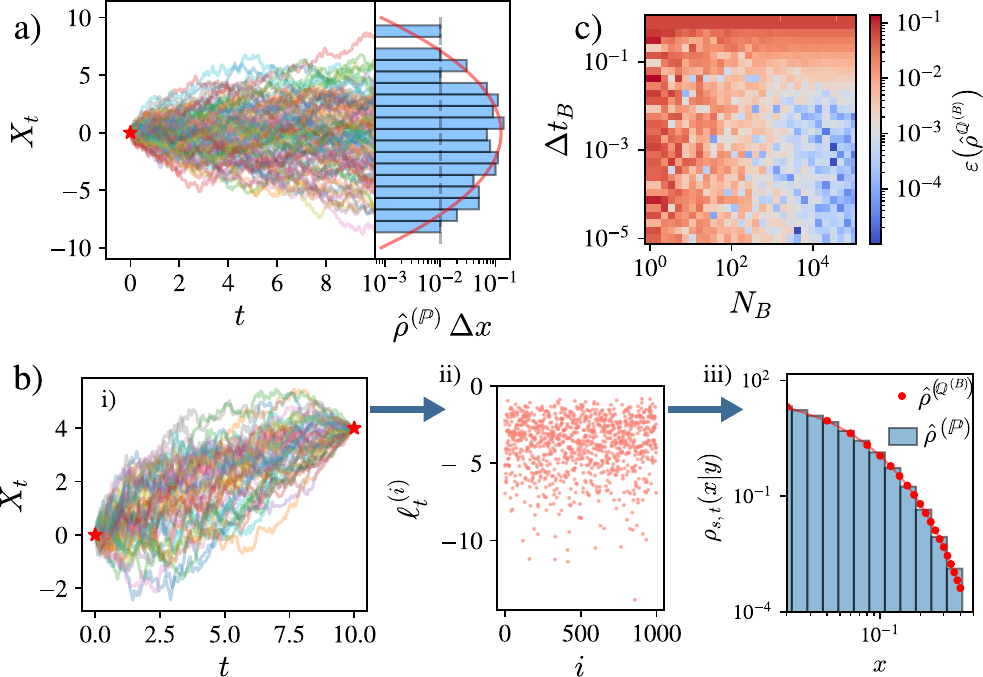}
\caption{\textbf{Propagator estimation using bridge change of measure.}   a) Trajectories of $100$ stochastic paths of a Wiener process (left), and its associated propagator estimation at time $t=10$ using the estimator of Eq.~\eqref{eq:MC_estimator_prop} (right). Note that 
    probabilities computed with this estimation method are lower bounded by the number of paths, signaled with a vertical dashed line in a), right.  Panel b) \textit{i}) shows $100$ Wiener bridges, connecting the origin with a specified final state, $X_T=4$, while \textit{ii}) shows the associated logarithm of the Radom-Nikodym derivative, computed through Eq.~\eqref{eq:RN_der_2} on every bridge. Finally, in \textit{iii}) we used bridge paths to evaluate Radon--Nikodym derivatives and estimate the propagator for the EN model~\ref{eq:general_model_environmental}. Each red dot represents the bridge change-of-measure estimator [Eq.~\eqref{eq:estimator_BCM}] with a different final condition, while the bars show the classical Monte Carlo estimator [Eq.~\eqref{eq:MC_estimator_prop}]. Both methods yield similar precision, but the bridge estimator required only \(10^3\) paths compared to \(10^5\) for standard Monte Carlo, making it far more efficient and eliminating the need for spatial discretization. Errors are on the order of the red circle size. In   c), relative errors in the propagator estimation for different values of the number of bridges ($N_B$) and bridge discretization, $\Delta t^{(B)}$. The target process is the  Ornstein-Uhlenbeck process, and we used Wiener bridges for the sampling. See parameter values in Appendix~\ref{AP_sec:parameters}.
    }
    \label{fig:BCM}
\end{figure*}

Suppose we wish to evaluate the propagator of a given process over a specific increment, from an initial state $y$ at time $t=0$, $(0, y)$, to a final state $x$ at time $t=T$, $(T, x)$.
That is, we aim to estimate
\begin{equation}\label{eq:propagator_as_expected_value}
\rho_{0,T} \left(x \mid y \right) = \mathbb{E}_\mathbb{P}\left[\delta(X_T - x)\right] = \int_0^t \delta(X_T - x) \, d\mathbb{P}(X_{[0,T]}),
\end{equation}
where $\mathbb{P}$ denotes the path measure of the process conditioned on the initial state $X_s = y$~(see Appendix~\ref{AP_sec:path_measures_Girsanov}), and $X_{[0,T]}$ is a realization of the process from $t=0$ to $t=T$.
A standard approach to estimate $\rho_{0,T}$ is to draw trajectories from $\mathbb{P}$ and approximate the Dirac delta in Eq.\eqref{eq:propagator_as_expected_value} with an indicator function over a small neighborhood of $x$, thereby obtaining a sample mean estimator for $\rho_{0,t}$:
\begin{equation}\label{eq:MC_estimator_prop}
    \hat{\rho}^{(\mathbb{P})} (x|y) = \frac{1}{\Delta x} \frac{1}{N}\sum_{i=1}^N \mathbb{I}\left(X^{(i)}_T\in\left[x,x+\Delta x\right]\right),
\end{equation}
and one can write
\begin{equation}
    \rho_{_{0,T}} \left(x\,|\,y\,\right) = \hat{\rho}^{(\mathbb{P})}(x|y)\pm\sqrt{\frac{\sigma^2\left(\hat{\rho}^{(\mathbb{P})} \right)}{N}+\left[\varepsilon_{\Delta x}\Delta x\right]^2},
\end{equation}
thus introducing a bias due to the state space discretization $\Delta x$, which is bounded by the proportionality constant $\varepsilon_{\Delta x}$. 
In Fig.~\ref{fig:BCM}-a, we illustrate this standard approach for estimating propagators via normalized histograms. Notably, this method can resolve probabilities down to the order of $1/N$, where $N$ is the number of sampled trajectories—corresponding to the smallest non-zero value that each bin can represent. An alternative approach consists of rewriting Eq.~\eqref{eq:propagator_as_expected_value} using a 
reweighting scheme or change of measure (see Appendix~\ref{AP_sec:path_measures_Girsanov}),
\begin{equation}\label{eq:propagator_as_Q_expectation}
    \rho_{_{0,T}} \left(x\,|\,y\,\right)=\E_\mathbb{Q}\left[L_T\, \delta(X_T-x)\right],
\end{equation}
where, $L_T$ is the Radon–Nikodym derivative ---analogous to that in Eq.\eqref{eq:RN_der}--- of the original process $\mathbb{P}$ with respect to a new one  $\mathbb{Q}$, connecting the initial point $(0, y)$ to the final point $(T, x)$~\footnote{Mathematically,  $\mathbb{P}$ needs to be absolutely continuous with respect to $\mathbb{Q}$, which implies that the support of the $\mathbb{Q}$-process includes (or coincides with) that of the $\mathbb{P}$-process~~\cite{Wolfgang2014,Aguilar2024,Asmussen2007}.}. This change of measure allows us to compute the expectation with respect to a different process ($\mathbb{P} \to \mathbb{Q}$) by reweighting the integrand ($\delta(X_T - x) \to L_T \delta(X_T - x)$). The corresponding Monte Carlo estimator of Eq.\eqref{eq:propagator_as_Q_expectation} is constructed analogously to Eq.\eqref{eq:MC_estimator_prop},
\begin{equation}\label{eq:Q-estimator}
    \hat{\rho}^{(\mathbb{Q})}  =  \frac{1}{N}\sum_{i=1}^N L_T^{(i)} \delta \left(X_T^{(i)}-x\right),
\end{equation}
where both $X^{(i)}_T$ and $L_T^{(i)}$ are sampled from $\mathbb{Q}$. There is considerable flexibility in choosing the reference measure $\mathbb{Q}$ \footnote{In particular, the absolute-continuity constraint can be relaxed at the cost of introducing bias~\cite{Aguilar2024}. Moreover, the auxiliary process need not belong to any particular class~\cite{Craigmile2023}}. This raises two key questions: how should $\mathbb{Q}$ be selected, and in what ways does an appropriate choice improve the estimation of the propagator?

A good choice for $\mathbb{Q}$ should satisfy the following criteria:
\begin{enumerate}
    \item The process $\mathbb{Q}$ should efficiently generate paths with fixed start and end points, since only these contribute to the sum in~\eqref{eq:Q-estimator}.

    \item Its Radon--Nikodym derivative, \(L = \frac{d\mathbb{P}}{d\mathbb{Q}}\), should be inexpensive to evaluate computationally.

    \item The generation of $\mathbb{Q}$-paths should be computationally efficient.

    \item The variance of the Radon-Nikodym derivative under $\mathbb{Q}$ should be minimized to reduce the number of paths needed for target error levels~\cite{Asmussen2007}. Ideally, $\sigma^2(\hat{\rho}^{(\mathbb{Q})}) < \sigma^2(\hat{\rho}^{(\mathbb{P})})$, lowering the computational cost for accurate estimation.
\end{enumerate}

 We propose using \emph{bridge processes}, i.e., stochastic processes conditioned on both initial and final points~\cite{Majumdar2015,Aguilar2022,Aguilar2024}, as the auxiliary process. By construction, bridges are particularly well-suited to satisfy all the desired criteria, making them an ideal choice for efficient and accurate propagator estimation.
 Any Markov process has an associated bridge process, which is the original process conditioned on passing through two points~\cite{Majumdar2015,Aguilar2022}. For instance, for a pure Brownian process
\begin{equation}\label{eq:SDE_brownian}
    dX_t = D\,dW_t,
\end{equation}
there is an associated Brownian-bridge process
\begin{equation}\label{eq:SDE_brownian_bridge}
    dX^{(B)}_t = \frac{x_T - X^{(B)}_t}{T - t}\,dt + D\,dW_t,\quad t\le T,
\end{equation}
where the drift forces paths to be at state \( (T, x_T) \), thereby generating constrained paths. In general, given a Markov process with measure \( \mathbb{Q} \), we can always find an associated bridge process, with measure \( \mathbb{Q}^{(B)} \), by means of an h-Doob transform (see Refs.~\cite{Asmussen2007,Chetrite2013, Chetrite2015} and Appendix~\ref{AP_sec:RN_wrt_bridge_measure}). 

Using stochastic bridges, point (1) is fully addressed, since all bridge paths start at $(0,y)$ and are constrained to end at $(T,x)$ (i.e., $x_T=x$). Thus, every generated path contributes to the statistics. In fact, stochastic bridges have been used to make other sampling methods, such as transition path sampling~\cite{Csajka1998}, rejection-free~\cite{orland_generating_2011,causer_rejection-free_2024}. Thus, the propagator estimator using a bridge process simplifies by removing the Dirac-delta condition, with the expectation of the Dirac-delta under $\mathbb{P}$ becoming the expectation of the Radon-Nikodym derivative under the bridge measure $\mathbb{Q^{(B)}}$
\begin{equation}\label{eq:QB_estimation}
    \rho_{_{0,T}} \left(x\,|\,y\,\right)=\E_\mathbb{P}\left[\delta(X_T-x)\right]=\E_{\mathbb{Q}^{(B)}}\left[L_T\right],
\end{equation}
so that the estimator reads
\begin{equation}\label{eq:estimator_BCM}
    \hat{\rho}^{(\mathbb{Q}^{(B)})}  =  \frac{1}{N_B}\sum_{i=1}^{N_B} L_T^{(i)},
\end{equation}
and its errors will be free from bias caused by state space discretization,
\begin{equation}\label{eq:statistical_errors_Qb_estimation}
    \rho_{_{s,t}} \left(x\,|\,y\,\right) = \hat{\rho}^{(\mathbb{Q}^{(B)})} \pm\sqrt{\frac{\sigma^2\left(\hat{\rho}^{(\mathbb{Q}^{(B)})} \right)}{N_B}}.
\end{equation}

The Radon-Nikodym derivative between the original process with measure \( \mathbb{P} \) and a bridge process with measure \( \mathbb{Q}^{(B)} \) is given by
\begin{equation}\label{eq:RN_bridge}
        L_T =\frac{d\mathbb{P}}{d\mathbb{Q}^{(B)}} =\frac{d\mathbb{P}}{d\mathbb{Q}}\frac{d\mathbb{Q}}{d\mathbb{Q}^{(B)}}= \frac{d\mathbb{P}}{d\mathbb{Q}}  \rho^{(\mathbb{Q})}_{0,T}(x | y),
\end{equation}
where we used the chain rule for Radom-Nikodym derivatives (see Appendix~\ref{AP_sec:path_measures_Girsanov}) as well as the relationship between unconstrained and constrained measures:
\begin{equation}\label{eq:theorem_change_of_measure_Q_and_Q_bridge}
    \frac{d\mathbb{Q}}{d\mathbb{Q}^{(B)}} = \rho^{(\mathbb{Q})}_{0,T}(x | y),
\end{equation}
that we proof as Theorem~\ref{AP_Th:RN_Q_Q^B} in Appendix~\ref{AP_sec:RN_wrt_bridge_measure}.
Notice that this last Radom-Nikodym derivative only depends on the initial and final states of the process and not on the entire path. On the other hand, the derivative $d\mathbb{P}/d\mathbb{Q}$ in Eq.~\eqref{eq:RN_bridge} can be computed by using Girsanov’s theorem~\cite{Klebaner2012}, as shown in Appendices~\ref{AP_sec:path_measures_Girsanov} and~\ref{AP_sec:RN_wrt_bridge_measure}. Therefore, Eq.\eqref{eq:RN_bridge} can be analytically evaluated, fulfilling requirement (2) for $\mathbb{Q}$.

Point (3) is also satisfied, since there are many methods to efficiently generate bridges ~\cite{Majumdar2015,Aguilar2022}. Furthermore, bridge trajectories for several processes can be generated exactly, reducing the need for finely discretized sampling times (see Appendix~\ref{AP_sec:exact_bridges}).

Finally, we
 discuss how to select the process $\mathbb{Q}$ to minimize its estimator variance. It is well known that an optimal change of measure exists --- precisely the bridge measure of the target process $\mathbb{P}^{(B)}$ --- for which the estimator in Eq.~\eqref{eq:estimator_BCM} has zero variance~~\cite{Asmussen2007}. From Eq.~\eqref{eq:RN_bridge}, replacing $\mathbb{Q}^{(B)}$ with $\mathbb{P}^{(B)}$ makes the estimator exactly equal to the propagator for each path generated by $\mathbb{P}^{(B)}$. Unfortunately, this optimal measure is not practically accessible since generating paths from $\mathbb{P}^{(B)}$ requires knowing the very propagator we aim to estimate. Nonetheless, this leads us to an educated choice: auxiliary processes whose bridges closely resemble those of the target process generally reduce statistical errors~~\cite{Asmussen2007}.

A step-by-step guide to using the bridge change of measure estimator for arbitrary Markov processes is provided in Appendix~\ref{AP_sec:guide_BCM}. Here, we summarize the method enumerating and discussing the key steps of the algorithm (as illustrated in Fig.~\ref{fig:BCM}-b). 

First, we generate \(N_B\) bridge paths with fixed endpoints (Fig.~\ref{fig:BCM}-b,i). To this end, we use bridges of simple processes—such as the Wiener bridge in Eq.\eqref{eq:SDE_brownian_bridge}—which can be generated exactly, that is, without bias errors (see Appendix~\ref{AP_sec:exact_bridges} and Refs.~\cite{Majumdar2015,Aguilar2024B}). Each of these bridges has an associated weight $L_T$, computed by the Radon-Nikodym derivatives in Eq.~\eqref{eq:RN_bridge} (Fig.~\ref{fig:BCM}-b,ii). Then, the Radon-Nikodym derivatives are averaged via Eq.\eqref{eq:estimator_BCM} to estimate the propagator (Fig.~\ref{fig:BCM}-b,iii). In Fig.~\ref{fig:BCM}-b,iii), we observe that the bridge change-of-measure method requires fewer paths than the standard Monte Carlo method [Eq.~\eqref{eq:MC_estimator_prop}] to achieve the same precision, making it both more efficient and more accurate (owing to the absence of bias errors). Crucially, this advantage stems from the bridge method’s ability to concentrate all sampling effort on a single endpoint, rather than distributing it across the entire propagator domain.

We emphasize that evaluating Eq.~\eqref{eq:RN_bridge} requires discretizing a path integral, which inevitably introduces bias. The crucial advantage of our method is that this bias diminishes as the resolution of the bridge‐path generation, $\Delta t_B$, becomes finer (and can in principle be made arbitrarily small). Importantly, $\Delta t_B$ and $N_B$ are not observable quantities, but computational tools that are independent of—and should not be confused with—the fixed data sampling interval and number of measures, $\Delta t$ and $M$ respectively. As shown in Fig.~\ref{fig:BCM}-c, the relative error in the propagator estimate decreases when $\Delta t_B$ is reduced and the number of bridge samples $N_B$ is increased.

\subsection{Using bridge change of measure for likelihood estimation}

To apply the bridge formalism to discrete time series, we reinterpret the generic interval $[0,T]$ as an arbitrary observation interval $[t_i,t_{i+1}]$. That is, for each pair of consecutive data points $(x_i, x_{i+1})$, we estimate the propagator by identifying $y \equiv x_i$, $x \equiv x_{i+1}$, and $T \equiv t_{i+1} - t_i$, and applying the previous construction independently on each interval. Hence, our method for accurate likelihood estimation consists of using the bridge change-of-measure technique to estimate all the propagators appearing in Eq.~\eqref{eq:prob_time_series}, $\{\rho_{t_i,t_{i+1}}(x_{i+1}\mid x_i)\}_{i=1,\dots,M-1}$. Once these likelihoods are available, inference can be performed in a principled manner, for instance within the MAP or MLE frameworks (Eqs.~\ref{eq:MAP} and~\ref{eq:gradient_of_likelihood} respectively).

In practice, parameter estimation proceeds in two stages. First, the analytical path-integral approximation of Sec.~\ref{sec:high_frec_inference} provides closed-form initial estimates at negligible computational cost. The second step is to use these initial estimations as starting points for numerical maximization of the exact likelihood (Eq.~\ref{eq:MAP}), performed, for example, via gradient descent in which each likelihood evaluation uses the bridge change-of-measure estimator.

When the dataset consists of multiple independent time series, a separate maximum likelihood estimate is obtained for each, and ensemble statistics across realizations quantify inferential uncertainty.

\section{Model distinguishability}\label{sec:distinguishability}

\begin{figure*}
    \centering
    \includegraphics[scale=1]{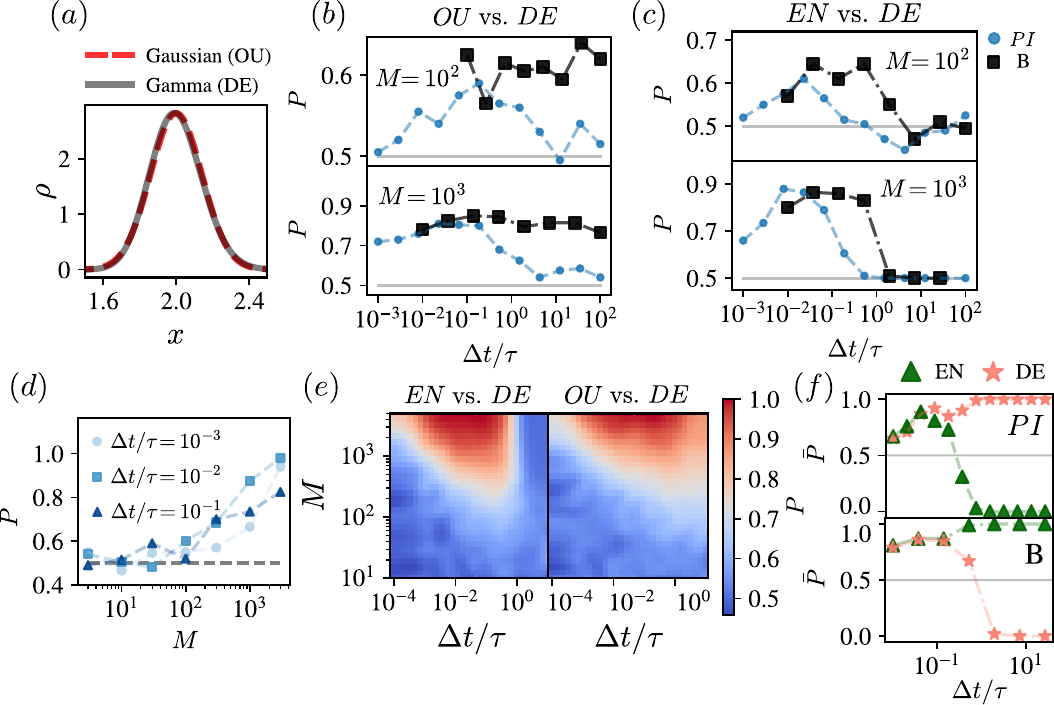}
    \caption{\textbf{Model distinguishability transition.} a) Stationary distributions for the OU and DE models (Eqs.\eqref{eq:OU_process} and \eqref{eq:general_model_demographic}, respectively) used to generate synthetic data.  b) Probability of correctly identifying the data-generating model (OU vs. DE) as a function of sampling time interval and for different number of measurements ($M=100$ top and $M=1000$ bottom), showing improved performance using the bridge change of measure MLE (B, black squares) over the path integral approximation (PI, blue dots)\rev{.}
    c), same as in panel b), but focusing in the distinguishability between EN and DE models (Eqs.\eqref{eq:general_model_environmental} and \eqref{eq:general_model_demographic}, respectively), sharing the same stationary distribution. In this case, there is a sharp drop in identification accuracy around $\Delta t = \tau$, where loss of temporal correlation limits model discrimination.   
      d) Probability of distinguishing DE and EN models for different sampling times and number of measures in high-frequency regime using path integral approximation.  e) Probability of distinguishing models as functions of $\Delta t$ and $M$ when competing models have the same stationary distribution (left panel for EN vs. DE models) and different stationary distributions (right panel for OU vs. DE models). f) Probability of correct generative model identification (EN model noted with green triangles and DE model noted with red stars) computed with path integral approximation (PI,top) and bridge change of measure (B,bottom); in the indistinguishable regime, $\Delta t\gg\tau$, inference methods systematically favor one model due to estimator-specific biases, despite theoretical equivalence, as both likelihoods are the same, Eq.~\eqref{eq:prob_time_series_uncorrelated}. 
}
\label{fig:model_distinguishability}
\end{figure*}

To ensure full control over the dynamics under study, we begin by testing model distinguishability on synthetic time series generated from known stochastic models. Given two candidate models, we evaluate which one maximizes the likelihood in Eq.~\eqref{eq:prob_time_series}, identifying it as the most probable generator of the observed data. In Appendix~\ref{AP_sec:BIC}, we show that this criterion follows from standard Bayesian model selection when comparing models of equal parametric complexity under a saddle-point approximation.

We begin by replicating the test described in Sec.\ref{sec:quadratic_variation}, generating time series with the OU and DE processes (Eqs.\eqref{eq:OU_process} and \eqref{eq:general_model_demographic}, respectively) and evaluating how well the MLE framework can identify the data-generating model. Although the models differ, their parameters are chosen so that their stationary distributions nearly overlap, as illustrated in Fig.\ref{fig:model_distinguishability}a, placing us in a parametric regime where the two remain indistinguishable at long sampling intervals when using the diffusion estimator (see Sec.\ref{sec:quadratic_variation}). Fig.~\ref{fig:model_distinguishability}b shows the probability of correctly identifying the data-generating model as a function of the sampling interval and the number of measurements, using both the path integral approximation and the bridge change-of-measure likelihood estimation. The path-integral approximation performs poorly once the sampling interval exceeds the process’s autocorrelation time, due to biases introduced by the approximation. To address this limitation, we use the path-integral result ---although insufficient on its own for model distinguishability--- as an informed initial estimate. From this starting point, we refine the parameters by numerically maximizing the exact likelihood [Eq.~\eqref{eq:prob_time_series}] with a gradient-based optimization algorithm. During this stage, bridge sampling introduced in Sec.\ref{sec:change of measure} is employed to efficiently compute the likelihood and its gradients. This procedure enables accurate and efficient likelihood approximation at arbitrary $\Delta t$. As also shown in Fig.\ref{fig:model_distinguishability}b, this approach allows the MLE framework to distinguish between OU and DE dynamics even at large $\Delta t$, thereby overcoming the shortcomings of the diffusion estimator discussed in Sec.\ref{sec:quadratic_variation}. This improvement arises because the MLE method leverages the full path statistics, capturing both dynamical and stationary information, whereas the diffusion estimator relies solely on two moments of the stationary distribution when $\Delta t$ is large (Eq.\eqref{eq:limiting_form_of_QV_estimator}).

The next test compares the EN and DE models (Eqs.~\eqref{eq:general_model_environmental} and \eqref{eq:general_model_demographic}), which share the same stationary distribution, to examine how low-frequency sampling affects model distinguishability. Fig.~\ref{fig:model_distinguishability}c shows that, although the bridge change of measure improves model differentiation over a wider range of sampling intervals than the path integral approximation, a sharp ``distinguishability transition" still occurs around $\Delta t = \tau$. This transition occurs because, for $\Delta t > \tau$, the time series becomes effectively uncorrelated, and each data point essentially represents an independent sample from the stationary distribution in which time plays no role. As a result, it becomes impossible to distinguish between models that share identical stationary fluctuations in the absence of temporal correlations. In mathematical terms, the likelihood of the time series simplifies a lot, since the probability of sampling this data becomes simply the product of stationary distributions evaluated at data points,
\begin{equation}\label{eq:prob_time_series_uncorrelated}
    L\left(\vec{x}\Big|\,\vec{\Theta}\,\right) = \rho_{_0}\left(x_0\Big|\,\vec{\Theta}\,\right)\prod_{i=1}^{M}\rho \left(x_{t_{i}},\,\vec{\Theta}\,\right),
\end{equation}
which is equal in both models.

So far, we have emphasized the impact of the sampling interval $(\Delta t)$ on model distinguishability. Another critical factor, however, is the \emph{number of observations} in the time series, which also strongly affects our ability to discriminate between models. This effect is particularly pronounced in the \emph{high-frequency regime}: even when small $\Delta t$ values generate distinct path likelihoods for models with similar stationary distributions, the strong correlation between consecutive points can produce large statistical fluctuations, limiting inference accuracy. As a result, \emph{sampling noise can mask underlying model differences}, rendering them effectively indistinguishable. This effect is illustrated in Fig.~\ref{fig:model_distinguishability}-d: synthetic trajectories sampled at high frequency become indistinguishable when the number of observations is too small, reducing the probability of correctly identifying the generative model to $50\%$, equivalent to random guessing.
Enhancing model distinguishability requires balancing the sampling interval and the number of measurements. In Fig.\ref{fig:model_distinguishability}e, we show the probability of correctly distinguishing between models as a function of both $\Delta t$ and $M$, in cases where the competing models share the same (EN-DE) or different (OU-DE) stationary distributions. At short $\Delta t$, more measurements are required because statistical errors increase due to data correlations \cite{Toral2014}. At large $\Delta t$, models with the same stationary distribution become indistinguishable, while for models with different stationary distributions, $\Delta t$ becomes irrelevant, since the data is fully uncorrelated. In Appendix~\ref{AP_sec:data_collapse}, we show data collapses of the distinguishability phase diagrams. These reveal a simple organizing principle: in the long-time regime ($\Delta t \gg \tau$), the probability of distinguishing between models depends only on the number of measurements $M$, whereas in the short-time regime ($\Delta t \ll \tau$), it is controlled by the effective observation time, given by the product $M \,\Delta t$. These arguments indicate that the presence of a non-distinguishable region is a generic feature that does not rely on prior knowledge of the true model, although the exact boundaries of this region may depend on the specific problem. Overall, our results echo familiar information-loss and aliasing effects from signal processing: when $\Delta t \gtrsim \tau$, coarse sampling suppresses dynamical information and limits what can be reliably inferred ~\cite{unser_2000,oppenheim_2014}. Unlike classical aliasing, however ---where lost frequency content cannot be recovered--- the impact here depends on  whether models differ in their stationary properties, so coarse sampling does not universally preclude model discrimination.

It is particularly informative to examine the behavior of inference methods in the regime of large sampling intervals, where the EN and DE models become indistinguishable. Although the likelihood becomes trivial in this regime [Eq.~\eqref{eq:prob_time_series_uncorrelated}], inference methods still produce biased outcomes due to unavoidable numerical and sampling errors. These small deviations lead to differences in the estimated likelihoods, even when the models are theoretically equivalent. As a result, the inference process consistently favors one model over the other---despite the absence of any genuine difference in the underlying dynamics.
To illustrate this, we compute the conditioned probabilities $\bar{P}(\text{EN})$ and $\bar{P}(\text{DE})$ that the method correctly identifies time series generated exclusively by the EN or DE model, respectively. These probabilities differ from the previously considered quantity $P$, where time series could belong to either model. For instance, $P=0.5$ can arise in different ways: $\bar{P}(\text{EN}) = \bar{P}(\text{DE}) = 0.5$ corresponds to random guessing, while $\bar{P}(\text{DE})=1$ and $\bar{P}(\text{EN})=0$ indicates a systematic bias toward DE, producing false DE-positives when data are generated by EN. As expected, in the indistinguishable regime, the inference method consistently favors one model over the other (see Fig.~\ref{fig:model_distinguishability}f). This yields correct classifications when the favored model is the true one, and incorrect classifications otherwise. Notably, the choice of favored model depends on the likelihood estimation method: under the path integral likelihood approximation, the DE model is systematically preferred, while the bridge-based likelihood estimation consistently favors the EN model. This discrepancy arises because each method introduces its own characteristic bias in the likelihood calculation. As a result, even in regimes where the models are theoretically indistinguishable, biases in the inference procedure can misleadingly favor one model over the other.

\section{Are stochastic models of complex phenomena distinguishable ?}\label{sec:data}

\begin{figure*}
    \centering
    \includegraphics[scale=0.9]{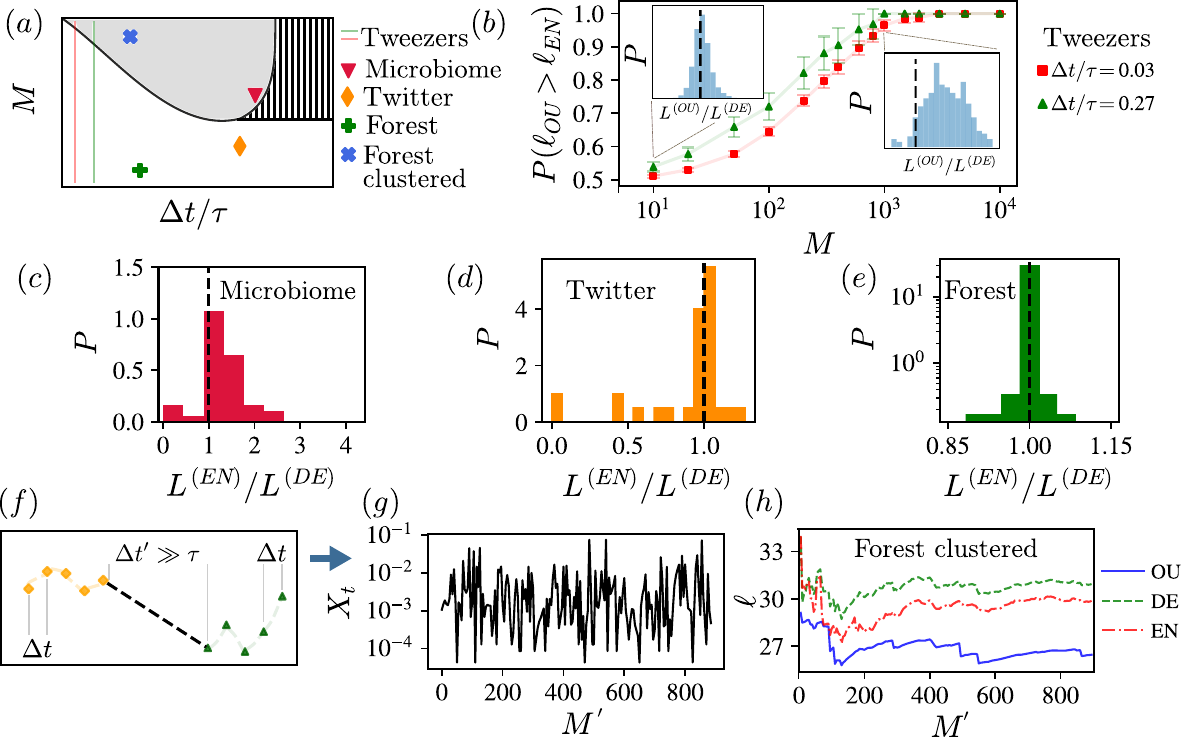}
    \caption{\textbf{Distinguishability transitions in real datasets}. a) Sketch of the distinguishability diagram from Fig.~\ref{fig:model_distinguishability}(e). The shaded region indicates where models are always distinguishable, the dashed region where only models with different stationary distributions can be distinguished, and the empty region where models cannot be differentiated. Symbols mark the approximate positions of the different datasets within the diagram. b) Probability of correctly identifying the additive nature of fluctuations in trajectory of particle trapped in optical tweezers for different numbers of measurements (x-axis) and sampling intervals (colors and symbols). c) Normalized histogram of likelihood ratios between the EN and DE models for microbial time series, showing a preference for environmental noise. d–e) Histograms of likelihood ratios for Twitter (d) and forest (e) data. In both cases, the dataset size is too small for reliable model distinguishability, as reflected in the distributions centered around $L^{(EN)}/L^{(DE)} = 1$. f) Sketch of the time-separation protocol used to merge time series from different tree species in the BCI data, based on a neutral-theory approach. g) Merged time series of neutral BCI data. h) Log-likelihood of the merged series as a function of the number of data points for the OU, EN, and DE models, indicating a preference for demographic noise. Dataset details are provided in Appendix~\ref{AP_sec:parameters}.
}
    \label{fig:data}
\end{figure*}

We next apply our framework of parametric inference and model distinguishability to four real-world datasets: trajectories of optically trapped particles, human microbiome dynamics, topic mentions in social media, and tropical-forest tree-species time series. These examples highlight the framework’s broad applicability to diverse complex systems. As we will see in the following analyses, we find varying degrees of distinguishability between competing models across datasets. Figure~\ref{fig:data}(a) provides an overview, sketching the approximate positions of the four datasets within the distinguishability diagram of Fig.~\ref{fig:model_distinguishability}.

We begin by applying our formalism to optical tweezers data 
~\cite{Ibanez2024}. Specifically, we analyze the position along one axis of a trapped particle of diameter $2.83 \,\,\mu\text{m}$, recorded over a $10\,\,\text{s}$ window at a sampling frequency of $50 \,\,\text{kHz}$. During the measurement, the trapping potential is kept constant, so the resulting time series is stationary. This dataset provides ideal conditions for inference: it is abundant ($M\sim10^5$) and strongly correlated ($\Delta t/\tau\sim0.01$). Under such conditions, the path-integral approximation of the likelihood and the parameter estimates of Sec.~\ref{sec:high_frec_inference} are essentially exact. When comparing the likelihoods of the models described in Sec.~\ref{sec:models}, the framework correctly identifies the additive-noise model, consistent with the expectation that fluctuations are thermally driven. This setting offers a natural test for the theory in Sec.~\ref{sec:distinguishability}, which predicts a transition in model distinguishability as data quality varies. Figure~\ref{fig:data}(a) illustrates this transition: when the dataset is downsampled —reducing the number of measurements while keeping the sampling interval fixed— the probability of correctly distinguishing additive from multiplicative noise evolves from $P=1.0$ (certain identification of additive noise with sufficient data) to $P=0.5$ (random guessing under scarce data). The figure also shows that in the high-frequency regime, fixing the number of measurements while increasing the sampling interval (by removing intermediate points) improves model distinguishability, since reduced correlations strengthen the inference. This behavior is in full agreement with the theoretical predictions of Sec.~\ref{sec:distinguishability}.

We now turn to datasets where the underlying dynamics are widely considered multiplicative, though the precise form of the diffusion function remains debated~\cite{grilli2020macroecological,Odwyer2023universal,azaele2023growth}. Distinguishing between models is more challenging here than in the optical tweezers data for two reasons. First, the data are of lower quality, as the time series are less correlated and relatively scarce. Second, we are comparing multiplicative models with common stationary distributions. As shown in Fig.~\ref{fig:model_distinguishability}(e), this situation leads to an abrupt transition in distinguishability around $\Delta t \approx \tau$, which can hinder our ability to determine the nature of the noise.

Our first dataset with multiplicative noise comes from human microbiome measurements in Ref.~\cite{caporaso2011moving}. It consists of Operational Taxonomic Unit (OTU) abundances—proxies for microbial species—sampled almost daily from the tongue, palm, and feces of two individuals over one year. From these data, we construct time series of relative OTU abundances, treating them as different realizations of the process $X_t$. We apply our framework of parametric inference and model distinguishability to the $N$ microbiome time series that pass the augmented Dickey–Fuller statistical test of stationarity~\cite{azaele2023growth}. For each series, and for both the DE and EN models, we estimate parameters using the procedure described in Sec.~\ref{sec:optimal_dt} and compute path likelihoods via the bridge change-of-measure method (Sec.~\ref{sec:change of measure}). This yields, for each time series $i = 1, \dots, N$, a pair of likelihoods $L^{(i,\mathrm{DE})}$ and $L^{(i,\mathrm{EN})}$. In Fig.~\ref{fig:data}(c), we show the histogram of likelihood ratios $L^{(i,\mathrm{EN})}/L^{(i,\mathrm{DE})}$. Most of the weight lies above one, indicating that $83\%$ of the time series favor the EN model, consistent with environmental fluctuations as the dominant noise source. However, this result must be interpreted cautiously. Because the data lie near the boundary between distinguishable and indistinguishable regimes ($\Delta t \approx \tau$), the likelihoods of the DE and EN models are both close to that of the common stationary-distribution sampling,Eq.~\ref{eq:prob_time_series_uncorrelated}. In this regime, inference is highly sensitive to biases in the framework, as illustrated in Fig.~\ref{fig:model_distinguishability}(f). This fact explains the diversity of works that assign either demographic or environmental fluctuations to the same microbiome data~\cite{grilli2020macroecological,Odwyer2023universal,azaele2023growth,Camacho2025}. Consequently, while daily sampling captures mesoscopic fluctuations around ecological equilibria in most microbiome time series, it may not resolve the correlations required for robust inference. Increasing the sampling frequency to the hourly scale—feasible at sites such as the palm and tongue—could substantially improve our ability to uncover the underlying stochastic dynamics~\cite{Chen2025}.

We now study two datasets together: topic references in social media and forest dynamics. Although these systems are very different, they share the feature of having only a limited number of temporal observations per species or topic, which constrains model distinguishability. On the social media side, we analyze time series of Twitter (X) hashtag mentions during major events~\cite{Zubiaga2018}. The dataset contains the timestamps and hashtags of tweets collected over several months covering events such as elections, epidemic outbreaks, and extreme weather. Tweets containing the same hashtag are aggregated into 24-h bins, yielding time series of hashtag abundances. This binning removes day–night cycles that would otherwise dominate the temporal signal. References to topics of this kind have previously been modeled as multiplicative processes~\cite{plata2021neutral}. On the ecological side, we use the Barro Colorado Island (BCI) forest dataset~\cite{condit2019complete}, which records the relative abundances of $N=178$ tree species sampled every $\Delta t = 5$ years over a 20-year period. In both datasets, the number of measurements per species or hashtag is small ($M=5$ for forest data and $M < 100$ for hashtag data). Consequently, we do not expect to distinguish reliably between competing models. As discussed in Sec.~\ref{sec:distinguishability}, when $M$ is small, statistical fluctuations in the log-likelihood dominate over genuine differences between models. Indeed, when we evaluate the likelihoods of the DE and EN models for each time series (as in the microbiome analysis) and construct a histogram of likelihood ratios, we find that the distribution is sharply peaked around $L^{(EN)}/L^{(DE)} = 1$, with more than $90\%$ of series falling in this region, as shown in Figs.~\ref{fig:data}(d),(e). This indicates that the two models are equally likely to explain the observed data.

n order to study a time series with better statistical support, we revisit the BCI forest dataset~\cite{azaele2016statistical}. While this dataset does not provide sufficient resolution to discriminate among species-level models, it remains suitable for analysis under a neutral-theory framework, in which all species are assumed to be statistically equivalent and each observed time series is treated as an independent realization of a shared underlying dynamical process. The neutral interpretation is intended to provide a global, minimal description that abstracts away species-specific contributions while capturing macroscopic statistical patterns, in the spirit of equilibrium models in statistical mechanics. In this framework, species-specific and aggregated representations can be considered as alternative perspectives for modeling, each emphasizing different levels of detail. Regarding the inference procedure, rather than assigning distinct parameters and likelihoods to each series, the neutral perspective interprets them collectively as multiple realizations of the same process. Accordingly, we aim to construct a unified model that captures the collective dynamics of all species simultaneously, rather than fitting separate models to each one. To this end, 
as illustrated in Figs.~\ref{fig:data}(f),(g),
we merge the individual species' time series into a single aggregated dataset. The merged data are treated as population time series consisting of two distinct temporal scales: a short timescale $\Delta t$, corresponding to the original sampling interval within each species' time series, and a long timescale $\Delta t'$, which separates one species' time series from the next one. We also assume that $\Delta t' \gg \tau$, where $\tau$ denotes the system's correlation time.  Under this assumption, transitions between species-specific time series can be treated as independent draws from the stationary distribution:
\begin{equation}
    \rho_{t,t+\Delta t'}(x^{(j+1)}_{t_1} \mid x^{(j)}_{t_M}) = \rho(x^{j+1}_{t_1}),
\end{equation}
where, $x^{(j)}_{t_i}$ denotes the $i^{\text{th}}$ measured relative abundance of species $j$, 
where $i = 1, \dots, M$ and $j = 1, \dots, N$. 
This construction generates a single time series with $M' =N \, M$ points, 
obtained from the $N$ original time series of $M$ points each.
This separation of temporal scales allows us to factor the full likelihood of the concatenated dataset into a product of correlated transitions within each species' time series and independent stationary transitions between them:
\begin{equation}\label{eq:likelihood_neutral_BCI}
    L(\vec{x}) = \prod_{j=1}^{N} \rho(x^{(j)}_{t_1}) \prod_{i=1}^{M-1} \rho_{t_i,t_{i}+\Delta t}(x^{(j)}_{t_{i+1}} \mid x^{(j)}_{t_i}),
\end{equation}
where $j$ indexes the species (or time series), and $i$ indexes time. In Eq.~\eqref{eq:likelihood_neutral_BCI}, $\rho(\cdot)$ and $\rho_{t,t+\Delta t}(\cdot\mid \cdot)$  corresponds to the stationary distribution [Eq.~\eqref{eq:st_distrib}] and the propagator respectively. In Fig.\ref{fig:data}h, we evaluate Eq.\eqref{eq:likelihood_neutral_BCI} for the OU, EN, and DE models as we sequentially add single-species time series to the merged neutral time series. The likelihoods of the EN and DE models consistently exceed those of the OU model, indicating that EN and DE dynamics were more likely to have generated the observed data. Ruling out the OU model highlights the multiplicative nature of the dynamics and is consistent with the fact that this model lacks a microscopic biological interpretation, although it can effectively describe population dynamics near equilibrium~\cite{Ross2006}. While the EN and DE models yield similar likelihoods in the low-data regime, as more data become available the DE model becomes statistically distinguishable: its log-likelihood exceeds that of the EN model by a gap larger than the typical magnitude of likelihood fluctuations, indicating that model selection increasingly favors demographic fluctuations.

\section{Conclusions}

In this work, we used the maximum likelihood estimator (MLE) to address problems of parametric inference and model distinguishability in one-dimensional stochastic Markovian models. Our approach does not aim to generate new classes of stochastic processes; rather, it evaluates the plausibility of a prescribed set of candidate models in light of observed data. We focus on models grounded in theoretical principles—those that are interpretable, expressed in meaningful physical units, and consistent with known symmetries. Within this framework, path inference enables a systematic comparison of competing minimal descriptions, combining interpretability with the ability to account for specific datasets.

Our main contribution to broadening the applicability of path inference is the development of a Monte Carlo scheme that enables the evaluation of propagators (and thus path likelihoods) with arbitrary precision. Although other methods for high-precision propagator evaluation exist~\cite{AitSahalia2002,AitSahalia2010,Albert2016,Craigmile2023}, our approach is relatively straightforward to implement, as it leverages the generation of stochastic bridges—a sampling technique that has seen extensive recent development~\cite{Majumdar2015,Aguilar2022}. Moreover, the method is well grounded in probability theory, allowing us to derive optimal sampling results that enhance computational efficiency. In addition to this methodological advance, we clarified the applicability and limitations of existing inference techniques by deriving exact expression for the bias in the diffusion function estimator and demonstrating the equivalence of the Girsanov~\cite{Klebaner2012,Pavliotis2014} and Gaussian (Euler–Maruyama) ~\cite{sorensen2004,Iacus2008,Craigmile2023} MLE approaches at finite sampling intervals. Importantly, while our methods cannot overcome limitations inherent to data quality or model assumptions, they are designed to extract as much information as possible from the available data. More broadly, our work underscores the importance of aligning methodological innovations with the specific demands of inference tasks—balancing accuracy, interpretability, and robustness—as inference continues to evolve as a central tool in scientific investigation.

Our primary goal was to apply these technical developments to address the challenges of parametric estimation and model distinguishability in  problems with fluctuating steady states. By applying our inference framework to fully controllable synthetic datasets, we demonstrated that inference outcomes can be misleading even in the absence of high statistical variability (e.g., in time series with many data points and without measurement errors). In particular, in the low-frequency regime, the diffusion function estimator tends toward an incorrect limiting form, and likewise, path likelihood inference can produce misleading signals when equivalent models in the stationary regime are compared. These findings underscore the need for caution when interpreting inference results, particularly in regimes with low information content—uncorrelated or with a small number of data points—as methodological artifacts may dominate over genuine signals from the data. Beyond these words of caution, there is substantial opportunity to use path inference in meaningful applications. One of the main messages from our work is that inference applied to simple, interpretable models can yield practical insights for experimental design. In particular, we can identify optimal sampling intervals that minimize errors in parameter estimation and design protocols that enhance model distinguishability. This suggests that even preliminary data can be leveraged to guide experimental design, improving the quality and informativeness of data acquisition.

We verified that the distinguishability regimes predicted by our theoretical analysis of synthetic datasets also emerge in real time series. The optical tweezers dataset is particularly rich, with abundant and correlated measurements that allow for nearly flawless inference. In such experiments, inference techniques can be applied for real-time calibration and precise control of the setup~\cite{Maragakis2008}. By degrading the quality of the optical tweezers time series (through downsampling), we confirmed the existence of a distinguishability transition tuned by data quality: the probability of distinguishing between models depends on the balance between sampling interval and number of measurements. This balance is nontrivial, since distinguishability deteriorates due to statistical errors when the dataset or the sampling times are too small, and it also breaks down when the sampling interval is so large that the data become fully uncorrelated. These sources of uncertainty became apparent when distinguishing between environmental and demographic fluctuations in the remaining datasets. Individual-species forest data and hashtag references suffer from low statistics due to limited sample sizes, while human microbiome data are heavily affected by biases introduced by temporal decorrelation. Nevertheless, our analyses reveal a clear dominance of demographic fluctuations in neutral forest data, which are sufficiently correlated to allow inference. Overall, these findings underscore the potential of our framework to disentangle the contributions of different sources of stochasticity when data quality is adequate, and to identify the minimum requirements that data acquisition should meet to ensure model distinguishability. 

A promising direction for future research is to extend this approach to high-dimensional models—for instance, by inferring interaction couplings in ecological or epidemic networks—which could provide deeper insights into system-level dynamics. An important next step would also be to generalize the framework beyond Markovian dynamics, allowing for memory effects and non-local temporal dependencies that are common in biological, ecological, and socio-economic systems~\cite{lacasa_2018}. In addition, our results could be combined with recent advances on learning closed-form stochastic dynamics in the presence of observational noise~\cite{fajardo_2023}, enabling a unified treatment that accounts explicitly for measurement errors, finite resolution, or partial observability, and thereby further enhancing the applicability of the framework to experimental data. Finally, while our analysis focused on summarizing inference through maximum likelihood estimates under uniform priors—an approach well-suited to the scope and goals of this work—an important extension would be to move toward a fully Bayesian treatment that incorporates informative, non-flat priors and propagates full posterior distributions over both parameters and models~\cite{toni_approximate_2009}. Such an extension would allow for the comparison of models with different parametric complexity, and better capture the effects of prior knowledge in data-limited regimes. Addressing these extensions would substantially broaden the scope of the method and clarify the interplay between learnability and information loss.

Importantly, the fundamental limits of inference that we observe in simple models are not expected to disappear in more complex ones; in fact, increased model complexity often introduces additional challenges such as overfitting and parameter sloppiness, which can obscure rather than clarify the underlying dynamics~\cite{Gutenkunst2007}. In this regard, simple models serve an essential role: they provide a controlled theoretical setting to rigorously evaluate inference methods, helping to identify conditions under which inference succeeds or fails, and to diagnose sources of bias, uncertainty, or instability. This reinforces the importance of strong theoretical foundations—not only for interpreting results, but also for assessing and improving the robustness of inference strategies themselves~\cite{Castro2020}.

\section{Acknowledgements} 
    We are grateful to Miguel Ibañez García and Raúl A. Rica for providing the optical tweezers data from one of their experiments. We thank Amos Maritan, Samir Suweis and Sandro Meloni for useful discussion. We also thank Alice Bertelli and Radu Diaconescu for proofreading the manuscript. This work has been supported by Grant No. 
    PID2023-149174NB-I00 financed by MICIU/AEI/10.13039/501100011033 and ERDF funds. It has also been partially funded by
    project Ref. PID2020$-$113681GB$-$I00 of the Spanish Ministry and Agencia Estatal de Investigación. 
    JA and SA acknowledge financial support under the National Recovery and Resilience Plan (NRRP), Mission 4, Component 2, CUP
2022WPHMXK, Investment 1.1, funded by the European Union – NextGenerationEU – Project Title:  ``Emergent Dynamical Patterns of Disordered
Systems with Applications to Natural Communities” . SA acknowledges the support of the NBFC to the University of Padova, funded by the Italian Ministry of University and Research, PNRR, Missione 4 Componente 2, ``Dalla ricerca all’impresa'', Investimento 1.4, Project CN00000033. The authors acknowledge the support of the Forest Global Earth Observatory (ForestGEO) of the Smithsonian Tropical Research Institute and the primary granting agencies that have supported the BCI plot. 

\section{Data and Code Availability Statement}

The code and data used in this work is publicly available at~\cite{jvrglr2025limits}.

\bibliography{refs}

@article{aguade2025,
   author = {Guim Aguadé‐Gorgorió and Ismaël Lajaaiti and Jean‐Francois Arnoldi and Sonia Kéfi},
   doi = {10.1111/oik.10980},
   issn = {0030-1299},
   issue = {1},
   journal = {Oikos},
   month = {1},
   publisher = {John Wiley and Sons Inc},
   title = {Unpacking sublinear growth: diversity, stability and coexistence},
   volume = {2025},
   url = {https://nsojournals.onlinelibrary.wiley.com/doi/10.1111/oik.10980},
   year = {2025},
}

@article{Aguilar2022,
   author = {Javier Aguilar and Joseph W. Baron and Tobias Galla and Raúl Toral},
   doi = {10.1103/PhysRevE.105.064138},
   issn = {2470-0045},
   issue = {6},
   journal = {Physical Review E},
   month = {6},
   pages = {064138},
   pmid = {35854535},
   publisher = {American Physical Society},
   title = {Sampling rare trajectories using stochastic bridges},
   volume = {105},
   url = {https://link.aps.org/doi/10.1103/PhysRevE.105.064138},
   year = {2022},
}

@article{Aguilar2023a,
   author = {Javier Aguilar and Beatriz Arregui García and Raúl Toral and Sandro Meloni and José J. Ramasco},
   doi = {10.1038/s42005-023-01302-0},
   issn = {2399-3650},
   issue = {1},
   journal = {Communications Physics},
   month = {7},
   pages = {187},
   publisher = {Springer US},
   title = {Endemic infectious states below the epidemic threshold and beyond herd immunity},
   volume = {6},
   url = {https://www.nature.com/articles/s42005-023-01302-0},
   year = {2023},
}

@article{Aguilar2024,
   author = {Javier Aguilar and Riccardo Gatto},
   doi = {10.1103/PhysRevE.109.034113},
   issn = {2470-0045},
   issue = {3},
   journal = {Physical Review E},
   month = {3},
   pages = {034113},
   pmid = {38632818},
   publisher = {American Physical Society},
   title = {Unified perspective on exponential tilt and bridge algorithms for rare trajectories of discrete Markov processes},
   volume = {109},
   url = {https://link.aps.org/doi/10.1103/PhysRevE.109.034113},
   year = {2024},
}

@article{Aguilar2024B,
   author = {Javier Aguilar and José J. Ramasco and Raúl Toral},
   doi = {10.1038/s42005-024-01648-z},
   issn = {2399-3650},
   issue = {1},
   journal = {Communications Physics},
   month = {5},
   pages = {155},
   publisher = {Nature Research},
   title = {Biased versus unbiased numerical methods for stochastic simulations},
   volume = {7},
   url = {https://www.nature.com/articles/s42005-024-01648-z},
   year = {2024},
}

@article{Arnold2000,
   author = {Peter Arnold},
   doi = {10.1103/PhysRevE.61.6099},
   issn = {1063-651X},
   issue = {6},
   journal = {Physical Review E},
   month = {6},
   pages = {6099-6102},
   title = {Symmetric path integrals for stochastic equations with multiplicative noise},
   volume = {61},
   url = {https://link.aps.org/doi/10.1103/PhysRevE.61.6099},
   year = {2000},
}

@article{AitSahalia2002,
   author = {Yacine Ait-Sahalia},
   doi = {10.1111/1468-0262.00274},
   issn = {0012-9682},
   issue = {1},
   journal = {Econometrica},
   month = {1},
   pages = {223-262},
   publisher = {Blackwell Publishing Ltd},
   title = {Maximum Likelihood Estimation of Discretely Sampled Diffusions: A Closed-form Approximation Approach},
   volume = {70},
   url = {http://doi.wiley.com/10.1111/1468-0262.00274},
   year = {2002},
}

@article{AitSahalia2010,
   author = {Yacine Aït-Sahalia and Robert L. Kimmel},
   doi = {10.1016/j.jfineco.2010.05.004},
   issn = {0304405X},
   issue = {1},
   journal = {Journal of Financial Economics},
   month = {10},
   pages = {113-144},
   title = {Estimating affine multifactor term structure models using closed-form likelihood expansions},
   volume = {98},
   year = {2010},
}

@misc{Akjouj2024,
   author = {Imane Akjouj and Matthieu Barbier and Maxime Clenet and Walid Hachem and Mylène Maïda and François Massol and Jamal Najim and Viet Chi Tran},
   doi = {10.1098/rspa.2023.0284},
   issn = {14712946},
   issue = {2285},
   journal = {Proceedings of the Royal Society A: Mathematical, Physical and Engineering Sciences},
   month = {3},
   publisher = {Royal Society Publishing},
   title = {Complex systems in ecology: a guided tour with large Lotka–Volterra models and random matrices},
   volume = {480},
   year = {2024},
}

@article{Albert2016,
   author = {Carlo Albert and Simone Ulzega and Ruedi Stoop},
   doi = {10.1103/PhysRevE.93.043313},
   issn = {24700053},
   issue = {4},
   journal = {Physical Review E},
   month = {4},
   publisher = {American Physical Society},
   title = {Boosting Bayesian parameter inference of nonlinear stochastic differential equation models by Hamiltonian scale separation},
   volume = {93},
   year = {2016},
}

@book{Asmussen2007,
   author = {Søren Asmussen and Peter W. Glynn},
   city = {New York, NY},
   doi = {10.1007/978-0-387-69033-9},
   isbn = {978-0-387-30679-7},
   journal = {Analysis},
   pages = {476},
   publisher = {Springer New York},
   title = {Stochastic Simulation: Algorithms and Analysis},
   volume = {57},
   url = {http://link.springer.com/10.1007/978-0-387-69033-9},
   year = {2007},
}

@article{Azaele2006,
   author = {Sandro Azaele and Simone Pigolotti and Jayanth R. Banavar and Amos Maritan},
   doi = {10.1038/nature05320},
   issn = {14764687},
   issue = {7121},
   journal = {Nature},
   month = {12},
   pages = {926-928},
   pmid = {17167485},
   publisher = {Nature Publishing Group},
   title = {Dynamical evolution of ecosystems},
   volume = {444},
   year = {2006},
}

@article{azaele2016statistical,
    title = {Statistical mechanics of ecological systems: Neutral theory and beyond},
  author = {Azaele, Sandro and Suweis, Samir and Grilli, Jacopo and Volkov, Igor and Banavar, Jayanth R. and Maritan, Amos},
  journal = {Rev. Mod. Phys.},
  volume = {88},
  issue = {3},
  pages = {035003},
  numpages = {31},
  year = {2016},
  month = {Jul},
  publisher = {American Physical Society},
  doi = {10.1103/RevModPhys.88.035003},
  url = {https://link.aps.org/doi/10.1103/RevModPhys.88.035003}
}

@article{azaele2023growth,
    author = {Brigatti, E. and Azaele, S.},
	date = {2025/01/22},
	doi = {10.1038/s41598-024-82882-x},
	id = {Brigatti2025},
	isbn = {2045-2322},
	journal = {Scientific Reports},
	number = {1},
	pages = {2789},
	title = {Growth-rate distributions of gut microbiota time series},
	url = {https://doi.org/10.1038/s41598-024-82882-x},
	volume = {15},
	year = {2025}
}

@misc{jvrglr2025limits,
  author       = {Javier Aguilar},
  title        = {The Limits of Inference in Complex Systems: When Stochastic Models Become Indistinguishable},
  year         = {2025},
  howpublished = {\url{https://github.com/jvrglr/The_Limits_of_Inference_in_Complex_Systems-When_Stochastic_Models_Become_Indistinguishable}},
  note         = {GitHub repository}
}

@article{Bernardi2020,
   author = {Davide Bernardi and Benjamin Lindner},
   doi = {10.1103/PhysRevE.101.062132},
   issn = {24700053},
   issue = {6},
   journal = {Physical Review E},
   month = {6},
   pmid = {32688497},
   publisher = {American Physical Society},
   title = {Receiver operating characteristic curves for a simple stochastic process that carries a static signal},
   volume = {101},
   year = {2020},
}

@article{berg-sorensen_2004,
	title = {Power spectrum analysis for optical tweezers},
	volume = {75},
	issn = {0034-6748, 1089-7623},
	url = {https://pubs.aip.org/rsi/article/75/3/594/460552/Power-spectrum-analysis-for-optical-tweezers},
	doi = {10.1063/1.1645654},
	number = {3},
	urldate = {2026-02-26},
	journal = {Review of Scientific Instruments},
	author = {Berg-S{\o}rensen, Kirstine and Flyvbjerg, Henrik},
	month = mar,
	year = {2004},
	pages = {594--612},
}

@article{bassett_2019,
	title = {Maximum a posteriori estimators as a limit of {Bayes} estimators},
	volume = {174},
	issn = {0025-5610, 1436-4646},
	url = {http://link.springer.com/10.1007/s10107-018-1241-0},
	doi = {10.1007/s10107-018-1241-0},
	number = {1-2},
	urldate = {2026-02-21},
	journal = {Mathematical Programming},
	author = {Bassett, Robert and Deride, Julio},
	month = mar,
	year = {2019},
	pages = {129--144},
}

@article{boyce1992,
 ISSN = {00664162},
 URL = {http://www.jstor.org/stable/2097297},
 author = {Mark S. Boyce},
 journal = {Annual Review of Ecology and Systematics},
 pages = {481--506},
 publisher = {Annual Reviews},
 title = {Population Viability Analysis},
 urldate = {2025-07-11},
 volume = {23},
 year = {1992}
}

@book{burnham_2010,
  address   = {New York, NY},
  edition   = {2},
  title     = {Model Selection and Multimodel Inference: A Practical Information-Theoretic Approach},
  isbn      = {978-0-387-95364-9, 978-1-4419-2973-0},
  publisher = {Springer},
  author    = {Burnham, Kenneth P. and Anderson, David R.},
  year      = {2010}
}

@article{Harris2024,
   author = {Brendan Harris and Leonardo L. Gollo and Ben D. Fulcher},
   doi = {10.1103/PhysRevX.14.031021},
   issn = {2160-3308},
   issue = {3},
   journal = {Physical Review X},
   month = {8},
   pages = {031021},
   publisher = {American Physical Society},
   title = {Tracking the Distance to Criticality in Systems with Unknown Noise},
   volume = {14},
   url = {https://link.aps.org/doi/10.1103/PhysRevX.14.031021},
   year = {2024},
}

@article{Ibanez2024,
	author = {Ib{\'a}{\~n}ez, M. and Dieball, C. and Lasanta, A. and Godec, A. and Rica, R. A.},
	date = {2024/01/01},
	doi = {10.1038/s41567-023-02269-z},
	id = {Ib{\'a}{\~n}ez2024},
	isbn = {1745-2481},
	journal = {Nature Physics},
	number = {1},
	pages = {135--141},
	title = {Heating and cooling are fundamentally asymmetric and evolve along distinct pathways},
	url = {https://doi.org/10.1038/s41567-023-02269-z},
	volume = {20},
	year = {2024}}

@article{Bianconi2023,
doi = {10.1088/2632-072X/ac7f75},
year = {2023},
month = {jan},
publisher = {IOP Publishing},
volume = {4},
number = {1},
pages = {010201},
author = {Bianconi, Ginestra and Arenas, Alex and Biamonte, Jacob and Carr, Lincoln D and Kahng, Byungnam and Kertesz, Janos and Kurths, Jürgen and Lü, Linyuan and Masoller, Cristina and Motter, Adilson E and Perc, Matjaž and Radicchi, Filippo and Ramaswamy, Ramakrishna and Rodrigues, Francisco A and Sales-Pardo, Marta and San Miguel, Maxi and Thurner, Stefan and Yasseri, Taha},
title = {Complex systems in the spotlight: next steps after the 2021 Nobel Prize in Physics},
journal = {Journal of Physics: Complexity}
}

@article{CAI2018,
title = {Environmental variability in a stochastic epidemic model},
journal = {Applied Mathematics and Computation},
volume = {329},
pages = {210-226},
year = {2018},
issn = {0096-3003},
doi = {https://doi.org/10.1016/j.amc.2018.02.009},
url = {https://www.sciencedirect.com/science/article/pii/S0096300318301036},
author = {Yongli Cai and Jianjun Jiao and Zhanji Gui and Yuting Liu and Weiming Wang}
}

@article{caporaso2011moving,
  title={Moving pictures of the human microbiome},
  author={Caporaso, J Gregory and Lauber, Christian L and Costello, Elizabeth K and Berg-Lyons, Donna and Gonzalez, Antonio and Stombaugh, Jesse and Knights, Dan and Gajer, Pawel and Ravel, Jacques and Fierer, Noah and others},
  journal={Genome biology},
  volume={12},
  pages={1--8},
  year={2011},
  publisher={Springer}
}

@article{Camacho2025,
   author = {José Camacho-Mateu and Aniello Lampo and Mario Castro and José A. Cuesta},
   doi = {10.1103/PhysRevE.111.044404},
   issn = {2470-0045},
   issue = {4},
   journal = {Physical Review E},
   month = {4},
   pages = {044404},
   publisher = {American Physical Society},
   title = {Microbial populations hardly ever grow logistically and never sublinearly},
   volume = {111},
   url = {https://link.aps.org/doi/10.1103/PhysRevE.111.044404},
   year = {2025},
}

@article{Castro2020,
   author = {Mario Castro and S\'a Ul Ares and Jos\'e Jos\'e and Jos\'e A. Cuesta and Susanna Manrubia},
   doi = {10.1073/pnas.2007868117/-/DCSupplemental.y},
   issue = {42},
   pages = {26190-26196},
   publisher = {PNAS},
   title = {The turning point and end of an expanding epidemic cannot be precisely forecast},
   volume = {117},
   year = {2020},
   journal = {Proceedings of the National Academy of Sciences},
}

@article{causer_rejection-free_2024,
	title = {Rejection-free quantum {Monte} {Carlo} in continuous time from transition path sampling},
	volume = {109},
	issn = {2469-9950, 2469-9969},
	url = {https://link.aps.org/doi/10.1103/PhysRevB.109.024307},
	doi = {10.1103/PhysRevB.109.024307},
	number = {2},
	urldate = {2026-03-22},
	journal = {Physical Review B},
	author = {Causer, Luke and Sfairopoulos, Konstantinos and Mair, Jamie F. and Garrahan, Juan P.},
	month = jan,
	year = {2024},
	pages = {024307},
}

@article{Csajka1998,
   author = {Christoph Dellago and Peter G. Bolhuis and Félix S. Csajka and David Chandler},
   doi = {10.1063/1.475562},
   issn = {0021-9606},
   issue = {5},
   journal = {The Journal of Chemical Physics},
   month = {2},

   pages = {1964-1977},
   title = {Transition path sampling and the calculation of rate constants},
   volume = {108},
   url = {https://pubs.aip.org/jcp/article/108/5/1964/182970/Transition-path-sampling-and-the-calculation-of},
   year = {1998},
}

@article{chan_empirical_1992,
  author       = {Chan, K. C. and Karolyi, G. A. and Longstaff, F. A. and Sanders, A. B.},
  title        = {An Empirical Comparison of Alternative Models of the Short‑Term Interest Rate},
  journal      = {Journal of Finance},
  volume       = {47},
  number       = {3},
  pages        = {1209--1227},
  year         = {1992},
  doi          = {10.1111/j.1540-6261.1992.tb04011.x},
}

@article{Chetrite2013,
   author = {Raphaël Chetrite and Hugo Touchette},
   doi = {10.1103/PhysRevLett.111.120601},
   issn = {0031-9007},
   issue = {12},
   journal = {Physical Review Letters},
   month = {9},
   pages = {120601},
   title = {Nonequilibrium Microcanonical and Canonical Ensembles and Their Equivalence},
   volume = {111},
   url = {https://link.aps.org/doi/10.1103/PhysRevLett.111.120601},
   year = {2013},
}

@article{Chetrite2015,
   author = {Raphaël Chetrite and Hugo Touchette},
   doi = {10.1007/s00023-014-0375-8},
   issn = {1424-0637},
   issue = {9},
   journal = {Annales Henri Poincaré},
   month = {9},
   pages = {2005-2057},
   publisher = {Springer},
   title = {Nonequilibrium Markov Processes Conditioned on Large Deviations},
   volume = {16},
   url = {http://link.springer.com/10.1007/s00023-014-0375-8},
   year = {2015},
}

@misc{Craigmile2023,
   author = {Peter Craigmile and Radu Herbei and Ge Liu and Grant Schneider},
   doi = {10.1002/wics.1585},
   issn = {19390068},
   issue = {2},
   journal = {Wiley Interdisciplinary Reviews: Computational Statistics},
   month = {3},
   publisher = {John Wiley and Sons Inc},
   title = {Statistical inference for stochastic differential equations},
   volume = {15},
   year = {2023},
}

@article{Chen2025,
   author = {Xiaowen Chen and Kyle Crocker and Seppe Kuehn and Aleksandra M. Walczak and Thierry Mora},
   doi = {10.1103/8gkj-rdzy},
   issn = {2835-8279},
   issue = {2},
   journal = {PRX Life},
   month = {6},
   pages = {023019},
   publisher = {American Physical Society (APS)},
   title = {Inferring Resource Competition in Microbial Communities from Time Series},
   volume = {3},
   url = {https://link.aps.org/doi/10.1103/8gkj-rdzy},
   year = {2025},
}

@article{Cocco2009,
   author = {Simona Cocco and Stanislas Leibler and Rémi Monasson},
   doi = {10.1073/pnas.0906705106},
   issn = {0027-8424},
   issue = {33},
   journal = {Proceedings of the National Academy of Sciences},
   month = {8},
   pages = {14058-14062},
   title = {Neuronal couplings between retinal ganglion cells inferred by efficient inverse statistical physics methods},
   volume = {106},
   url = {https://pnas.org/doi/full/10.1073/pnas.0906705106},
   year = {2009},
}

@book{Cocco2022,
   author = {Simona Cocco and Rémi Monasson and Francesco Zamponi},
   doi = {10.1093/oso/9780198864745.001.0001},
   isbn = {0198864744},
   month = {9},
   publisher = {Oxford University PressOxford},
   title = {From Statistical Physics to Data-Driven Modelling},
   url = {https://academic.oup.com/book/44725},
   year = {2022},
}

@article{Coyte2015,
   author = {Katharine Z. Coyte and Jonas Schluter and Kevin R. Foster},
   doi = {10.1126/science.aad2602},
   issn = {0036-8075},
   issue = {6261},
   journal = {Science},
   month = {11},
   pages = {663-666},
   title = {The ecology of the microbiome: Networks, competition, and stability},
   volume = {350},
   url = {https://www.science.org/doi/10.1126/science.aad2602},
   year = {2015},
}

@article{condit2019complete,
  title={Complete data from the Barro Colorado 50-ha plot: 423617 trees, 35 years},
  author={Condit, Richard and P{\'e}rez, Rolando and Aguilar, Salomon and Lao, Suzanne and Foster, Robin and Hubbell, Stephen},
  journal={(No Title)},
  year={2019},
  doi={https://doi.org/10.15146/5xcp-0d46 },
  publisher={Dryad}
}

@article{deBuyl,
  title={Stochastic logistic models reproduce experimental time series of microbial communities},
  author={Descheemaeker, Lana and De Buyl, Sophie},
  journal={Elife},
  volume={9},
  pages={e55650},
  year={2020},
  publisher={eLife Sciences Publications, Ltd}
}

@article{didomenico2024,
   author = {Laura Di Domenico and Eugenio Valdano and Vittoria Colizza},
   doi = {10.1103/PhysRevResearch.6.023265},
   issn = {2643-1564},
   issue = {2},
   journal = {Physical Review Research},
   month = {6},
   pages = {023265},
   publisher = {American Physical Society},
   title = {Limited data on infectious disease distribution exposes ambiguity in epidemic modeling choices},
   volume = {6},
   url = {https://link.aps.org/doi/10.1103/PhysRevResearch.6.023265},
   year = {2024},
}

@article{diSanto,
  title={Simple unified view of branching process statistics: Random walks in balanced logarithmic potentials},
  author={di Santo, Serena and Villegas, Pablo and Burioni, Raffaella and Mu{\~n}oz, Miguel A},
  journal={Physical Review E},
  volume={95},
  number={3},
  pages={032115},
  year={2017},
  doi={https://doi.org/10.1103/PhysRevE.95.032115},
  publisher={APS}
}

@article{dornicPRL,
  title = {Integration of Langevin Equations with Multiplicative Noise and the Viability of Field Theories for Absorbing Phase Transitions},
  author = {Dornic, Ivan and Chat\'e, Hugues and Mu\~noz, Miguel A.},
  journal = {Phys. Rev. Lett.},
  volume = {94},
  issue = {10},
  pages = {100601},
  numpages = {4},
  year = {2005},
  month = {Mar},
  publisher = {American Physical Society},
  doi = {10.1103/PhysRevLett.94.100601},
  url = {https://link.aps.org/doi/10.1103/PhysRevLett.94.100601}
}

@article{Dubkov2000,
   author = {A. A. Dubkov and A. N. Malakhov and A. I. Saichev},
   doi = {10.1007/BF02677200},
   issn = {0033-8443},
   issue = {4},
   journal = {Radiophysics and Quantum Electronics},
   month = {4},
   pages = {335-346},
   title = {Correlation time and structure of the correlation function of nonlinear equilibrium brownian motion in arbitrary-shaped potential wells},
   volume = {43},
   url = {http://link.springer.com/10.1007/BF02677200},
   year = {2000},
}

@article{fajardo_2023,
	title = {Fundamental limits to learning closed-form mathematical models from data},
	volume = {14},
	issn = {2041-1723},
	url = {https://www.nature.com/articles/s41467-023-36657-z},
	doi = {10.1038/s41467-023-36657-z},
	number = {1},
	urldate = {2026-02-04},
	journal = {Nature Communications},
	author = {Fajardo-Fontiveros, Oscar and Reichardt, Ignasi and De Los R{\'i}os, Harry R. and Duch, Jordi and Sales-Pardo, Marta and Guimer{\`a}, Roger},
	month = feb,
	year = {2023},
	pages = {1043},
}

@article{Feller1951,
   author = {William Feller},
   doi = {10.2307/1969318},
   issn = {0003486X},
   issue = {1},
   journal = {The Annals of Mathematics},
   month = {7},
   pages = {173},
   title = {Two Singular Diffusion Problems},
   volume = {54},
   url = {https://www.jstor.org/stable/1969318?origin=crossref},
   year = {1951},
}

@article{Ferretti2020,
  author={Ferretti, Federica and Chardès, Victor and Mora, Thierry and Walczak, Aleksandra M and Giardina, Irene},
   doi = {10.1103/PhysRevX.10.031018},
   issn = {2160-3308},
   issue = {3},
   journal = {Physical Review X},
   month = {7},
   pages = {031018},
   publisher = {American Physical Society},
   title = {Building General Langevin Models from Discrete Datasets},
   volume = {10},
   url = {https://link.aps.org/doi/10.1103/PhysRevX.10.031018},
   year = {2020},
}

@article{Ferguson2005,
	author = {Ferguson, Neil M. and Cummings, Derek A. T. and Cauchemez, Simon and Fraser, Christophe and Riley, Steven and Meeyai, Aronrag and Iamsirithaworn, Sopon and Burke, Donald S.},
	date = {2005/09/01},
	doi = {10.1038/nature04017},
	id = {Ferguson2005},
	isbn = {1476-4687},
	journal = {Nature},
	number = {7056},
	pages = {209--214},
	title = {Strategies for containing an emerging influenza pandemic in Southeast Asia},
	url = {https://doi.org/10.1038/nature04017},
	volume = {437},
	year = {2005}}

@article{florin_1998,
  title   = {Photonic force microscope calibration by thermal noise analysis},
  journal = {{Applied Physics A: Materials Science \& Processing}},
  volume  = {66},
  number  = {7},
  pages   = {S75--S78},
  year    = {1998},
  month   = {March},
  doi     = {10.1007/s003390051103},
  author  = {Florin, E.-L. and Pralle, A. and Stelzer, E. H. K. and H{\"o}rber, J. K. H.}
}

@article{Fulcher2013,
   author = {Ben D. Fulcher and Max A. Little and Nick S. Jones},
   doi = {10.1098/rsif.2013.0048},
   issn = {1742-5689},
   issue = {83},
   journal = {Journal of The Royal Society Interface},
   month = {6},
   pages = {20130048},
   pmid = {23554344},
   publisher = {Royal Society},
   title = {Highly comparative time-series analysis: the empirical structure of time series and their methods},
   volume = {10},
   url = {https://royalsocietypublishing.org/doi/10.1098/rsif.2013.0048},
   year = {2013},
}

@article{Fulcher2017,
   author = {Ben D. Fulcher and Nick S. Jones},
   doi = {10.1016/j.cels.2017.10.001},
   issn = {24054712},
   issue = {5},
   journal = {Cell Systems},
   month = {11},
   pages = {527-531.e3},
   pmid = {29102608},
   publisher = {Cell Press},
   title = {hctsa : A Computational Framework for Automated Time-Series Phenotyping Using Massive Feature Extraction},
   volume = {5},
   url = {https://linkinghub.elsevier.com/retrieve/pii/S2405471217304386},
   year = {2017},
}

@article{franosch_resonances_2011,
	title = {Resonances arising from hydrodynamic memory in {Brownian} motion},
	volume = {478},
	copyright = {http://www.springer.com/tdm},
	issn = {0028-0836, 1476-4687},
	url = {https://www.nature.com/articles/nature10498},
	doi = {10.1038/nature10498},
	number = {7367},
	urldate = {2026-02-26},
	journal = {Nature},
	author = {Franosch, Thomas and Grimm, Matthias and Belushkin, Maxim and Mor, Flavio M. and Foffi, Giuseppe and Forr{\'o}, L{\'a}szl{\'o} and Jeney, Sylvia},
	month = oct,
	year = {2011},
	pages = {85--88},
}

@article{Frishman2020,
   author = {Anna Frishman and Pierre Ronceray},
   doi = {10.1103/PhysRevX.10.021009},
   issn = {21603308},
   issue = {2},
   journal = {Physical Review X},
   month = {6},
   publisher = {American Physical Society},
   title = {Learning Force Fields from Stochastic Trajectories},
   volume = {10},
   year = {2020},
}

@article{gerardos2025principled,
	title = {Principled {Model} {Selection} for {Stochastic} {Dynamics}},
	volume = {135},
	issn = {0031-9007, 1079-7114},
	url = {https://link.aps.org/doi/10.1103/ltdt-hvh7},
	doi = {10.1103/ltdt-hvh7},
	number = {16},
	journal = {Physical Review Letters},
	author = {Gerardos, Andonis and Ronceray, Pierre},
	month = oct,
	year = {2025},
	pages = {167401},
}

@article{Gardiner2009,
   author = {C. W. Gardiner},
   doi = {10.1002/bbpc.19850890629},
   isbn = {9783540707127},
   issn = {0005-9021},
   issue = {6},
   journal = {Berichte der Bunsengesellschaft für physikalische Chemie},
   month = {6},
   pages = {721-721},
   title = {{C. W. Gardiner: Handbook of Stochastic Methods for Physics, Chemistry and the Natural Sciences , Springer‐Verlag, Berlin, Heidelberg, New York, Tokyo 1983. 442 Seiten, Preis: DM 115,‐}},
   volume = {89},
   url = {https://onlinelibrary.wiley.com/doi/10.1002/bbpc.19850890629},
   year = {1985},
}

@book{gelman1995bayesian,
  title={Bayesian data analysis},
  author={Gelman, Andrew and Carlin, John B and Stern, Hal S and Rubin, Donald B},
  year={1995},
  publisher={Chapman and Hall/CRC}
}

@article{Goldford2018,
   author = {Joshua E. Goldford and Nanxi Lu and Djordje Bajić and Sylvie Estrela and Mikhail Tikhonov and Alicia Sanchez-Gorostiaga and Daniel Segrè and Pankaj Mehta and Alvaro Sanchez},
   doi = {10.1126/science.aat1168},
   issn = {0036-8075},
   issue = {6401},
   journal = {Science},
   month = {8},
   pages = {469-474},
   title = {Emergent simplicity in microbial community assembly},
   volume = {361},
   url = {https://www.science.org/doi/10.1126/science.aat1168},
   year = {2018},
}

@article{Gutenkunst2007,
   author = {Ryan N. Gutenkunst and Joshua J. Waterfall and Fergal P. Casey and Kevin S. Brown and Christopher R. Myers and James P. Sethna},
   doi = {10.1371/journal.pcbi.0030189},
   issn = {15537358},
   issue = {10},
   journal = {PLoS Computational Biology},
   pages = {1871-1878},
   pmid = {17922568},
   publisher = {Public Library of Science},
   title = {Universally sloppy parameter sensitivities in systems biology models},
   volume = {3},
   year = {2007},
}

@article{grilli2020macroecological,
  title={Macroecological laws describe variation and diversity in microbial communities},
  author={Grilli, Jacopo},
  journal={Nature communications},
  volume={11},
  number={1},
  pages={4743},
  year={2020},
  publisher={Nature Publishing Group UK London}
}

@article{Guimera2020,
   author = {Roger Guimerà and Ignasi Reichardt and Antoni Aguilar-Mogas and Francesco A. Massucci and Manuel Miranda and Jordi Pallarès and Marta Sales-Pardo},
   doi = {10.1126/sciadv.aav6971},
   issn = {2375-2548},
   issue = {5},
   journal = {Science Advances},
   month = {1},
   pmid = {32064326},
   title = {A Bayesian machine scientist to aid in the solution of challenging scientific problems},
   volume = {6},
   url = {https://www.science.org/doi/10.1126/sciadv.aav6971},
   year = {2020},
}

@book{Henkel2008,
   author = {Malte Henkel and Haye Hinrichsen and Sven Lübeck},
   city = {Dordrecht},
   doi = {10.1007/978-1-4020-8765-3},
   isbn = {978-1-4020-8764-6},
   issn = {1864-5879},
   journal = {Journal of Chemical Information and Modeling},
   month = {11},
   pmid = {25246403},
   publisher = {Springer Netherlands},
   title = {Non-Equilibrium Phase Transitions},
   volume = {53},
   url = {http://link.springer.com/10.1007/978-1-4020-8765-3},
   year = {2008},
}

@article{Hatton2024,
   author = {Ian A. Hatton and Onofrio Mazzarisi and Ada Altieri and Matteo Smerlak},
   doi = {10.1126/science.adg8488},
   issn = {0036-8075},
   issue = {6688},
   journal = {Science},
   month = {3},
   pmid = {38484074},
   title = {Diversity begets stability: Sublinear growth and competitive coexistence across ecosystems},
   volume = {383},
   url = {https://www.science.org/doi/10.1126/science.adg8488},
   year = {2024},
}

@article{Hasselmann1976,
   author = {K. Hasselmann},
   doi = {10.3402/tellusa.v28i6.11316},
   issn = {1600-0870},
   issue = {6},
   journal = {Tellus A: Dynamic Meteorology and Oceanography},
   month = {1},
   pages = {473},
   publisher = {Stockholm University Press},
   title = {Stochastic climate models: Part I. Theory},
   volume = {28},
   url = {https://a.tellusjournals.se/article/10.3402/tellusa.v28i6.11316/},
   year = {1976},
}

@book{Iacus2008,
   author = {Stefano M. Iacus},
   city = {New York, NY},
   doi = {10.1007/978-0-387-75839-8},
   isbn = {978-0-387-75838-1},
   publisher = {Springer New York},
   title = {Simulation and Inference for Stochastic Differential Equations},
   volume = {1},
   url = {https://link.springer.com/10.1007/978-0-387-75839-8},
   year = {2008},
}

@article{Harunari2022,
   author = {Pedro E. Harunari and Annwesha Dutta and Matteo Polettini and Édgar Roldán},
   doi = {10.1103/PhysRevX.12.041026},
   issn = {2160-3308},
   issue = {4},
   journal = {Physical Review X},
   month = {12},
   pages = {041026},
   publisher = {American Physical Society},
   title = {What to Learn from a Few Visible Transitions’ Statistics?},
   volume = {12},
   url = {https://link.aps.org/doi/10.1103/PhysRevX.12.041026},
   year = {2022},
}

@article{hastings_2006,
  author = {Hastings, M. B.},
  title = {Community detection as an inference problem},
  journal = {Phys. Rev. E},
  volume = {74},
  number = {3},
  pages = {035102},
  month = sep,
  year = {2006},
  doi = {10.1103/PhysRevE.74.035102},
  url = {https://link.aps.org/doi/10.1103/PhysRevE.74.035102}
}

@article{Kiorbe_2009,
author = {Kiorboe, Thomas},
title = {A Mechanistic Approach to Plankton Ecology},
journal = {ASLO Web Lectures},
volume = {1},
number = {2},
pages = {1-91},
doi = {https://doi.org/10.4319/lol.2009.tkiorboe.2},
year = {2009}
}

@book{Klebaner2012,
   author = {Fima C Klebaner},
   doi = {10.1142/p821},
   isbn = {978-1-84816-831-2},
   month = {3},
   publisher = {IMPERIAL COLLEGE PRESS},
   title = {Introduction to Stochastic Calculus with Applications},
   url = {https://www.worldscientific.com/worldscibooks/10.1142/p821},
   year = {2012},
}

@article{lacasa_2018,
  author = {Lacasa, Lucas and Mari\~no, In\'es P. and M\'iguez, Joaquin and 
            Nicosia, Vincenzo and Rold\'an, \'Edgar and Lisica, Ana and 
            Grill, Stephan W. and G\'omez-Garde\~nes, Jes\'us},
  title = {Multiplex {Decomposition} of {Non}-{Markovian} {Dynamics} 
           and the {Hidden} {Layer} {Reconstruction} {Problem}},
  journal = {Phys. Rev. X},
  volume = {8},
  number = {3},
  pages = {031038},
  year = {2018},
  doi = {10.1103/PhysRevX.8.031038}
}

@book{Langouche1982,
   author = {F. Langouche and D. Roekaerts and E. Tirapegui},
   city = {Dordrecht},
   doi = {10.1007/978-94-017-1634-5},
   isbn = {978-90-481-8377-7},
   journal = {Functional Integration and Semiclassical Expansions},
   publisher = {Springer Netherlands},
   title = {Functional Integration and Semiclassical Expansions},
   url = {http://link.springer.com/10.1007/978-94-017-1634-5},
   year = {1982},
}

@article{Majumdar2015,
   author = {Satya N Majumdar and Henri Orland},
   doi = {10.1088/1742-5468/2015/06/P06039},
   issn = {1742-5468},
   issue = {6},
   journal = {Journal of Statistical Mechanics: Theory and Experiment},
   month = {6},
   pages = {P06039},
   publisher = {IOP Publishing},
   title = {Effective Langevin equations for constrained stochastic processes},
   volume = {2015},
   url = {https://iopscience.iop.org/article/10.1088/1742-5468/2015/06/P06039},
   year = {2015},
}

@article{martinez_survey_2017,
	title = {A {Survey} of {Link} {Prediction} in {Complex} {Networks}},
	volume = {49},
	issn = {0360-0300, 1557-7341},
	url = {https://dl.acm.org/doi/10.1145/3012704},
	doi = {10.1145/3012704},
	number = {4},
	urldate = {2026-03-29},
	journal = {ACM Computing Surveys},
	author = {Mart{\'i}nez, V{\'i}ctor and Berzal, Fernando and Cubero, Juan-Carlos},
	month = dec,
	year = {2017},
	pages = {1--33},
}

@article{Lawton1999,
   author = {John H. Lawton},
   doi = {10.2307/3546712},
   isbn = {201208:48:42},
   issn = {00301299},
   issue = {2},
   journal = {Oikos},
   month = {2},
   pages = {177},
   title = {Are There General Laws in Ecology?},
   volume = {84},
   url = {https://www.jstor.org/stable/3546712?origin=crossref},
   year = {1999},
}

@book{Lande2003,
    author = {Lande, Russell and Engen, Steinar and Saether, Bernt-Erik},
    title = {Stochastic Population Dynamics in Ecology and Conservation},
    publisher = {Oxford University Press},
    year = {2003},
    month = {04},
    isbn = {9780198525257},
    doi = {10.1093/acprof:oso/9780198525257.001.0001},
    url = {https://doi.org/10.1093/acprof:oso/9780198525257.001.0001},
}

@article{Lluis2024,
  title = {Effective theory of collective deep learning},
  author = {Arola-Fern\'andez, Llu\'{\i}s and Lacasa, Lucas},
  journal = {Phys. Rev. Res.},
  volume = {6},
  issue = {4},
  pages = {L042040},
  numpages = {8},
  year = {2024},
  month = {Nov},
  publisher = {American Physical Society},
  doi = {10.1103/PhysRevResearch.6.L042040},
  url = {https://link.aps.org/doi/10.1103/PhysRevResearch.6.L042040}
}

@article{Lucarini2019,
   author = {Valerio Lucarini and Tamás Bódai},
   doi = {10.1103/PhysRevLett.122.158701},
   issn = {0031-9007},
   issue = {15},
   journal = {Physical Review Letters},
   month = {4},
   pages = {158701},
   pmid = {31050495},
   publisher = {American Physical Society},
   title = {Transitions across Melancholia States in a Climate Model: Reconciling the Deterministic and Stochastic Points of View},
   volume = {122},
   url = {https://link.aps.org/doi/10.1103/PhysRevLett.122.158701},
   year = {2019},
}

@article{Maragakis2008,
    author = {Maragakis, Paul and Ritort, Felix and Bustamante, Carlos and Karplus, Martin and Crooks, Gavin E.},
    title = {Bayesian estimates of free energies from nonequilibrium work data in the presence of instrument noise},
    journal = {The Journal of Chemical Physics},
    volume = {129},
    number = {2},
    pages = {024102},
    year = {2008},
    month = {07},
    issn = {0021-9606},
    doi = {10.1063/1.2937892},
    url = {https://doi.org/10.1063/1.2937892},
}

@article{McClelland2025,
    author = {James L. McClelland },
    title = {Profile of John Hopfield and Geoffrey Hinton: 2024 Nobel laureates in Physics},
    journal = {Proceedings of the National Academy of Sciences},
    volume = {122},
    number = {16},
    pages = {e2423094122},
    year = {2025},
    doi = {10.1073/pnas.2423094122},
    URL = {https://www.pnas.org/doi/abs/10.1073/pnas.2423094122},
    eprint = {https://www.pnas.org/doi/pdf/10.1073/pnas.2423094122},
}

@article{Manabe1967,
   author = {Syukuro Manabe and Richard T. Wetherald},
   doi = {10.1175/1520-0469(1967)024<0241:TEOTAW>2.0.CO;2},
   issn = {0022-4928},
   issue = {3},
   journal = {Journal of the Atmospheric Sciences},
   month = {5},
   pages = {241-259},
   title = {Thermal Equilibrium of the Atmosphere with a Given Distribution of Relative Humidity},
   volume = {24},
   url = {http://journals.ametsoc.org/doi/10.1175/1520-0469(1967)024<0241:TEOTAW>2.0.CO;2},
   year = {1967},
}

@article{Morita2016,
   author = {Satoru Morita},
   doi = {10.1038/srep22506},
   issn = {2045-2322},
   issue = {1},
   journal = {Scientific Reports},
   month = {3},
   pages = {22506},
   publisher = {Nature Publishing Group},
   title = {Six Susceptible-Infected-Susceptible Models on Scale-free Networks},
   volume = {6},
   url = {https://www.nature.com/articles/srep22506},
   year = {2016},
}

@article{Newman2002,
  title = {Spread of epidemic disease on networks},
  author = {Newman, M. E. J.},
  journal = {Phys. Rev. E},
  volume = {66},
  issue = {1},
  pages = {016128},
  numpages = {11},
  year = {2002},
  month = {Jul},
  publisher = {American Physical Society},
  doi = {10.1103/PhysRevE.66.016128},
  url = {https://link.aps.org/doi/10.1103/PhysRevE.66.016128}
}

@article{Odwyer2023universal,
author = {Ashish B. George  and James O’Dwyer },
title = {Universal abundance fluctuations across microbial communities, tropical forests, and urban populations},
journal = {Proceedings of the National Academy of Sciences},
volume = {120},
number = {44},
pages = {e2215832120},
year = {2023},
doi = {10.1073/pnas.2215832120},
URL = {https://www.pnas.org/doi/abs/10.1073/pnas.2215832120},
}

@article{orland_generating_2011,
	title = {Generating transition paths by {Langevin} bridges},
	volume = {134},
	issn = {0021-9606, 1089-7690},
	url = {https://pubs.aip.org/jcp/article/134/17/174114/699536/Generating-transition-paths-by-Langevin-bridges},
	doi = {10.1063/1.3586036},
	number = {17},
	urldate = {2026-03-22},
	journal = {The Journal of Chemical Physics},
	author = {Orland, Henri},
	month = may,
	year = {2011},
	pages = {174114},
}

@article{Onsager1953,
   author = {L. Onsager and S. Machlup},
   doi = {10.1103/PhysRev.91.1505},
   issn = {0031-899X},
   issue = {6},
   journal = {Physical Review},
   month = {9},
   pages = {1505-1512},
   title = {Fluctuations and Irreversible Processes},
   volume = {91},
   url = {https://link.aps.org/doi/10.1103/PhysRev.91.1505},
   year = {1953},
}

@article{peixoto_2019,
  author = {Peixoto, Tiago P.},
  title = {Network {Reconstruction} and {Community} {Detection} from {Dynamics}},
  journal = {Phys. Rev. Lett.},
  volume = {123},
  number = {12},
  pages = {128301},
  month = sep,
  year = {2019},
  doi = {10.1103/PhysRevLett.123.128301},
  url = {https://link.aps.org/doi/10.1103/PhysRevLett.123.128301}
}

@book{Pavliotis2014,
   author = {Grigorios A. Pavliotis},
   city = {New York, NY},
   doi = {10.1007/978-1-4939-1323-7},
   isbn = {978-1-4939-1322-0},
   publisher = {Springer New York},
   title = {Stochastic Processes and Applications},
   volume = {60},
   url = {https://link.springer.com/10.1007/978-1-4939-1323-7},
   year = {2014},
}

@article{Picot2023,
   author = {Aurore Picot and Shota Shibasaki and Oliver J Meacock and Sara Mitri},
   doi = {10.1016/j.mib.2023.102354},
   issn = {13695274},
   journal = {Current Opinion in Microbiology},
   month = {10},
   pages = {102354},
   pmid = {37421708},
   publisher = {Elsevier Ltd},
   title = {Microbial interactions in theory and practice: when are measurements compatible with models?},
   volume = {75},
   url = {https://linkinghub.elsevier.com/retrieve/pii/S1369527423000917},
    year = {2023},
}

@article{Pasqualini2025,
  author  = {Pasqualini, Jacopo and Maritan, Amos and Rinaldo, Andrea and Facchin, Sonia and Savarino, Edoardo and Altieri, Ada and Suweis, Samir},
  title   = {Microbiomes Through the Looking Glass},
  journal = {eLife},
  year    = {2025},
  month   = {4},
  doi     = {10.7554/eLife.105948},
  url     = {https://doi.org/10.7554/eLife.105948}
}

@article{plata2021neutral,
  title={Neutral theory for competing attention in social networks},
  author={Plata, Carlos A and Pigani, Emanuele and Azaele, Sandro and Calleja-Solanas, Violeta and Palazzi, Mar{\'\i}a J and Sol{\'e}-Ribalta, Albert and Borge-Holthoefer, Javier and Meloni, Sandro and Suweis, Samir},
  journal={Physical Review Research},
  volume={3},
  number={1},
  pages={013070},
  year={2021},
  publisher={APS}
}

@article{Risken1991,
   author = {H. Risken and T. K. Caugheyz},
   doi = {10.1115/1.2897281},
   issn = {0021-8936},
   issue = {3},
   journal = {Journal of Applied Mechanics},
   pages = {860-860},
   title = {The Fokker-Planck Equation: Methods of Solution and Application, 2nd ed.},
   volume = {58},
   year = {1991},
}

@article{Ross2006,
   author = {J.V. Ross and T. Taimre and P.K. Pollett},
   doi = {10.1016/j.tpb.2006.08.001},
   issn = {00405809},
   issue = {4},
   journal = {Theoretical Population Biology},
   month = {12},
   pages = {498-510},
   pmid = {16984803},
   title = {On parameter estimation in population models},
   volume = {70},
   url = {https://linkinghub.elsevier.com/retrieve/pii/S0040580906001079},
   year = {2006},
}

@article{scanagatta_survey_2019,
	title = {A survey on {Bayesian} network structure learning from data},
	volume = {8},
	issn = {2192-6352, 2192-6360},
	url = {http://link.springer.com/10.1007/s13748-019-00194-y},
	doi = {10.1007/s13748-019-00194-y},
	number = {4},
	urldate = {2026-03-29},
	journal = {Progress in Artificial Intelligence},
	author = {Scanagatta, Mauro and Salmer{\'o}n, Antonio and Stella, Fabio},
	month = dec,
	year = {2019},
	pages = {425--439},
}

@article{serGiacomi2018,
   author = {Enrico Ser-Giacomi and Lucie Zinger and Shruti Malviya and Colomban De Vargas and Eric Karsenti and Chris Bowler and Silvia De Monte},
   doi = {10.1038/s41559-018-0587-2},
   issn = {2397334X},
   issue = {8},
   journal = {Nature Ecology and Evolution},
   month = {8},
   pages = {1243-1249},
   pmid = {29915345},
   publisher = {Nature Publishing Group},
   title = {Ubiquitous abundance distribution of non-dominant plankton across the global ocean},
   volume = {2},
   year = {2018},
}

@article{sorensen2004,
   author = {Helle Sørensen},
   doi = {10.1111/j.1751-5823.2004.tb00241.x},
   issn = {0306-7734},
   issue = {3},
   journal = {International Statistical Review},
   month = {12},
   pages = {337-354},
   publisher = {International Statistical Institute},
   title = {Parametric Inference for Diffusion Processes Observed at Discrete Points in Time: a Survey},
   volume = {72},
   url = {https://onlinelibrary.wiley.com/doi/10.1111/j.1751-5823.2004.tb00241.x},
   year = {2004},
}

@article{Hoshino2020,
   author = {Tatsuhiko Hoshino and Hideyuki Doi and Go-Ichiro Uramoto and Lars Wörmer and Rishi R. Adhikari and Nan Xiao and Yuki Morono and Steven D’Hondt and Kai-Uwe Hinrichs and Fumio Inagaki},
   doi = {10.1073/pnas.1919139117},
   isbn = {1919139117},
   issn = {0027-8424},
   issue = {44},
   journal = {Proceedings of the National Academy of Sciences},
   month = {11},
   pages = {27587-27597},
   title = {Global diversity of microbial communities in marine sediment},
   volume = {117},
   url = {https://pnas.org/doi/full/10.1073/pnas.1919139117},
   year = {2020},
}

@article{Nature1998,
  title={Nature of different types of absorbing states},
  author={Mu\~noz, Miguel A},
  journal={Physical Review E},
  volume={57},
  number={2},
  pages={1377},
  year={1998},
  publisher={APS}
}

@book{Toral2014,
   author = {Raul Toral and Pere Colet},
   city = {Weinheim, Germany},
   doi = {10.1002/9783527683147},
   editor = {Raúl Toral and Pere Colet},
   isbn = {9783527411498},
   issn = {0007-1250},
   issue = {479},
   journal = {The British Journal of Psychiatry},
   month = {7},
   pages = {1009-1010},
   publisher = {Wiley},
   title = {Stochastic Numerical Methods},
   volume = {111},
   url = {https://onlinelibrary.wiley.com/doi/book/10.1002/9783527683147},
   year = {2014},
}

@article{unser_2000,
	title = {Sampling-50 years after {Shannon}},
	volume = {88},
	copyright = {https://ieeexplore.ieee.org/Xplorehelp/downloads/license-information/IEEE.html},
	issn = {0018-9219, 1558-2256},
	url = {http://ieeexplore.ieee.org/document/843002/},
	doi = {10.1109/5.843002},
	number = {4},
	urldate = {2026-02-17},
	journal = {Proceedings of the IEEE},
	author = {Unser, M.},
	month = apr,
	year = {2000},
	pages = {569--587},
}

@book{vankampen2007spp,
  author = {Kampen, NG Van},
  description = {{Stochastic processes in physics and chemistry}},
  publisher = {North Holland},
  title = {Stochastic processes in physics and chemistry},
  year = 2007
}

@article{VanKampen1981,
   author = {N. G. van Kampen},
   doi = {10.1007/BF01007642},
   issn = {15729613},
   issue = {1},
   journal = {Journal of Statistical Physics},
   pages = {175-187},
   title = {Itô versus Stratonovich},
   volume = {24},
   year = {1981},
}

@article{Varadhan1985,
   author = {S. R. S. Varadhan and M. I. Freidlin and A. D. Wentzell},
   doi = {10.2307/2287939},
   isbn = {9783642258466},
   issn = {01621459},
   issue = {390},
   journal = {Journal of the American Statistical Association},
   month = {6},
   pages = {490},
   title = {Random Perturbations of Dynamical Systems.},
   volume = {80},
   url = {https://www.jstor.org/stable/2287939?origin=crossref},
   year = {1985},
}

@book{oppenheim_2014,
	address = {Harlow},
	edition = {Third edition, Pearson New international edition},
	series = {Always learning},
	title = {Discrete-time signal processing},
	isbn = {978-1-292-02572-8 978-1-292-03815-5},
	publisher = {Pearson},
	author = {Oppenheim, Alan V. and Schafer, Roland W.},
	year = {2014},
}

@article{Pastor2015,
  title = {Epidemic processes in complex networks},
  author = {Pastor-Satorras, Romualdo and Castellano, Claudio and Van Mieghem, Piet and Vespignani, Alessandro},
  journal = {Rev. Mod. Phys.},
  volume = {87},
  issue = {3},
  pages = {925--979},
  numpages = {55},
  year = {2015},
  month = {Aug},
  publisher = {American Physical Society},
  doi = {10.1103/RevModPhys.87.925},
  url = {https://link.aps.org/doi/10.1103/RevModPhys.87.925}
}

@article{Sireci,
  title={Environmental fluctuations explain the universal decay of species-abundance correlations with phylogenetic distance},
  author={Sireci, Matteo and Mu{\~n}oz, Miguel A and Grilli, Jacopo},
  journal={Proceedings of the National Academy of Sciences},
  volume={120},
  number={37},
  pages={e2217144120},
  year={2023},
  publisher={National Acad Sciences}
}

@article{Ravishankara2022,
   author = {A. R. Ravishankara and David A. Randall and James W. Hurrell},
   doi = {10.1073/pnas.2120669119},
   issn = {10916490},
   issue = {2},
   journal = {Proceedings of the National Academy of Sciences of the United States of America},
   month = {1},
   pmid = {34996875},
   publisher = {NLM (Medline)},
   title = {Complex and yet predictable: The message of the 2021 Nobel Prize in Physics},
   volume = {119},
   year = {2022},
}

@book{rolski1999,
   city = {Hoboken, NJ, USA},
   doi = {10.1002/9780470317044},
   editor = {Tomasz Rolski and Hanspeter Schmidli and Volker Schmidt and Jozef Teugels},
   isbn = {9780470317044},
   month = {2},
   publisher = {{John Wiley \& Sons, Inc.}},
   title = {Stochastic Processes for Insurance \& Finance},
   url = {http://doi.wiley.com/10.1002/9780470317044},
   year = {1999},
}

@article{Roldan2023,
   author = {Edgar Roldán and Izaak Neri and Raphael Chetrite and Shamik Gupta and Simone Pigolotti and Frank Jülicher and Ken Sekimoto},
   doi = {10.1080/00018732.2024.2317494},
   issn = {0001-8732},
   issue = {1-2},
   journal = {Advances in Physics},
   month = {4},
   pages = {1-258},
   publisher = {Taylor and Francis Ltd.},
   title = {Martingales for physicists: a treatise on stochastic thermodynamics and beyond},
   volume = {72},
   url = {https://www.tandfonline.com/doi/full/10.1080/00018732.2024.2317494},
   year = {2023},
}

@article{Rossi2010,
   author = {Giuliano De Rossi},
   doi = {10.1007/s10614-010-9208-0},
   issn = {0927-7099},
   issue = {1},
   journal = {Computational Economics},
   month = {6},
   pages = {1-16},
   title = {Maximum Likelihood Estimation of the Cox–Ingersoll–Ross Model Using Particle Filters},
   volume = {36},
   url = {http://link.springer.com/10.1007/s10614-010-9208-0},
   year = {2010},
}

@article{toni_approximate_2009,
	title = {Approximate {Bayesian} computation scheme for parameter inference and model selection in dynamical systems},
	volume = {6},
	issn = {1742-5689, 1742-5662},
	url = {https://royalsocietypublishing.org/doi/10.1098/rsif.2008.0172},
	doi = {10.1098/rsif.2008.0172},
	number = {31},
	journal = {Journal of The Royal Society Interface},
	author = {Toni, Tina and Welch, David and Strelkowa, Natalja and Ipsen, Andreas and Stumpf, Michael P.H},
	month = feb,
	year = {2009},
	pages = {187--202},
}

@article{van_dongen_2006,
	title = {Prior specification in {Bayesian} statistics: {Three} cautionary tales},
	volume = {242},
	copyright = {https://www.elsevier.com/tdm/userlicense/1.0/},
	issn = {00225193},
	shorttitle = {Prior specification in {Bayesian} statistics},
	url = {https://linkinghub.elsevier.com/retrieve/pii/S0022519306000609},
	doi = {10.1016/j.jtbi.2006.02.002},
	number = {1},
	urldate = {2026-02-04},
	journal = {Journal of Theoretical Biology},
	author = {Van Dongen, Stefan},
	month = sep,
	year = {2006},
	pages = {90--100},
}

@article{volpe_torque_2006,
	title = {Torque {Detection} using {Brownian} {Fluctuations}},
	volume = {97},
	copyright = {http://link.aps.org/licenses/aps-default-license},
	issn = {0031-9007, 1079-7114},
	url = {https://link.aps.org/doi/10.1103/PhysRevLett.97.210603},
	doi = {10.1103/PhysRevLett.97.210603},
	number = {21},
	urldate = {2026-02-26},
	journal = {Physical Review Letters},
	author = {Volpe, Giovanni and Petrov, Dmitri},
	month = nov,
	year = {2006},
	pages = {210603},
}

@article{volpe_brownian_2007,
	title = {Brownian motion in a nonhomogeneous force field and photonic force microscope},
	volume = {76},
	copyright = {http://link.aps.org/licenses/aps-default-license},
	issn = {1539-3755, 1550-2376},
	url = {https://link.aps.org/doi/10.1103/PhysRevE.76.061118},
	doi = {10.1103/PhysRevE.76.061118},
	number = {6},
	urldate = {2026-02-26},
	journal = {Physical Review E},
	author = {Volpe, Giorgio and Volpe, Giovanni and Petrov, Dmitri},
	month = dec,
	year = {2007},
	pages = {061118},
}

@book{Wio2013,
   author = {Horacio S Wio},
   doi = {10.1142/8695},
   isbn = {978-981-4447-99-7},
   issue = {April},
   journal = {Angewandte Chemie International Edition, 6(11), 951–952.},
   month = {3},
   pages = {15-38},
   publisher = {WORLD SCIENTIFIC},
   title = {Path Integrals for Stochastic Processes},
   volume = {13},
   url = {https://www.worldscientific.com/worldscibooks/10.1142/8695},
   year = {2013},
}

@book{Wolfgang2014,
   author = {Wolfgang Paul and Jörg Baschnagel},
   city = {Heidelberg},
   doi = {10.1007/978-3-319-00327-6},
   isbn = {978-3-319-00326-9},
   journal = {Introduction to Imprecise Probabilities},
   month = {8},
   pages = {258-278},
   publisher = {Springer International Publishing},
   title = {Stochastic Processes},
   url = {https://link.springer.com/10.1007/978-3-319-00327-6},
   year = {2013},
}

@article{Zubiaga2018,
   author = {Arkaitz Zubiaga},
   doi = {10.1002/asi.24026},
   issn = {2330-1635},
   issue = {8},
   journal = {Journal of the Association for Information Science and Technology},
   month = {8},
   pages = {974-984},
   publisher = {John Wiley and Sons Inc.},
   title = {A longitudinal assessment of the persistence of twitter datasets},
   volume = {69},
   url = {https://asistdl.onlinelibrary.wiley.com/doi/10.1002/asi.24026},
   year = {2018},
}

\clearpage

\onecolumngrid
\appendix
\setcounter{page}{1}
\setcounter{figure}{0}
\setcounter{equation}{0}
\setcounter{section}{0}
\renewcommand{\thesubsection}{\thesection.\arabic{subsection}}
\renewcommand{\thefigure}{A \arabic{figure}}

\section{Parameters used in figures}\label{AP_sec:parameters}
\begin{itemize}
    \item \underline{Figure~\ref{fig:Estimator_CV_B}:} In a),b),c): $\mu^\text{(OU)}=\mu^\text{(DE)}=k^\text{(OU)}=k^\text{(DE)}=1$, $D^\text{(OU)}=D^\text{(DE)}=0.1$. In d) $v=10^{-3}$, $k^\text{(OU)}=k^\text{(DE)}=1$, $\mu^\text{(OU)} = 2$, $\mu^\text{(DE)} = \mu^\text{(OU)}+\Delta\mu$, $D^\text{(OU)}= \sqrt{2\,v\,k}$,  $D^\text{(DE)}= \sqrt{2\,v\,k/\mu^\text{(OU)}}$.
    
    \item \underline{Figure~\ref{fig:Estimation_mu_k_D}:} All time series were generated with parameters $\mu = 2$, $k = 1$, and $D = 0.2$. $\tau$ was obtained analytically for the DE and OU models, and numerically for the EN model (see Sec.~\ref{AP_sec:correlation time}), obtaining $\tau^\text{(OU)}=\tau^\text{(DE)}=1$, and $\tau^\text{(EN)}\approx1.09$. CP refers to the contact process model~\cite{Henkel2008},
    \begin{equation}\label{eq:CP_model} dX_t = k\,X_t\,(\mu - X_t) + D\sqrt{X_t}\,dW_t\quad(\text{CP}). \end{equation} 

    \item \underline{Figure~\ref{fig:BCM}:} In a), trajectory generated with Wiener process ($dX_t=dW_t$). In b), trajectories are generated using Wiener bridge (Eq.~\eqref{eq:SDE_brownian_bridge}) with $D=1$, fixed final point in $x_T=4$ at $T=10$. In c), we estimated the propagator for the OU process with $\mu = 2$, $k=1$, $D=0.2$.

    \item \underline{Figure \ref{fig:model_distinguishability}:} All time series were generated with parameters $\mu = 2$, $k = 1$, and $D = 0.2$. We use the same values for $\tau$ reported for Figure 3.
    
    \item \underline{Figure \ref{fig:data}:} In c, dataset exacted from Ref.~\cite{caporaso2011moving} using all OTUs and all sites in individuals M3 and F4 that pass the stationary test. In d, data extracted from~\cite{Zubiaga2018}using top $10$ most cited hashtags within the events Brexit, Panamapapers, Nepalearthquake, Ferguson, Ebola and Euro2012. In e, data set extracted from Ref.~\cite{condit2019complete}, using data for all species from 1995 to 2015.
\end{itemize}

\section{Biased errors in diffusion function estimation}\label{AP_sec:B_err_df}

As discussed in the main text, the bias in the estimator of the diffusion function [Eq.~\eqref{eq:learned_diff_eq}] can be analyzed by examining the expected value of the estimator. This expectation is not affected by statistical (sampling) errors and deviates from the true value $B(x)$ solely due to discretization bias:
\begin{equation}\label{AP_eq:mean_dif_estimator}
    \bar{B}(x)=\mathbb{E}\left[\hat{B}(x)\right] = \frac{1}{\Delta t}\mathbb{E}\left[\left(X_{t+\Delta t}-x\right)^2\Big|X_t=x\right].
\end{equation}
The limit $\Delta t\to \infty$ of the above expression can be computed for any stationary model with well-defined first and second centered stationary moments ($\E(X_\infty) $ and $\text{Var}(X_\infty)$ respectively). In this limit, conditioned expectations are simply replaced by the expectations at infinity,
\begin{equation}\label{AP_eq:limiting_form}
    \lim_{\Delta t\to\infty} \bar{B}(x)= \frac{1}{\Delta t}\mathbb{E}\left[\left(X_\infty-x\right)^2\right]= \frac{1}{\Delta t}\left[\text{Var}(X_\infty)+\left(\E(X_\infty)-x\right)^2\right].
\end{equation}
Eq.~\eqref{eq:limiting_form_of_QV_estimator} is obtained from Eq.~\eqref{AP_eq:limiting_form} replacing the stationary centered moments by its estimators using the time series data.

In the OU and DE models, Eq.~\eqref{AP_eq:mean_dif_estimator} can be evaluated at arbitrary sampling times. Indeed, the Itô process with linear drift,
\begin{equation}
    dX_t = k(\mu-X_t)\,dt +B(X_t)\,dW_t,
\end{equation}
has solution
\begin{equation}
    X_t = \mu +(X_0-\mu)\,e^{-k t} + e^{-k t}\int_0^t e^{ks} B(X_s)\,dW_s, 
\end{equation}
which can be proven using the change of variables $V_t =(X_t-\mu)\,e^{kt}$ together with Ito's formula for change of variables. From the solution one can readily obtain the first moment,
\begin{equation}\label{AP_eq:first_moment_linear_SDE}
    \E(X_t| X_0=x_0) = \mu +(x_0-\mu)\,e^{-kt}.
\end{equation}
The second moment is, in general, not accessible since its equation is not closed
\begin{equation}\label{AP_eq:second_moment_linear_SDE}
    \text{Var}\left(X_t\big|X_0=x_0\right) = e^{-2kt}\int_0^te^{2ks} \,\E(B^2(X_s)| X_0=x_0)\,ds,
\end{equation}
where we used $dW_s \,dW_{s'}=ds\,\delta(s-s')$. However, in the OU and DE processes, the above integral can be evaluated, obtaining
\begin{equation}\label{AP_eq:second_moment_OU}
    \text{Var}\left(X^{(\text{OU})}_t\big|X^{(\text{OU})}_0=x_0\right) = \frac{D^2}{2k} \left(1-e^{-2kt}\right), 
\end{equation}
and,
\begin{equation}\label{AP_eq:second_moment_DE}
    \text{Var}\left(X^{(\text{DE})}_t\big|X^{(\text{DE})}_0=x_0\right) = \frac{\mu\,D^2}{2k} \left[\left(1-e^{-2kt}\right)^2+\frac{2\,x_0}{\mu}\left(e^{-kt}-e^{-2kt}\right)\right], 
\end{equation}
Substituting Eqs.~\eqref{AP_eq:second_moment_DE}, \eqref{AP_eq:second_moment_OU}, and ~\eqref{AP_eq:first_moment_linear_SDE} into Eq.~\eqref{AP_eq:mean_dif_estimator} we get the analytical expressions used to make Fig.~\ref{fig:Estimator_CV_B}-c).

\section{Heuristic introduction to path measures and Girsanov theorem}\label{AP_sec:path_measures_Girsanov}

For the sake of self-consistency, we summarize here the theory of probability measures over continuous paths and provide a heuristic derivation of Girsanov's theorem and its generalization. Deeper discussions on this topic can be found in refs.~\cite{Asmussen2007,Klebaner2012,Pavliotis2014,Wolfgang2014,Wio2013,Aguilar2024}.

Let us first briefly recall the concept of probability measures and Radon-Nikodym derivatives for one random variable, as these naturally generalize to analogous objects in the context of stochastic processes. Let $X$ be a continuous random variable taking values in $\mathbb{R}$ and assume there exists a probability density function $p(\cdot)$, meaning that
\begin{equation}
    p(x) = \lim_{\Delta x\to 0} \frac{1}{\Delta x}\,\text{Prob.}\left[X\in[x,x+\Delta x]\right].
\end{equation}

What is the probability measure associated with $X$, denoted here as $\mathbb{P}$? While we refer to~\cite{Klebaner2012,Pavliotis2014} for a formal yet simple definition of the measure associated with a random variable, for computational purposes we simply state here that the measure is the (normalized and nonnegative) function that weights the values of the random variable's domain when computing expected values. Let $Z = z(X)$, then
\begin{equation}\label{eq:expected_value_rv}
    \E_\mathbb{P}\left(Z\right) = \int z(x)\,d\mathbb{P}(x).
\end{equation}

What is the form of $d\mathbb{P}(x)$? The probability measure is characterized by its ratio with respect to other measures. In the context of random variables, we typically use the ratio between the random variable's measure and the Lebesgue measure, which coincides with the probability density function $p(x)$,
\begin{equation}\label{eq:def_of_prob_density_as_RN}
    \frac{d\mathbb{P}(x)}{dx} = p(x),
\end{equation}
from which we can rewrite Eq.~\eqref{eq:expected_value_rv} in the well-known form
\begin{equation}\label{eq:expected_value_rv_2}
    \E\left(Z\right) = \int Z(x)\,p(x)\,dx.
\end{equation}

Ratios of measures such as the one in Eq.~\eqref{eq:def_of_prob_density_as_RN} can be generalized to ratios of any measures and these are referred to as \textit{Radon-Nikodym derivatives}~\cite{Klebaner2012,Wolfgang2014,Pavliotis2014}. The existence of such derivatives between arbitrary measures,
\begin{equation}
    L = \frac{d\mathbb{P}}{d\mathbb{Q}},
\end{equation}
is guaranteed under the \textit{absolute continuity} condition, denoted as $d\mathbb{P} \ll d\mathbb{Q}$, which means that the support of $\mathbb{P}$ must be contained within the support of $\mathbb{Q}$.

Radon–Nikodym derivatives can be used to compute expected values via the change-of-measure technique, which involves replacing the probability measure used to compute the expected value ($\mathbb{P} \to \mathbb{Q}$) at the cost of modifying the random variable ($Z \to LZ$).
\begin{equation}\label{eq:expected_value_rv_3}
    \E_\mathbb{P}\left(Z\right) = \E_\mathbb{Q}\left(L\,Z\right) = \int Z(x)\,L(x)\,d\mathbb{Q}(x) = \int Z(x)\,\frac{p(x)}{q(x)}\,q(x)\,dx,
\end{equation}
where $q(x) = d\mathbb{Q}(x)/dx$.

Using these concepts at the level of random variables may seem excessive, as we simply recover the standard definition of a probability density function, and the Radon-Nikodym derivative is just a ratio of probability densities. However, this framework becomes essential when applied to stochastic processes, where the probability density of entire paths is typically not accessible. In such cases, one can still derive integral representations of Radon-Nikodym derivatives, enabling rigorous comparisons between different stochastic models.

Now consider a stochastic Markov process $X_t$ with propagator 
\begin{equation}
    \rho_{s,t}(x|y) = \lim_{\Delta x\to 0} \frac{1}{\Delta x}\,\text{Prob.}\left[X_{t}\in[x,x+\Delta x]\,\big|\,X_s = y\right].
\end{equation}
For a discrete sample $\vec{x}$ with $M$ components $\{x_{t_1},\dots,x_{t_M}\}$, the joint probability --- or likelihood --- is
\begin{equation}\label{AP_eq:prob_time_series}
    L\left(\vec{x}\right) = \prod_{j=1}^{M-1}\rho_{_{t_j,t_{j+1}}} \left(x_{t_{j+1}}\Big|x_{t_{j}}\right),
\end{equation}
where, to ease notation, we omit the explicit dependence of the likelihood on the model and its parameters, in contrast to Eq.~\eqref{eq:prob_time_series}. Let us denote the initial and final times of the time series by $t_1 = 0$ and $t_M = T$, and assume a homogeneous partition $t_{j+1} - t_j = \Delta t$ for $j=1,2,\dots,M-1$, so that $M - 1 = T/\Delta t$. With this partition, the time between intermediate jumps becomes infinitesimal as $\Delta t \to 0$, while the path duration $T = (M - 1)\Delta t$ remains constant, and the time-series probability becomes a path measure:
\begin{equation}\label{AP_eq:def_as_limit_of_path_measure}
    d\mathbb{P} = \lim_{\substack{\Delta t \to 0 \\ M \to \infty}} \prod_{j=1}^{M-1}\rho_{_{t_j,t_{j+1}}} \left(x_{t_{j+1}}\Big|x_{t_{j}}\right)\, dx_j.
\end{equation}
The path measure weights continuous paths in the calculation of expectations. \rev{In particular, defining the path functional $Z=z(X_{[0,T]})$, where $X_{[0,T]}$ denotes a continuous path of duration $T$, $X_{[0,T]}=\{X_t\}_{t\in[0,T]}$} ,
\rev{\begin{equation}\label{eq:expected_value_path}
    \E\left(Z\right) = \int z(x_{[0,T]})\,d\mathbb{P}(x_{[0,T]}).
\end{equation}}

The limit in Eq.~\eqref{AP_eq:def_as_limit_of_path_measure} is a formal expression that is commonly used in applications such as average calculations~\cite{Wio2013} and serves as the starting point for saddle-point approximations at the path level~\cite{Varadhan1985}. However, in inference applications, one needs to evaluate Eq.~\eqref{AP_eq:def_as_limit_of_path_measure} under different parameters and models, which, strictly speaking, cannot be done. This becomes evident when considering a specific process as an example, together with the Gaussian (or Euler–Maruyama) approximation of the propagator. Let us define the stochastic process $X_t$ with
\rev{\begin{equation}\label{AP_eq:fully_general_1d_SDE}
    dX_t = A(X_t)\,dt + dW_t,
\end{equation}}
then the corresponding propagator reads
\rev{\begin{equation}\label{AP_eq:Gaussian_increments}
    \lim_{\Delta t\to 0} \rho_{t,t+\Delta t} \left(x \mid y \right) = \lim_{\Delta t\to 0} \frac{1}{\sqrt{2\pi\Delta t}} \exp\left[-\frac{(x - y - A(y)\Delta t)^2}{2\Delta t}\right].
\end{equation}}

Inserting Eq.~\eqref{AP_eq:Gaussian_increments} into Eq.~\eqref{AP_eq:def_as_limit_of_path_measure}, we obtain (replacing $x \to x_{j+1}$ and $y \to x_{j}$)
\rev{\begin{equation}\label{AP_eq:Gaussian_increments_lim_dt_0}
    d\mathbb{P} = \lim_{\substack{\Delta t \to 0 \\ M \to \infty}} \prod_{k=1}^{M} \frac{dx_k}{\sqrt{2\pi\Delta t}} \exp\left[ -\frac{1}{2} \sum_{j=1}^{M} \frac{(x_{j+1} - x_j)^2}{\Delta t} \right] \exp\left[ \sum_{j=1}^M (x_{j+1} - x_j) A(x_j) - \frac{1}{2} \sum_{j=1}^M A^2(x_j)\Delta t \right].
\end{equation}}

Which we rewrite as the product of a regular and a  singular term,
\rev{\begin{equation}
    d\mathbb{P} = L_{T} \,d\mathbb{W}.
\end{equation}}
Using the Riemann-Stieltjes and It\^o integrals~\cite{Klebaner2012}, the first term reads
\rev{\begin{equation}
    L_T =\lim_{\substack{\Delta t \to 0 \\ M \to \infty}}\exp\left[ \sum_{j=1}^M (x_{j+1} - x_j) A(x_j) - \frac{1}{2} \sum_{j=1}^M A^2(x_j)\Delta t \right]= \exp\left[ \int_0^T A(x_t)\,dx_t - \frac{1}{2} \int_0^T A^2(x_t)\,dt \right],
\end{equation}}
which constitutes the regular part of Eq.~\eqref{AP_eq:Gaussian_increments_lim_dt_0}, as it is a well-defined quantity: nothing prevents these integrals from existing for any realization \( x_{[0,T]} \).

In contrast, 
\begin{equation}\label{AP_label:Wiener_measure}
    d\mathbb{W} = \lim_{\substack{\Delta t \to 0 \\ M \to \infty}} \prod_{i=k}^M\frac{ dx_k}{\sqrt{2\pi\Delta t}} \exp\left[ -\frac{1}{2} \sum_{j=1}^M \frac{(x_{j+1} - x_j)^2}{\Delta t} \right],
\end{equation}
which we identify as the Wiener measure, cannot be evaluated directly, since the sum in Eq.~\ref{AP_label:Wiener_measure} corresponds to the quadratic variation, which converges along stochastic paths~\cite{Klebaner2012}, whereas the normalization factor diverges.

In summary, the path measure, which we wrote as \rev{$d\mathbb{P} = L_T d\mathbb{W}$}, cannot be evaluated explicitly~\cite{Roldan2023}, although it can be formally defined in specific contexts.

Even though path measures cannot be evaluated directly, their ``ratios"—known as Radon-Nikodym derivatives—are well defined. In particular, 
\rev{\begin{equation}\label{AP_eq:Girsanov}
    \frac{d\mathbb{P}}{d\mathbb{W}} = L_T = \exp\left[ \int_0^T A(x_t)\,dx_t - \frac{1}{2} \int_0^T A^2(x_t)\,dt \right],
\end{equation}}
which is a statement of Girsanov’s theorem: the Radon-Nikodym derivative between the additive process defined in Eq.~\eqref{AP_eq:fully_general_1d_SDE} and the Wiener process is given by Eq.~\eqref{AP_eq:Girsanov}.

The possibility of computing Radon-Nikodym derivatives between path measures allows using changes of measure within the context of real-valued continuous stochastic processes, just as we did before in the case of random variables
\rev{\begin{equation}\label{eq:expected_value_rv_4}
\E_\mathbb{P}\left(Z\right) = \E_\mathbb{Q}\left(L_T\,Z\right) = \int L_T\,z(x_{[0,T]})\,d\mathbb{Q}(x_{[0,T]}),
\end{equation}}
where the existence of \rev{$L_T=d\mathbb{P}/d\mathbb{Q}$} is ensured if absolute continuity, $d\mathbb{P}\ll d\mathbb{Q}$, holds; that is, if the support of the $\mathbb{Q}$-processes contains that of the $\mathbb{P}$-processes. For real-valued additive processes, Radon-Nikodym derivatives can be computed using a combination of the chain rule and Eq.~\eqref{AP_eq:Girsanov},
\rev{\begin{equation}\label{AP_eq:chain_rule_RN}
    L_T=\frac{d\mathbb{P}}{d\mathbb{Q}}=\frac{d\mathbb{P}}{d\mathbb{W}} \left(\frac{d\mathbb{Q}}{d\mathbb{W}}\right)^{-1}.
\end{equation}}
For non-additive processes, the same reasoning leads to an expression for the Radon-Nikodym derivative. In particular, consider that the path measure $\mathbb{P}$ is associated now to the general process
\rev{\begin{equation}
    dX_t = A(X_t)\,dt+B(X_t)\,dW_t,
\end{equation}}
while $\mathbb{V}$ is the path measure for the diffusion
\begin{equation}\label{AP_eq:dritless_process}
    dX_t = B(X_t)\,dW_t.
\end{equation}
Then, the Radon-Nikodym derivative reads
\rev{\begin{equation}\label{AP_eq:generalized_Girsanov}
    L_T=\frac{d\mathbb{P}}{d\mathbb{V}}  = \exp\left[ \int_0^T \frac{A(x_t)}{B^2(x_t)}\,dx_t - \frac{1}{2} \int_0^T \frac{A^2(x_t)}{B^2(x_t)}\,dt \right],
\end{equation}}
which can also be derived through the Onsager-Machlup functional~\cite{Onsager1953,Langouche1982,Arnold2000,Wio2013}. Naturally, the derivative between two measures ($\mathbb{P}$ and $\mathbb{Q}$) that share the same diffusion function, and with $\mathbb{P}\ll\mathbb{Q}$, generalizes Eq.~\eqref{AP_eq:chain_rule_RN_V2}
\rev{\begin{equation}\label{AP_eq:chain_rule_RN_V2}
    L_T=\frac{d\mathbb{P}}{d\mathbb{Q}}=\frac{d\mathbb{P}}{d\mathbb{V}} \left(\frac{d\mathbb{Q}}{d\mathbb{V}}\right)^{-1}.
\end{equation}}
Eqs.~\eqref{AP_eq:generalized_Girsanov} and~\eqref{AP_eq:chain_rule_RN_V2} allows us to evaluate the Radon-Nykodym derivatives present in the bridge change of measure estimator of the propagator (Eqs.~\eqref{eq:QB_estimation} and~\eqref{eq:estimator_BCM} in the main text).
\section{Path integral estimation of drift function parameters}\label{AP_sec:MLE_Girsanov}

In this section, we make explicit the maximum likelihood estimation (MLE) of parameters in the drift function of stochastic differential equations, using the path integral representation derived from Girsanov’s theorem. \rev{We first consider a model with a general dependence on the process, but in which only the drift term depends linearly on an $N$-component vector of target parameters $\vec{\Theta}$. That is,
$
A(X_t,\vec{\Theta}) = \vec{\Theta} \cdot \vec{g}(X_t)$,  while $B(X_t,\vec{\Theta}) = B(X_t)$},
\begin{equation}\label{AP_eq:GAM}
    dX_t  = \vec{\Theta} \cdot \vec{g}(X_t)\,dt\,+B(X_t)\,dW_t.
\end{equation}
\rev{Hence,} each component of $\vec{\Theta}$ is a parameter to be determined and each component of $\vec{g}(\cdot)$ is an arbitrary function non-dependent on $\vec{\Theta}$. For example, consider the process,
\begin{equation}\label{AP_eq:lamperti_transformed_model}
    dX_t = \alpha X_t^{w_\alpha} + \beta X_t^{w_\beta}+B(X_t)\,dW_t,
\end{equation}
where ${w_\alpha}$ and ${w_\beta}$ are considered here as known numbers, and $B(X_t)$ is also considered as known (in applications, parameters of the noise function will be determined by other means, see Appendix~\ref{AP_sec:QV_calculations}). In this example, the components of the vectors $\vec{\Theta}$ and $\vec{g}(\cdot)$ read
\begin{equation}
    \Theta_1=\alpha,\quad\Theta_2=\beta,\quad g_1(x)=x^{w_\alpha},\quad g_2(x)=x^{w_\beta}.
\end{equation}
The family of process in Eq.\eqref{AP_eq:lamperti_transformed_model} encompasses all distinct models considered in this work.

According to Girsanov's theorem [Eq.~\eqref{AP_eq:generalized_Girsanov}], the Radon-Nykodim derivative between the path-measure of the process in Eq.~\eqref{AP_eq:GAM} and its associated process without drift [Eq.~\eqref{AP_eq:dritless_process}] reads 
\rev{\rev{\begin{equation}\label{AP_eq:Girsanov_GAM}
    L_T = \exp\left[ \int_0^T \frac{\vec{\Theta} \cdot \vec{g}(x_t)}{B^2(x_t)}\,dx_t - \frac{1}{2} \int_0^T \left(\frac{\vec{\Theta} \cdot \vec{g}(x_t)}{B(x_t)}\right)^2\,dt \right],
\end{equation}}
where $x_t$ denotes a particular path with $t\in[0,T]$. Maximizing $L_T$ is equivalent to maximizing the normalized log-likelihood, which has the advantage of being intensive in the path duration, $T$:
\begin{equation}\label{AP_eq:ell_GAM}
    \ell_T = \frac{1}{T}\log\left(L_T\right) =  \frac{1}{T}\int_0^T \frac{\vec{\Theta} \cdot \vec{g}(x_t)}{B^2(x_t)}\,dx_t - \frac{1}{2T} \int_0^T \left(\frac{\vec{\Theta} \cdot \vec{g}(x_t)}{B^2(x_t)}\right)^2\,dt ,
\end{equation}}
Since the normalized log-likelihood is quadratic in the parameters, it has only one extrema which fulfills
\rev{\begin{equation}
    \nabla_{\vec{\Theta}}\ell_T = \vec{0}.
\end{equation}}
Rewriting Eq.~\eqref{AP_eq:ell_GAM} as 
\rev{\begin{equation}
    \ell_T = \sum_{i=1}^{N}\left(\Theta_i\,[g_i]_{_B}-\frac{1}{2}\Theta_i^2\langle g_i^2\rangle_{_B}-\Theta_i\sum_{j>i} \Theta_j\langle g_i g_j\rangle_{_B}\right),
\end{equation}}
where we introduced the short notation for process and time integrals
\begin{equation}\label{AP_eq:temporal_and_process_int}
    \langle f \rangle_{_B} = \frac{1}{T} \int_0^T  \frac{f(x_t)}{B^2(x_t)}\,dt, \qquad [f]_{_B} = \frac{1}{T} \int_0^T \frac{f(x_t)}{B^2(x_t)}\,dx_t.
\end{equation}
Deriving the above expression and equating to zero we find
\rev{\begin{equation}
    \partial_{\Theta_i} \ell_T = 0\to [g_i]_{_B} = \sum_{j=1}^N \Theta_j \langle g_ig_j\rangle_{_B},
\end{equation}}
which forms a linear equation for $\vec{\Theta}$. In vectorial form it reads
\begin{equation}
    \mathbf{A} \,\vec{\Theta} = \vec{[g]}_{_B},
\end{equation}
with $\left(\mathbf{A}\right)_{i,j} = \langle g_i g_j \rangle_{_B}$ a symmetric matrix, and $\left(  \vec{[g]}_{_B} \right)_i = [g_i]_{_B}$. Therefore, if the matrix $\mathbf{A}$ is invertible, the parameter vector reads
\begin{equation}\label{AP_eq:general_MLE}
    \vec{\Theta}=\mathbf{A}^{-1} \vec{[g]}_{_B}.
\end{equation}
\rev{Given the quadratic form of the log-likelihood in Eq.~\eqref{AP_eq:ell_GAM}, the extremum in Eq.~\eqref{AP_eq:general_MLE} corresponds to a local maximum. }

In particular, if $N=2$, and $Z = \langle g_1^2\rangle_{_B} \langle g_2^2\rangle_{_B} -\langle g_1g_2\rangle_{_B}^2\neq0$,  we find
\begin{equation}\label{AP_eq:alpha_and_beta_estimators}
\begin{cases}
    \Theta_1 = \frac{\langle g^2_2\rangle_{_B}[ g_1]_{_B}-\langle g_2\,g_1\rangle_{_B}[ g_2]_{_B}}{Z}, \\
    \Theta_2 = \frac{\langle g^2_1\rangle_{_B}[ g_2]_{_B}-\langle g_2\,g_1\rangle_{_B}[ g_1]_{_B}}{Z}.
\end{cases}  
\end{equation}
Using the above expression in the model of Eq.\eqref{AP_eq:lamperti_transformed_model}, i.e. $\Theta_1 = \alpha$, $\Theta_2 = \beta$, $g_1(x)=x^{w_\alpha}$, $g_2(x) = x^{w_\beta}$, we obtain 
\begin{equation}\label{eq:MLE_estimator_continuous_paths}
    \begin{cases}
        \hat{\alpha} = \dfrac{\langle X^{w_\alpha + w_\beta} \rangle_{_B}\, [X^{w_\beta}]_{_B} - \langle X^{2w_\beta} \rangle_{_B}\, [X^{w_\alpha}]_{_B}}{Z}, \\ \\
        \hat{\beta} = \dfrac{\langle X^{w_\alpha + w_\beta} \rangle_{_B}\, [X^{w_\alpha}] - \langle X^{2w_\alpha} \rangle_{_B}\, [X^{w_\beta}]_{_B}}{Z},
    \end{cases}
\end{equation}
with normalization factor
\begin{equation}
    Z = \langle X^{w_\alpha + w_\beta} \rangle_{_B}^2 - \langle X^{2w_\alpha} \rangle_{_B}\, \langle X^{2w_\beta} \rangle_{_B}.
\end{equation}

We used Eq.~\eqref{eq:MLE_estimator_continuous_paths} to compute the estimators $\hat{\mu}$ and $\hat{k}$ for the models shown in Fig.~\eqref{fig:Estimation_mu_k_D}. For all models, $\hat{k} = -\hat{\beta}$ and $\hat{\mu} = \hat{\alpha}/\hat{\beta}$, while the values of $w_\alpha$ and $w_\beta$ depend on the specific model. The remaining challenge is how to evaluate integrals of the form in Eq.~\eqref{AP_eq:temporal_and_process_int} using discrete data. We address this using a combination of exact and approximate methods, described in the following subsections.

\subsection{Exact calculation of It\^o integrals} 
Using It\^o's lemma, the process integrals (noted as $[\cdot]_B$) can be written as temporal integrals (noted as $\langle\cdot\rangle_{_B}$). Let $f$ be an arbitrary function and say that we aim to compute
\rev{\begin{equation}
    [f]_{_B}=\frac{1}{T}\int_{0}^{T} \tilde{f}(x_t)\,dx_t,
\end{equation}}
where \rev{$\tilde{f}(x)=f(x)/B^2(x)$}, and paths \rev{evaluated in the integral} correspond to the general process
\begin{equation}
    dX_t = A(X_t)\,dt+B(X_t)\,dW_t.
\end{equation}
Let us use the change of variables
\rev{\begin{equation}
    \tilde{F}_t = \tilde{F}(X_t),
\end{equation}}
where we defined
\rev{\begin{equation}
     \tilde{F}(x)= \int^x \tilde{f}(s)\,ds,
\end{equation}}
and where $\int^x$ stands for anti-derivative, meaning that we do not need to specify the particular primitive function of \rev{$\tilde{f}$}.  It\^o's formula for the change of variables reads\rev{
\begin{equation}
    d\tilde{F}_t = \tilde{f}(X_t)\,dX_t+\frac{1}{2} \tilde{f}'(X_t)B^2(X_t)\, dt,
\end{equation}}
where \rev{$\tilde{f}'_i(X_t)=\partial_s\tilde{f}(s)\big|_{s=X_t} $}. Therefore,
\rev{\begin{equation}\label{AP_eq:process_integral_formula}
    [f]_{_B} = \frac{1}{T} \int^{T}_{0} d\tilde{F}_t -\frac{1}{2T}\int^{T}_{0}\tilde{f}'(X_t)B^2(X_t)dt=\frac{\tilde{F}(X_{t_f})-\tilde{F}(X_{t_0})}{T}-\frac{1}{2}\langle \tilde{f}'_i\rangle_B.
\end{equation}}

When the above expression is used to evaluate the process integrals in Eq.~\eqref{eq:MLE_estimator_continuous_paths}, it can reduce computational costs, as it generally decreases the number of integrals to be computed. Moreover, in some cases Eq.~\eqref{AP_eq:process_integral_formula} can be evaluated exactly, thereby reducing numerical errors.

\subsection{Integral discretization} 
In general, the integrals in Eq.~\eqref{eq:MLE_estimator_continuous_paths} cannot be evaluated exactly and need to be discretized, introducing bias errors in the path integral MLE framework described above. The discretization should be consistent with the Itô interpretation of stochastic integrals in the infinitesimal limit. Given a test function \rev{$f(\cdot)$}, the discretized forms read:
\rev{\begin{equation}
\left[f\right]_{_B} = \frac{1}{T}\int_0^T \frac{f\left(x_t\right)}{B^2\left(x_t\right)}\,dx_t \approx \frac{1}{T} \sum_{j=1}^{M-1} \frac{f\left(x_{t_j}\right)}{B^2\left(x_{t_j}\right)}\left(x_{t_{j+1}} - x_{t_j}\right),
\end{equation}
and
\begin{equation}
\langle f \rangle_{_B} = \frac{1}{T}\int_0^T \frac{f\left(x_t\right)}{B^2\left(x_t\right)}\,dt \approx \frac{1}{T} \sum_{j=1}^{M-1} \frac{f\left(x_{t_j}\right)}{B^2\left(x_{t_j}\right)}\left(t_{j+1} - t_j\right).
\end{equation}}
When the time series is sampled at uniform intervals, i.e., $t_{i+1} - t_i = \Delta t$ for all $i$ then $T=(M-1)\Delta t$, and the time integral simplifies to the empirical mean:

\rev{\begin{equation}
\langle f \rangle_{_B} \approx \frac{1}{M-1} \sum_{j=1}^{M-1} \frac{f\left(x_{t_j}\right)}{B^2\left(x_{t_j}\right)}.
\end{equation}}

\section{Estimation of noise parameter through quadratic variation}\label{AP_sec:QV_calculations}
The estimation of $D$ is carried out using quadratic variation applied to the Cole–Hopf–Lamperti-transformed process~\cite{Pavliotis2014}. Specifically, consider the stochastic differential equation
\begin{equation}
    dX_t = A(X_t)\,dt + D\,\tilde{B}(X_t)\,dW_t,
\end{equation}
where $\tilde{B}$ here is considered to be a known function. Define the transformed process $Y_t = h(X_t)$, where
\begin{equation}
    h(x) = \int^x \frac{1}{\tilde{B}(s)}\,ds,\quad h'(x) = \frac{1}{\tilde{B}(x)},\quad h''(x)= -\frac{\tilde{B}'(x)}{\tilde{B}^2(x)}.
\end{equation}
Then, in virtue of It\^o's formula, the process $Y_t$  satisfies
\begin{equation}\label{AQ_eq:transformed_process_V2}
    dY_t = h'(X_t)\,dX_t+\frac{1}{2} \tilde{B}^2(X_t) h''(X_t) = \tilde{A}(Y_t)\,dt + D\,dW_t,
\end{equation}
where we introduced the Lamperti-transformed drift that encodes the diffusion function in the original process
$$
\tilde{A}(x) = \frac{A(h^{-1}(x))}{\tilde{B}(h^{-1}(x))} - \frac{\tilde{B}'(h^{-1}(x))}{2},
$$
with $h^{-1}(\cdot)$ noting the inverse function of $h(\cdot)$.

As discussed in Sec.~\ref{sec:quadratic_variation}, the parameter $D$ in Eq.~\eqref{AQ_eq:transformed_process_V2} satisfies
\begin{equation}\label{AP_eq:quadratic_variation_additive}
    D^2 = \lim_{\Delta t \to 0} \frac{1}{\Delta t} \left\langle \left(Y_{t+\Delta t} - Y_t\right)^2 \right\rangle.
\end{equation}
But actually, we can make an stronger claim, since the quadratic variation computed over a path and the diffusion function are related in a strong sense~\cite{Klebaner2012}---i.e. over each stochastic path rather than only over averages---. Therefore, we can write
\begin{equation}\label{AP_eq:QV_additive}
    D^2=\frac{1}{T}\lim_{\Delta t\to0}\sum_{i=1}^{M-1} (Y_{t_{i+1}}-Y_{t_i}),
\end{equation}
where the above expression is valid over \rev{continuous stochastic paths, and where $t_{i+1}-t_i=\Delta t$, and $T=\Delta t\,(M-1)$ is finite.}

Given a time series $\{x_{t_i}\}_{i=1,\dots,M}$, the transformed series $\{y_{t_i}\}_{i=1,\dots,M}$ is obtained by applying $h$ to each data point:
\begin{equation}
    y_{t_i} = h(x_{t_i}).
\end{equation}

Then, the quadratic variation estimator of $D$ is computed by removing the limit in Eq.~\eqref{AP_eq:QV_additive}:
\begin{equation}\label{AP_eq:QV_estimator_D}
    \hat{D} = \sqrt{ \frac{1}{\Delta t (M-1)} \sum_{i=1}^{M-1} \left(y_{t_{i+1}} - y_{t_i} \right)^2 }.
\end{equation}

Table~\ref{AP_tab:h_transformations} provides the explicit expressions for the transformation $h(\cdot)$ for the different models considered in the text.
\begin{table}[h]
\centering
\begin{tabular}{|c|c|c|}
\hline
\textbf{Model} & $\,\tilde{B}(x)\,$ & $\,h(x)\,$ \\
\hline
Demographic and contact process models [Eqs.~\eqref{eq:general_model_demographic},\eqref{eq:CP_model}] & $\sqrt{x}$ & $2\sqrt{x}$ \\
\hline
Ornstein-Uhlenbeck process [Eq.~\eqref{eq:OU_process}] & $1$ & $x$ \\
\hline
Environmental model [Eq.~\eqref{eq:general_model_environmental}] & $x$ & $\log(x)$ \\
\hline
General gamma model for $\theta\neq1$ [Eq.~\eqref{eq:general_model_theta}] & $x^{\frac{\theta+1}{2}}$ & $\frac{1}{1-\theta}x^{\frac{1-\theta}{2}}$ \\
\hline
\end{tabular}
\caption{Lamperti transformations $h(x)$ used for different models.}
\label{AP_tab:h_transformations}
\end{table}

\subsection{MLE estimation of $D$}\label{AP_sec:MLE_Gauss}
\begin{figure}
    \centering
    \includegraphics[width=0.5\linewidth]{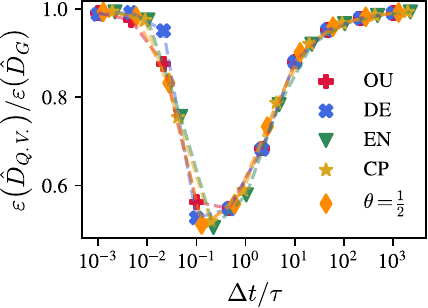}
    \caption{ relative errors of the estimation of D using the path-integral approximation $\varepsilon(\hat{D}_G)$ are in general
bigger than the same estimation using the quadratic variation method $\varepsilon(\hat{D}_{Q.V,})$. This is true for synthetic time-series generated
for all models in Sec.~\ref{sec:high_frec_inference}. }
    \label{AP_fig:D_estimation}
\end{figure}

One could proceed as we did with the parameters of the drift function in section~\ref{AP_sec:MLE_Girsanov} and obtain an estimator for $D$ minimizing the time series likelihood using the Gaussian (Euler-Maruyama) approximation for the propagator. Once this is done, the resulting expression would read
\rev{\begin{equation}
    \partial_D\ell_T = -(M-1)+\frac{1}{D^2}\sum_{i=1}^{M-1}\left[-\frac{(x_{t_{i+1}} - x_{t_{i}} - A(x_{t_{i}})\Delta t)^2}{2\Delta t \,\tilde{B}^2(x_{t_{i}})}\right]=0,
\end{equation}}
therefore,
\begin{equation}\label{AP_eq:Gaussian_estimator_of_D}
    \hat{D}^{(G)} = \sqrt{\frac{1}{M-1}\sum_{i=1}^{M-1}\frac{(x_{t_{i+1}} - x_{t_{i}} - A(x_{t_{i}})\Delta t)^2}{2\Delta t \,\tilde{B}^2(x_{t_{i}})}}.
\end{equation}
The estimator in Eq.~\eqref{AP_eq:Gaussian_estimator_of_D} converges to the quadratic variation estimator [Eq.~\eqref{AP_eq:QV_estimator_D}] in the limit $\Delta t \to 0$, where the latter becomes exact. However, Figure~\ref{AP_fig:D_estimation} shows that the estimator in Eq.~\eqref{AP_eq:QV_estimator_D} is more precise than the one in Eq.~\eqref{AP_eq:Gaussian_estimator_of_D} when evaluated in time series with finite $\Delta t$.

\section{Computation of 
${d\mathbb{P}}/{d\mathbb{Q}^{(B)}}$
}\label{AP_sec:RN_wrt_bridge_measure}

In this section, we give analytical form to the Radon-Nikodym derivative between a target measure $\mathbb{P}$ and the measure of a bridge process, $\mathbb{Q}^{(B)}$. Let the two measures $\mathbb{Q}$ and $\mathbb{P}$ be such that $\mathbb{P}\ll\mathbb{Q}$, meaning that the support of $\mathbb{Q}$ contains that of $\mathbb{P}$. We also assume that both measures are absolutely continuous with respect to the Wiener measure, $\mathbb{P}\ll\mathbb{W}$, and $\mathbb{Q}\ll\mathbb{W}$. In other words, we consider the Wiener process and other two \emph{additive} processes (the $\mathbb{P}$-process and the $\mathbb{Q}$-process) that take values in a subset of the reals.

The $\mathbb{Q}$-process has an associated bridge. The bridges of the $\mathbb{Q}$-process are realizations constrained to fixed initial and final points. That is, in the bridge process the values at the initial and final times ($t=0$ and $t=T$, respectively) are not random: for all realizations, the states at these times are always the same (denoted $X_0 = x_0$ and $X_T = x_T$, respectively). The bridge defines itself a measure, here called $\mathbb{Q}^{(B)}$. Although the unconstrained ($\mathbb{Q}$)  and bridge ($\mathbb{Q}^{(B)}$) measures are related, they assign different weights to paths, hence their path measures are different in general. For example, taking as the $\mathbb{Q}$-process a Brownian motion with $X_0=0$ and
\begin{equation}
    dX_t = D\,dW_t,
\end{equation}
all paths with $X_T\in\mathbb{R}$ will have some non-null measure. Instead, its associated bridge with measure $\mathbb{Q^{(B)}}$ is described through the SDE
\begin{equation}
    dX_t = \frac{X_t-x_T}{T-t}dt+D\,dW_t,
\end{equation}
and the set of all paths with $X_T\neq x_T$ has zero measure.

Typically, the bridge process is characterized through its infinitesimal generator, which can be obtained as a Doob-h transform of the generator of the unconditioned process~\cite{Asmussen2007,Chetrite2013,Chetrite2015}. Instead, we characterize the bridge measure with the following theorem:

\begin{theorem}\label{AP_Th:RN_Q_Q^B}
    Let $\mathbb{Q}$ be a Markov path measure with fixed initial condition $X_0=x_0$, and propagator $\rho^{(\mathbb{Q})}$, meaning that $\E_{\mathbb{Q}}\left[\delta(X_t-x)|X_0=x_0\right]=\rho^{(\mathbb{Q})}_{0,t}(x|x_0)$. Then, the Radon-Nikodym derivative between the unconstrained process $\mathbb{Q}$ and the bridge measure constraining paths to $X_0=x_0$ and $X_T=x_T$ reads
    \begin{equation}\label{AP_eq:RN_Q_Q^B}
    \frac{d\mathbb{Q}(x_{[0,T]})}{d\mathbb{Q}^{(B)}(x_{[0,T]})}=\rho^{(\mathbb{Q})}_{0,T}(x_T|x_0).
    \end{equation}
\end{theorem}
\begin{proof}
Let us first remind that the propagator of the bridge process is that of the unconstrained process adding the condition $X_T=x_T$. \rev{Considering an intermediate state $y$ at time $s$,
\begin{equation}\label{AP_eq:prop_bridge}
\rho^{(\mathbb{Q^{(B)}})}_{s,t}(x|y)=\rho^{(\mathbb{Q})}_{s,t,T}(x|y,x_T),\quad\forall\, 0\le s \le t\le T,
\end{equation}
with 
\begin{equation}
    \rho^{(\mathbb{Q})}_{s,t,T}(x|y,x_T)=\E_{\mathbb{Q}}\left[\delta(X_t-x)|X_s=y,X_T=x_T\right].
\end{equation}
Then, using the Markov property together with the definition of conditioned probabilities:
\begin{equation}\label{AP_eq:prop_conditioned}
    \rho^{(\mathbb{Q^{(B)}})}_{s,t}(x|y)=\rho^{(\mathbb{Q})}_{s,t,T}(x|y,x_T) = \frac{\rho^{(\mathbb{Q})}_{s,t,T}(x,y,x_T)}{\rho^{(\mathbb{Q})}_{s,T}(y,x_T)}=\frac{\rho^{(\mathbb{Q})}_{s,t}(x|y)\rho^{(\mathbb{Q})}_{t,T}(x_T|x)}{\rho^{(\mathbb{Q})}_{s,T}(x_T|y)},
\end{equation}
where we used $\rho^{(\mathbb{Q})}_{s,t,T}(x_T|x,x_0)=\rho^{(\mathbb{Q})}_{t,T}(x_T|x)$.} As in Eq.~\eqref{AP_eq:def_as_limit_of_path_measure}, let us build the infinitesimal measures $d\mathbb{Q}$ and $d\mathbb{Q}^{(B)}$ through the infinitesimal partition $0=t_1<t_2<\dots<t_N=T$, with $t_{i+1}-t_i=\Delta t$, \rev{ $\forall i$}, and the product of propagators:
\begin{equation}
    d\mathbb{Q}[x_{[0,T]}|x_0] = \lim_{\Delta t\to0} \prod_{i=1}^N dx_i \,\rho^{(\mathbb{Q})}_{_{t_i,t_{i+1}}} \left(x_{t_{i+1}}\Big|x_{t_{i}}\right) ,
\end{equation}
and 
\begin{equation}
    d\mathbb{Q}^{(B)}[x_{[0,T]}|x_0] = \lim_{\Delta t\to0} \prod_{i=1}^N dx_i \,\rho^{(\mathbb{Q}^{(B)})}_{_{t_i,t_{i+1}}} \left(x_{t_{i+1}}\Big|x_{t_{i}}\right) ,
\end{equation}
\rev{Using the above expressions together with Eq.~\eqref{AP_eq:prop_conditioned} one finds Eq.\eqref{AP_eq:RN_Q_Q^B} through a telescope simplification.}
\end{proof}

It is remarkable that the Radon-Nikodym derivative in Eq.~\eqref{AP_eq:RN_Q_Q^B} depends only on the initial and final states of the process, not on the whole path as it usually happens, e.g. Eq.~\eqref{AP_eq:generalized_Girsanov}. Just as Girsanov's theorem allows computing Radon-Nikodym derivatives with respect to the Wiener measure, Theorem~\eqref{AP_Th:RN_Q_Q^B} allows us to compute Radon-Nikodym derivatives with respect to bridge measures. In particular, using the chain rule of Radon-Nikodym derivatives,
\begin{equation}\label{AP_eq:RN_P_Q^B}
    \frac{d\mathbb{P}}{d\mathbb{Q}^{(B)}} =  \frac{d\mathbb{P}}{d\mathbb{Q}}\frac{d\mathbb{Q}}{d\mathbb{Q}^{(B)}}=\frac{d\mathbb{P}}{d\mathbb{Q}}\rho^{(\mathbb{Q})}_{0,T}(x_T|x_0).
\end{equation}
\rev{The above discussion can be extended to the case of multiplicative processes. For the ratio $d\mathbb{P}/d\mathbb{Q}$ to exist, the $\mathbb{Q}$- and $\mathbb{P}$-processes must share the same diffusion (noise) coefficient~\cite{Klebaner2012}. If this condition holds, we can proceed with the computation by applying the chain rule with respect to their associated driftless reference process [Eq.~\eqref{AP_eq:dritless_process}].} 

\section{Exact sampling of stochastic bridges}\label{AP_sec:exact_bridges}

In this section, we demonstrate the generation of bridge processes when the propagator of its associated unconstrained process is known. Consider a process with path measure $\mathbb{Q}$ and initial condition $X_s = x_s$, for which the propagator $\rho_{s,t}^{(\mathbb{Q})}(x \mid x_s)$ is known. Then, according to Eqs.~\eqref{AP_eq:prop_bridge} and~\eqref{AP_eq:prop_conditioned}, the propagator of the bridge conditioned to end at $X_T = x_T$ is given by:
\begin{equation}\label{AP_eq:prop_bridge_V2}
    \rho_{s,t}^{(\mathbb{Q}^{(B)})}(x_t \mid x_s) = \frac{\rho^{(\mathbb{Q})}_{s,t}(x_t \mid x_s)\rho^{(\mathbb{Q})}_{t,T}(x_T \mid x_t)}{\rho^{(\mathbb{Q})}_{s,T}(x_T \mid x_s)}, \quad \forall\, s \le t \le T.
\end{equation}

When the propagator of the $\mathbb{Q}$-process is known, it is, in principle, always possible to devise an exact sampling method to generate random variables from Eq.~\eqref{AP_eq:prop_bridge_V2}. For instance, if sampling from $\rho^{(\mathbb{Q})}_{s,t}(x \mid x_s)$ is feasible, one may implement a proposal-rejection sampling scheme~\cite{Toral2014}: propose values of $x$ by sampling from $\rho^{(\mathbb{Q})}_{s,t}(x \mid x_s)$, and accept each proposal with probability density proportional to $\rho^{(\mathbb{Q})}_{t,T}(x_T \mid x)$, normalized by $Z = \int \rho^{(\mathbb{Q})}_{t,T}(x_T \mid x)\, dx$.

Simpler schemes can be devised for specific models in which the product $\rho^{(\mathbb{Q})}_{s,t}(x \mid x_s)\rho^{(\mathbb{Q})}_{t,T}(x_T \mid x)$ simplifies to a known distribution for which efficient sampling routines already exist. We now discuss two of such cases, which will be of use in our bridge change of measure estimator of the propagator.

\subsection{Gaussian bridges}
It is known that the Ornstein-Uhlenbeck and Wiener bridges are Gaussian processes~\cite{Majumdar2015}, so that the bridge propagator is a Gaussian
\begin{equation}
    \rho^{(\mathbb{Q^{(B)}})}_{s,t}(x \mid x_s)=G(x;\mu^{(B)}(x_s,x_T,x,s,T),\sigma^{(B)}(x_s,x_T,x,s,T)).
\end{equation}
With
\begin{equation}
    G(x;\mu,\sigma)=\sqrt{\frac{1}{2\pi\sigma^2}}\exp\left[-\frac{\left(x-\mu\right)^2}{2\pi\sigma^2}\right].
\end{equation}
In particular, the first and second centered moments for the Wiener and OU  bridges read
\begin{equation}
    \mu^{(B)}(t)=
    \begin{cases}
    \left[x_s(T-t)-x_T(t-s)\right](T-s)^{-1},\quad(\text{WI})
    \\ 
        \mu +\left[\left(x_s-\mu\right)\sinh\left[k(T-t)\right]+(x_T-\mu)\sinh\left[k(t-s)\right]\right]\sinh\left[k(T-s)\right]^{-1},\quad(\text{OU})
    \end{cases}
\end{equation}
and
\begin{equation}
    \left(\sigma^{(B)}\right)^2(t)=
    \begin{cases}
    D^2\left[(T-t)(t-s)\right]\,(T-s)^{-1},\quad(\text{WI})
    \\ 
        \frac{D^2}{k}\left[\sinh\left[k(T-t)\right]\sinh\left[k(t-s)\right]\right]\sinh\left[k(T-s)\right]^{-1}.\quad(\text{OU})
    \end{cases}
\end{equation}
Therefore, given the OU or Wiener bridges in state $X_s=x_s$, the state $x_{s+\Delta t}$ is simulated as
\begin{equation}\label{AP_eq:Gaussian_bridge}
    x_{s+\Delta t} = \mu^{(B)}(\Delta t)+\sigma^{(B)}(\Delta t)\,\mathcal{N}(0,1),
\end{equation}
where $\mathcal{N}(0,1)$ is a Gaussian random variable with zero mean and unit variance. We note that Eq.~\eqref{AP_eq:Gaussian_bridge} generates exact realizations of the bridge for any $\Delta t$.

\subsection{Geometric Brownian bridges}
Considering as the $\mathbb{Q}$-process the geometric Brownian motion, 
\begin{equation}
    dX_t = kX_t\,dt+DX_t\,dW_t,
\end{equation}
its propagator reads
\begin{equation}\label{AP_eq:prop_GB}
    \rho^{(\mathbb{Q})}_{s,t}(x|x_s)=\sqrt{\frac{1}{2\pi D^2 x^2(t-s)}}\exp\left[-\frac{\left(\log(x/x_s)-(k-\frac{1}{2}D^2)(t-s)\right)^2}{2D^2(t-s)}\right].
\end{equation}
Inserting Eq.~\eqref{AP_eq:prop_GB} into Eq.~\eqref{AP_eq:prop_bridge_V2} and after a series of manipulations we find

\begin{equation}\label{AP_eq:prop_GB_bridge}
    \rho^{(\mathbb{Q^{(B)}})}_{s,t}(x \mid x_s) = \sqrt{\frac{1}{2 \pi \sigma_B^2 \,x^2}}\exp\left[-\frac{\left(\log(x)-\mu_B\right)^2}{2\,\sigma_B^2}\right],
\end{equation}
with
\begin{equation}
    \mu_B (t)= \log(x_T)\frac{t-s}{T-s}+\log(x_0)\frac{T-t}{T-s},
\end{equation}
and
\begin{equation}
    \sigma^2_B(t)=D^2\frac{(T-t)(t-s)}{T-s}.
\end{equation}
The distribution in Eq.~\eqref{AP_eq:prop_GB_bridge} is a log-normal distribution, and therefore, given that the bridge is in position $X_s=x_s$, new states of the geometric Brownian bridge can be sampled as
\begin{equation}\label{AP_eq:geometric_bridge}
    x_{s+\Delta t} = \exp\left(\mu^{(B)}(\Delta t)+\sigma^{(B)}(\Delta t)\,\mathcal{N}(0,1)\right),
\end{equation}
which also produces exact instances of the geometric bridge for any $\Delta t$.

\section{Propagator estimation with bridge change of measure}\label{AP_sec:guide_BCM}
We first provide a detailed example of how to use the bridge change of measure for propagator estimation, and then describe the general algorithm. \rev{As in the main text, we describe how to estimate the propagator over a specific increment, from an initial state $y$ at time $t=0$, $(0, y)$, to a final state $x$ at time $t=T$, $(T, x)$. The extension to a generic interval—such as the $i$-th pair of consecutive observations in a time series—is straightforward, upon the substitutions $x \to x_{i+1}$, $y \to x_i$, and $T \to t_{i+1} - t_i$.}. As a case study, we consider the Ornstein–Uhlenbeck process, for which this sampling method is unnecessary since the exact form of the propagator is known. We use this example both as an illustration and to assess the precision of our method. In particular, we consider
\begin{equation}\label{eq:SDE_OU_V2}
    dX_t = k\left(\mu-X_t\right)\,dt+dW_t,
\end{equation}
with propagator,
\rev{\begin{equation}
    \rho_{0,T}(x|y)=G\left(x,\mu+(y-\mu)e^{kT},\frac{1}{2k}\left(1-e^{2kT}\right)\right).
\end{equation}}
We first try our method using the Wiener process as the auxiliary process. Therefore, to match with the notation of the previous sections, the measure of the target process ($\mathbb{P}$) is the Ornstein-Uhlenbeck measure, the unconstrained auxiliary process is the Wiener process (therefore $\mathbb{Q}=\mathbb{W}$) , and the bridge measure ($\mathbb{Q}^{(B)}= \mathbb{W}^{(B)}$) is the one associated with the Wiener bridge. The processes associated to the $\mathbb{W}$ and  $\mathbb{W}^{(B)}$ processes are, respectively, Eqs.~\eqref{eq:SDE_brownian} and~\eqref{eq:SDE_brownian_bridge} setting $D=1$.

We can generate an ensemble of $N_B$ trajectories of the Wiener bridges connecting the initial state ($s,y$) and ($t,x$) with time discretization $\Delta t^{(B)}$. These trajectories are sampled directly from the propagator of the Wiener bridge, and therefore trajectories are sampled efficiently and without errors (see details in Appendix~\ref{AP_sec:exact_bridges}). Computing the estimator in Eq.~\eqref{eq:estimator_BCM}
requires evaluating the Radom-Nikodym derivative $d \mathbb{P}/d \mathbb{Q}^{(B)}$, which we do using Eq.~\eqref{eq:RN_bridge} (see Appendix~\ref{AP_sec:RN_wrt_bridge_measure}),
\rev{\begin{equation}\label{AP_eq:RN_P_Wiener}
    L_T=\frac{d\mathbb{P}(x^{(B)}_{[0,T]})}{d\mathbb{Q}^{(B)}(x^{(B)}_{[0,T]})} =\frac{d\mathbb{P}}{d\mathbb{Q}}\frac{d\mathbb{Q}}{d\mathbb{Q}^{(B)}}=\frac{d\mathbb{P}}{d\mathbb{Q}}\rho^{(\mathbb{Q})}_{0,T}(x|y)= 
    \frac{d\mathbb{P}(x^{(B)}_{[0,T]})}{d\mathbb{W}(x^{(B)}_{[0,T]})} G\left(x,y,T\right).
\end{equation}}
Where \rev{$x^{(B)}_{[0,T]}$} is a continuous path generated with the bridge process. 

The Radon-Nikodym derivative $d\mathbb{P}/d\mathbb{W}$ appearing in Eq.~\eqref{AP_eq:RN_P_Wiener}  can be computed through Eq.~\eqref{AP_eq:Girsanov}, where process integrals can be computed exactly, finding
\rev{\begin{equation}
    \frac{d\mathbb{P}(x^{(B)}_{[0,T]})}{d\mathbb{W}(x^{(B)}_{[0,T]})}  = 
    \exp\left[k\mu(x-y)-\frac{k}{2}\left(x^2-y^2\right)-\frac{k^2}{2  }\int_0^T\left(\mu-x^{(B)}_t\right)^2\,dt\right].
\end{equation}}
Paths generated with the bridge process are not really continuous, but rather discrete points where the bridge sampling time ($\Delta t^{(B)}$) can be as small as desired.
Therefore, the temporal integral in the above expression needs to be approximated over a discrete path, introducing biased errors in the method that scale with $\Delta t^{(B)}$, \rev{$\int_0^T\left(\mu-x^{(B)}_t\right)^2\,dt\approx\hat{I}$}, with
\begin{equation}
     \hat{I}=\Delta t^{(B)}\sum_j \left(\mu-x^{(B)}_{t_j}\right)^2.
\end{equation}
Therefore, we will use bridges to compute the random variable 
\rev{\begin{equation}
    L_T = 
    \exp\left[k\mu(x-y)-\frac{k}{2}\left(x^2-y^2\right)-\frac{\hat{I}k^2}{2  }\right]G\left(x,y,T\right).
\end{equation}}
Each of the $N_B$ realizations of the bridge produces a different value of $\hat{I}$, and therefore a different instance of \rev{$L_T$}. By averaging over these values, we obtain the propagator estimator \rev{using the bridge change of measure,}
\begin{equation}
    \hat{\rho}^{(\mathbb{Q}^B)} = \frac{1}{N_B}\sum_{i=1}^{N_B}L_T^{(i)}.
\end{equation}
This is the estimator used to produce Fig.~\ref{fig:BCM}. The above procedure remains the same for the propagator estimation of any additive process, simply changing Eq.~\eqref{AP_eq:RN_P_Wiener} by the more general expression
\rev{\begin{equation}\label{AP_eq:RN_bridge}
        L_T =\frac{d\mathbb{P}}{d\mathbb{Q}^{(B)}} = \frac{d\mathbb{P}}{d\mathbb{W}} \left[ \frac{d\mathbb{Q}}{d\mathbb{W}} \right]^{-1} \rho^{(\mathbb{Q})}_{0,T}(x | y),
\end{equation}}
and computing Radon-Nikodym derivatives with respect to the Wiener process using Eq.~\eqref{AP_eq:Girsanov}, while Radon-Nikodym derivatives with respect to the bridge are computed through Theorem~\ref{AP_Th:RN_Q_Q^B}.

\subsection{General algorithm}

The bridge change of measure technique can be extended to compute propagators of general multiplicative processes by means of noise transformation. Specifically, consider that we aim at computing the propagator for a \rev{$\mathbb{P}$-}process defined through the
stochastic differential equation
\begin{equation}\label{AP_eq:general_SDE}
    dX_t = A(X_t)\,dt + B(X_t)\,dW_t.
\end{equation}
Then, a we define a \rev{$\tilde{\mathbb{P}}$-}process with a desired power law as noise function,
\begin{equation}\label{AP_eq:transformed_general_SDE}
    dY_t = \tilde{A}(Y_t)\,dt + Y_t^\lambda\,dW_t,
\end{equation}
is obtained through 
transformation  \rev{$Y_t = \tilde{h}(X_t)$ , where
\begin{align}\label{AP_eq:Generalized_Lamperti_transformation}
    \tilde{h}(x) =
    \begin{cases}
    \left[(1-\lambda)\int^x B^{-1}(s)\,ds\right]^\frac{1}{1-\lambda},\quad &\text{if}\,\,\lambda\neq 1,  \\
    \exp\left[{\int^xB^{-1}(s)\,ds}\right],\quad &\text{if}\,\,\lambda=1.
    \end{cases}
\end{align}}
\begin{proof}
    We derive Eq.~\eqref{AP_eq:transformed_general_SDE} through Itô's rule,
\rev{\begin{equation}\label{AQ_eq:transformed_process}
    dY_t = \tilde{h}'(X_t)\,dX_t+\frac{1}{2} B^2(X_t) \tilde{h}''(X_t) dt= \tilde{A}(Y_t)\,dt + Y_t^\lambda dW_t,
\end{equation}}
where
\rev{\begin{equation}\label{AP_eq:Lamperti_transformation}
     \tilde{h}'(x) = \frac{\tilde{h}^\lambda(x)}{B(x)},\quad \tilde{h}''(x)= \frac{\lambda \tilde{h}^{\lambda-1}(x)B(x)-\tilde{h}^\lambda(x)B'(x)}{B^2(x)},\quad \tilde{A}(y) = \tilde{h}'\left((\tilde{h}^{-1}(y)\right)A\left((\tilde{h}^{-1}(y)\right) +\frac{1}{2} B^2\left((\tilde{h}^{-1}(y)\right) \tilde{h}''\left((\tilde{h}^{-1}(y)\right),
\end{equation}}
with \rev{$\tilde{h}^{-1}(\cdot)$ noting the inverse function of $\tilde{h}(\cdot)$ ($\tilde{h}(x)$ is invertible).}
\end{proof}

The transformation in Eq.~\eqref{AP_eq:Generalized_Lamperti_transformation} generalizes Lamperti's transformation used in Appendix~\ref{AP_sec:QV_calculations}, which is obtained as the particular case $\lambda=0$. The propagators of the original and transformed processes, here denoted as $\rho$ and $\tilde{\rho}$ respectively, are related through the Jacobian of the transformation of variables,
\rev{\begin{equation}\label{AP_eq:noise_transformation_inversion}
    \rho_{0,T}(x|y) =  \tilde{h}'(x) \,\tilde{\rho}_{0,T}(x|y) = \frac{\tilde{h}^\lambda(x)}{B(x)}\tilde{\rho}_{0,T}(x|y).
\end{equation}}
Now we have all the necessary ingredients to define the general algorithm to estimate the propagator of the process in Eq.~\eqref{AP_eq:general_SDE}.
\begin{algorithm}[H]
  \caption{Estimation of propagator $\rho_{0,T}(x|y)$ for the process in Eq.~\eqref{AP_eq:general_SDE}.}
  \label{EPSA}
   \begin{algorithmic}[1]
    \State Perform the Lamperti transformation on data, \rev{$\tilde{x}=\tilde{h}(x)$, and $\tilde{y}=\tilde{h}(y)$, where $\tilde{h}$} is given by Eq.~\eqref{AP_eq:Generalized_Lamperti_transformation}.
    \State Generate $N_B$ stochastic bridges with discretization $\Delta_B\le T$ connecting the points \rev{($0,\tilde{y}$) and ($T,\tilde{x}$)}. The idea is to generate bridges for a process that resembles as much as possible the target process while being easy to sample, in section~\ref{AP_sec:exact_bridges}, we summarize the exact generation of Wiener, Ornstein-Uhlenbeck, and geometric Brownian bridges. The form of the noise function in the bridge and target process must coincide. Therefore, for Wiener and Ornstein-Uhlenbeck, transformation of Eq.~\eqref{AP_eq:Generalized_Lamperti_transformation} must be used with $\lambda=0$, while for the geometric Brownian $\lambda=1$ must be used.
    \State \rev{Compute $L_T=d\mathbb{\tilde{P}}/d\mathbb{Q}^{(B)}=d\mathbb{\tilde{P}}/d\mathbb{Q}\,\rho^{(Q)}_{0,T}(\tilde{x}|\tilde{y})$ for each bridge. Here, $\tilde{P}$ corresponds to the measure of the transformed process [Eq.~\eqref{AP_eq:transformed_general_SDE}], $\mathbb{Q}^{(B)}$ is the measure of the bridge process, $\mathbb{Q}$ is the unconstrained measure from which the bridge process derives, and $\rho^{(Q)}$ is the propagator associated to the measure $\mathbb{Q}$ (section~\ref{AP_sec:RN_wrt_bridge_measure}). The Radon-Nikodym derivative  $d\mathbb{\tilde{P}}/d\mathbb{Q}$ is computed through discretization of Eq.~\eqref{AP_eq:chain_rule_RN_V2}. This step yields $n_B$ samples $\{L_T^{(i)}\}_{i=1,\dots,n_B}$.}
    \State Estimate the propagator of the transformed process as \rev{$\hat{\tilde{\rho}}_{0,T}(\tilde{x}|\tilde{y})=\sum_{i=1}^{N_B}L_T^{(i)}/N_B$}.
    \State Finally, obtain the propagator of the target process inverting the change of variable through Eq.~\eqref{AP_eq:noise_transformation_inversion}.
    
\noindent We note that, if the target process has the same noise function as the bridge process, there is no need of performing steps 1 and 5, since $\tilde{x}=x$, $\tilde{y}=y$, and $\hat{\tilde{\rho}}=\hat{\rho}$.  
   \end{algorithmic}
\end{algorithm}

\section{Probability of models: Bayesian Information Criterion}\label{AP_sec:BIC}

\rev{Within a Bayesian framework for model comparison, the probability of model $\mathcal{M}$ conditioned on data is defined through marginalization of the posterior over model parameters,
\begin{equation}\label{eq:prob_of_models}
\text{Prob.}\left(\mathcal{M}|\vec{x}\right)=\frac{1}{Z\left(\mathcal{M}\right)}\int d\vec{\Theta} \, L\left(\vec{x}|\vec{\Theta},\mathcal{M}\right)\pi_o\left(\vec{\Theta},\mathcal{M}\right),
\end{equation}
where we have changed the notation with respect to the main text to make explicit the dependence on the model of both the Likelihood and prior, and $Z\left(\mathcal{M}\right)$  is the model evidence, normalizing the posterior over the parameters of model $\mathcal{M}$. Hence, the most likely model generating data would be that maximizing Eq.~\eqref{eq:prob_of_models}. When the posterior is sufficiently peaked around its maximum — meaning it concentrates near the MAP estimate — a standard simplification is to evaluate the integrals in Eq.~\eqref{eq:prob_of_models} via a saddle-point approximation. Under this approximation, the MAP-optimal model becomes that minimizing a prior-weighted generalization of the Bayesian Information Criterion (BIC),
\begin{equation}\label{AP_eq:BIC}
\text{BIC}\left(\mathcal{M}|\vec{x}\right) = N_\mathcal{M}\log(M)-2\log\left[L\left(\vec{x}\,\bigg|\,\vec{\Theta}_{\text{MAP}},\mathcal{M}\right)\pi_o\left(\vec{\Theta}_{\text{MAP}},\mathcal{M}\right)\right],
\end{equation}
where $\vec{\Theta}_\text{MAP}$ is the MAP estimator maximizing the posterior [Eq.~\eqref{eq:MAP}], $M$ is the number of points in the time series, and $N_\mathcal{M}$ is the number of parameters of model $\mathcal{M}$. The BIC thus accounts for the probability of generating the observed time series under each model, weighted by prior information, while penalizing parametric complexity to guard against overfitting.}

\rev{Throughout this work, we have compared models within the same parametric class: one-dimensional Markov processes with three parameters. Therefore, the prefactor in Eq.~\eqref{AP_eq:BIC} penalizing parametric complexity is the same across all models and plays no role in the comparison, and the MAP-optimal model can be interpreted as the most probable model generating the data. As a further simplification, under flat priors the MAP and MLE estimators coincide, so maximizing the posterior reduces to maximizing the likelihood evaluated at the MLE. We therefore compared models by evaluating the likelihood at the MLE, as described in Sections~\ref{sec:distinguishability} and~\ref{sec:data} of the main text.}

\section{Data collapse of distinguishability transition diagram}\label{AP_sec:data_collapse}

Fig.~\ref{fig:model_distinguishability}-(e) shows the distinguishability transition diagram, which provides the probability of distinguishing between competing generative stochastic models. The level curves in the diagram appear to follow smooth functions of $\Delta t$ and $M$. The aim of this section is to analyze the basic properties of these level curves using heuristic arguments.

\begin{figure}
    \centering
\includegraphics{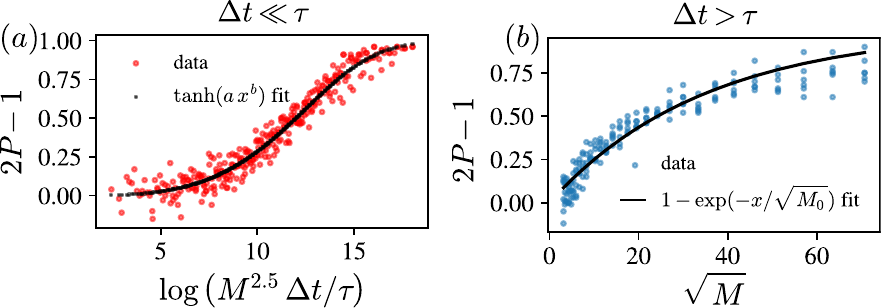}
    \caption{\textbf{Data collapse in distinguishability transition diagram.} (a) In the short-time regime ($\Delta t \ll \tau$), distinguishability is dominated by the total duration $T = \Delta t M$, with data collapsing under the generalized scaling $M^\lambda \Delta t$, with $\lambda=2.5$. (b) In the long-time regime ($\Delta t \gg \tau$), distinguishability is controlled by the stationary likelihood and scales wit $\sqrt{M}$, with a characteristic threshold $M_0 \sim 1000$ measurements required for reliable model differentiation.}
    \label{AP_fig:collapse}
\end{figure}

In the regime $\Delta t \ll\tau$, one would expect the continuous-trajectory inference to be essentially exact. Therefore, the capability of distinguishing between models should be dominated by the total duration $T=\Delta t \,M$. Fig.~\ref{AP_fig:collapse}-(a) partially confirms this reasoning, as we obtain through fitting that the data collapses to a curve with the generalized scaling $M^\lambda\,\Delta t$.

In the regime $\Delta t \gg \tau$, distinguishability is governed by the stationary likelihood of Eq.~\eqref{eq:prob_time_series_uncorrelated}, so the probability of telling models apart should be independent of $\Delta t$. Furthermore, by the central limit theorem, fluctuations in the likelihood scale as $1/\sqrt{M}$. This prediction is confirmed in Fig.~\ref{AP_fig:collapse}-(b): when $\Delta t \gg \tau$ and models differ in their stationary distributions, the probability of distinguishing them grows exponentially with the expected $\sqrt{M}$-scaling. The exponential fit also defines a characteristic scale $M_0 \sim 1000$, which represents the typical number of measurements required to reliably differentiate between models in this regime.

\section{Measures of correlation time}\label{AP_sec:correlation time}

Through the main text, we have emphasized the special role of the correlation time in determining the effectiveness of parametric inference and model distinguishability. In principle, the correlation time can be computed as~\cite{Dubkov2000}
\begin{equation}\label{eq:correlation_time}
    \tau = \int_0^{\infty}C\left(s\right)\,ds,
\end{equation}
where we make use of the stationary autocorrelation function 
\begin{equation}
    C\left(\Delta t\right) =\frac{\E\left(X_{t+\Delta t}\,X_t\right)-\mu^2}{\sigma^2},\,\forall t,
\end{equation}
where $\mu$ and $\sigma^2$ correspond to the mean and variance of the stationary distribution of the process, respectively. The rationale behind Eq.~\eqref{eq:correlation_time} is that the autocorrelation function typically exhibits exponential decay, $C\left(\Delta t\right) \sim \exp\left(-\Delta t/\tau\right)$, as can be derived using spectral theory~\cite{Gardiner2009,Risken1991}. Substituting expected values with sample means and replacing the integral in Eq.~\eqref{eq:correlation_time} with a sum over time delays present in the time series, one can construct an estimator for the correlation time, here denoted by $\hat{\tau}$. Due to the positivity of the correlation function and discretization errors, the correlation time tends to be overestimated, i.e., $\hat{\tau} > \tau$. In particular, the minimum autocorrelation time that can be estimated from real data is of the order of the sampling time,
\begin{equation}
    \hat{\tau} =\sum_{i=0} \hat{C}(i\,\Delta t)\,(i+1)\,\Delta t\ge\Delta t,
\end{equation}
reflecting the intuitive fact that we cannot retrieve information about temporal scales that are not sampled.

\begin{table}[h]
\centering
\begin{tabular}{| c | c | c | c |}
\hline
\textbf{Parameters} & $\tau_\text{EN}$  & $\tau_\text{CP}$ &  $\tau_{\theta=1/2}$  \\
\hline
$K=1$, $\mu=2$, $D=0.2$ & 0.45 & 0.51 & 0.78\\
\hline
$K=0.5$, $\mu=2$, $D=0.2$ & 1.09 & - & -\\
\hline
\end{tabular}
\caption{Measure of autocorrelation time in the contact process environmental, and generalized gamma  models [Eqs.~\eqref{eq:general_model_environmental},~\eqref{eq:CP_model}, and~\eqref{eq:general_model_theta} respectively].}
\label{AP_tab:tau_measure}
\end{table}

\end{document}